%% file: main.tex
\documentclass[dvipsnames,sigconf,pbalance]{acmart}

\input{commands}
\input{preamble}

\begin{document}
\input{sections/abstract}

\maketitle

\input{sections/body/introduction}
\input{sections/body/preliminaries}
\input{sections/body/overview}
\input{sections/body/protocol}
\input{sections/body/evaluation}
\input{sections/body/related}
\input{sections/body/discussion}
\input{sections/body/conclusion}

\begin{acks}
This work was supported in part by Grants from the
\grantsponsor{hk-rgc}{HK RGC}{} CRF Grant
(Project No. \grantnum{hk-rgc}{C5097-25G}) and GRF Grant
(Project No. \grantnum{hk-rgc}{PolyU 15226224}).
\end{acks}

\bibliographystyle{ACM-Reference-Format}
\bibliography{references}

\input{sections/considerations}

\appendix
\input{sections/appendices/full}
\input{sections/appendices/definition}
\input{sections/appendices/sybil}
\input{sections/appendices/security-uc}

\end{document}

%% file: commands.tex
\newtheoremstyle{defs}      
  {}                        
  {}                        
  {\itshape}                
  {}                        
  {\itshape}                
  {.}                       
  { }                       
  {\thmname{#1}\thmnumber{ #2}\thmnote{ (#3)}} 

\newtheoremstyle{thms}      
  {}                        
  {}                        
  {\itshape}                
  {}                        
  {\scshape}                
  {.}                       
  { }                       
  {\thmname{#1}\thmnumber{ #2}\thmnote{ (#3)}} 

\theoremstyle{defs}
\newtheorem{definition}{Definition}

\theoremstyle{thms}
\newtheorem{theorem}{Theorem}

\mathchardef\hph="2D

\newcommand{\issuer}{\ensuremath{\mathcal{I}}}
\newcommand{\holder}{\ensuremath{\mathcal{H}}}
\newcommand{\verifier}{\ensuremath{\mathcal{V}}}
\newcommand{\registry}{\ensuremath{\mathcal{R}}}

\newcommand{\env}{\ensuremath{\mathcal{E}}}
\newcommand{\simulator}{\ensuremath{\mathcal{S}}}
\newcommand{\adv}{\ensuremath{\mathcal{A}}}
\newcommand{\party}{\ensuremath{\mathcal{P}}}

\newcommand{\func}{\ensuremath{\mathcal{F}}}
\newcommand{\funb}{\ensuremath{\func_\sysname}}
\newcommand{\funl}{\ensuremath{\mathcal{L}}}
\newcommand{\prob}{\ensuremath{\Pi_\sysname}}

\newcommand{\exec}{\textsc{exec}}
\newcommand{\ideal}{\textsc{ideal}}
\newcommand{\negl}{\ensuremath{\mathsf{negl}}}
\newcommand{\poly}{\ensuremath{\mathsf{poly}}}

\newcommand{\Zp}{\ensuremath{\mathbb{Z}_p}}
\newcommand{\Fp}{\ensuremath{\mathbb{F}_p}}
\newcommand{\N}{\ensuremath{\mathbb{N}}}
\newcommand{\Gpa}{\ensuremath{\mathbb{G}_1}}
\newcommand{\Gpb}{\ensuremath{\mathbb{G}_2}}
\newcommand{\Pair}{\ensuremath{\mathrm{P}}}

\newcommand{\pk}{\ensuremath{\mathsf{pk}}}
\newcommand{\sk}{\ensuremath{\mathsf{sk}}}
\newcommand{\id}{\ensuremath{\mathsf{id}}}
\newcommand{\aid}{\ensuremath{\mathsf{aid}}}

\newcommand{\sid}{\ensuremath{\mathsf{sid}}}
\newcommand{\cred}{\ensuremath{\mathsf{cred}}}

\newcommand{\idi}{\ensuremath{\id^\issuer}}
\newcommand{\ski}{\ensuremath{\sk^\issuer}}
\newcommand{\pki}{\ensuremath{\pk^\issuer}}
\newcommand{\idh}{\ensuremath{\id^\holder}}
\newcommand{\skh}{\ensuremath{\sk^\holder}}

\newcommand{\skv}{\ensuremath{\sk^\verifier}}
\newcommand{\pkv}{\ensuremath{\pk^\verifier}}

\newcommand{\zk}{\Pi}
\newcommand{\zksetup}{\mathsf{Setup}}
\newcommand{\zkprove}{\mathsf{Prove}}
\newcommand{\zkverify}{\mathsf{Verify}}

\newcommand{\cmt}{\Gamma}
\newcommand{\cmtcommit}{\mathsf{Commit}}
\newcommand{\cmtopen}{\mathsf{Open}}

\newcommand{\mtinsert}{\mathsf{Insert}}
\newcommand{\mtauth}{\mathsf{Auth}}

\newcommand{\pubkeygen}{\mathsf{PKGen}}
\newcommand{\sign}{\mathsf{Sign}}
\newcommand{\versign}{\mathsf{VerSig}}

\newcommand{\relgen}{\mathcal{RG}}

\newcommand{\sample}{\leftarrow_{\$}}

\newcommand{\zkpok}{\mathsf{ZK}\hph\mathsf{PoK}}

\newcommand{\sysname}[0]{\textsc{Braid}\xspace}

\definecolor{beni}          {HTML} {FF6666}     
\definecolor{enji}          {HTML} {D65DB1}     
\definecolor{momo}          {HTML} {FF6F91}     
\definecolor{anzu}          {HTML} {FFB366}     
\definecolor{kogane}        {HTML} {FFC75F}     
\definecolor{ki}            {HTML} {F9F871}     
\definecolor{mizu}          {HTML} {40C7F4}     
\definecolor{ao}            {HTML} {5ED9C4}     
\definecolor{kon}           {HTML} {009F85}     
\definecolor{ruri}          {HTML} {2C73D2}     
\definecolor{usumurazaki}   {HTML} {B39CD0}     
\definecolor{murazaki}      {HTML} {845EC2}     

\newcommand{\protocol}[3][\linewidth]{
    \begin{boxedminipage}[t]{#1}
        \begin{center}
            \small{\textbf{#2}}
        \end{center}
        \footnotesize{#3}
    \end{boxedminipage}
}

\newcommand{\seq}[1]{
    \ensuremath{{\bm #1}}
}

\newcommand{\witness}[1]{
	\ensuremath{
		{\color{violet} #1}
	}
}

\newcommand{\method}[1]{
    {\color{ruri}\textbf{#1}}
}

\newcommand{\spar}[1]{%
    \noindent%
    \textbf{#1.}\xspace%
}

\newcommand{\cirn}[1]{%
    \raisebox{-0.7pt}{\ding{#1}}%
}

\newcommand{\fullcir}{\tikz[baseline=-0.68ex]\draw[color=.,fill=.] (0,0) circle (0.68ex);}
\newcommand{\emptycir}{\tikz[baseline=-0.68ex]\draw[color=.,fill=white] (0,0) circle (0.68ex);}
\newcommand{\halfcir}{\tikz[baseline=-0.68ex]{
    \draw[color=.,fill=white] (0,0) circle (0.68ex);
    \fill[color=.] (0,0.68ex) arc (90:270:0.68ex) -- cycle;
}}

\newsavebox{\badge}
\newsavebox{\blankfile}
\newsavebox{\filledfile}
\newsavebox{\badgedfile}
\newsavebox{\badgedfilledfile}
\newsavebox{\key}

\newsavebox{\justice}
\newsavebox{\holderalice}
\newsavebox{\holderblank}
\newsavebox{\computer}
\newsavebox{\card}
\newsavebox{\balloona}
\newsavebox{\balloonb}
\newsavebox{\balloonc}

%% file: preamble.tex
\setcopyright{cc}
\copyrightyear{2026}
\acmYear{2026}
\setcctype{by-nc-nd}

\acmConference[CCS '26] {Proceedings of the 2026 ACM SIGSAC Conference on Computer and Communications Security}{November 15--19, 2026}{The Hague, Netherlands.}
\acmBooktitle{Proceedings of the 2026 ACM SIGSAC Conference on Computer and Communications Security (CCS '26), November 15--19, 2026, The Hague, Netherlands}
\acmISBN{979-8-4007-2871-6/2026/11}
\acmDOI{10.1145/3830454.3832631}

\title[\sysname: Sybil-Resistant Decentralized Identity]{\sysname: Sybil-Resistant Decentralized Identity with Trustless Key Recovery and Non-Transferable Anonymous Credentials}

\author{Rui Song}
\email{csrsong@comp.polyu.edu.hk}
\affiliation{%
  \department{Department of Computing}
  \institution{The Hong Kong Polytechnic University}
  \city{Kowloon}
  \country{Hong Kong}}

\author{Tianyu Zheng}
\email{tian-yu.zheng@connect.polyu.hk}
\affiliation{%
  \department{Department of Computing}
  \institution{The Hong Kong Polytechnic University}
  \city{Kowloon}
  \country{Hong Kong}}

\author{Shang Gao}
\email{shanggao@polyu.edu.hk}
\affiliation{%
  \department{Department of Computing}
  \institution{The Hong Kong Polytechnic University}
  \city{Kowloon}
  \country{Hong Kong}}

\author{Guyue Li}
\email{guyuelee@seu.edu.cn}
\affiliation{%
  \department{School of Cyber Science and Engineering}
  \institution{Southeast University}
  \city{Nanjing}
  \country{China}}

\author{Xue Liu}
\email{xueliu@cs.mcgill.ca}
\affiliation{%
  \department{School of Computer Science}
  \institution{McGill University}
  \city{Montreal}
  \country{Canada}}

\author{Bin Xiao}
\correspondingauthor
\email{b.xiao@polyu.edu.hk}
\affiliation{%
  \department{Department of Computing}
  \institution{The Hong Kong Polytechnic University}
  \city{Kowloon}
  \country{Hong Kong}}

\renewcommand{\shortauthors}{Rui Song et al.}

\ccsdesc[500]{Security and privacy~Web protocol security}
\ccsdesc[300]{Security and privacy~Web application security}

\keywords{Decentralized identity; Anonymous credentials; Sybil resistance; Trustless key recovery; Zero-knowledge proofs}

\newcommand{\artifacturl}{\url{https://github.com/wurahara/Braid-Impl}}

\usepackage{bm}
\usepackage{tikz}
\usepackage{pgfplots}

\usepackage{amsthm}

\usepackage{amssymb}
\usepackage{amsfonts}

\usepackage{xspace}
\usepackage{xcolor}
\usepackage{pifont}

\usepackage{multirow}
\usepackage{booktabs}

\usepackage{graphicx}
\usepackage{textcomp}
\usepackage{enumitem}
\usepackage{boxedminipage}
\usepackage{threeparttable}

\pgfplotsset{compat=1.18}
\usetikzlibrary{shadows}
\usetikzlibrary{arrows}

\allowdisplaybreaks[4]


%% file: sections/abstract.tex
\begin{abstract}
  Decentralized identity frameworks grant users full sovereignty over their digital assets in the Web3 ecosystem. However, allowing arbitrary creation of identifiers makes the system susceptible to Sybil attacks and puts assets at risk when keys are lost or compromised. Moreover, the lack of identification prevents anonymous credential schemes from deterring malicious transfers. While existing solutions attempt to address these issues by linking identifiers to entities through trusted intermediaries, these entities are not always accessible and require costly offline interactions.

  In this work, we introduce \sysname, a decentralized identity system with Sybil resistance, trustless key recovery, and non-transferable anonymous credentials.\@ \sysname{} creates blockchain-based bindings between identifiers and gradually combines identifiers belonging to the same holder into a unified associated identifier. As all identifiers within an association are presumed to belong to one individual, any fraudulent activity can be detected. The association grows larger as interactions increase, substantially reducing the likelihood of successful Sybil attacks. This mechanism allows holders to recover identifiers with lost keys or partial key compromise by proving knowledge of specific association structures. Additionally, \sysname{} prevents unauthorized transfers through blockchain-based identifier-key bindings and proofs of ownership for credentials.
  Compared with representative prior approaches that provide only campaign-scoped Sybil protection or stateless anonymous-credential presentation, \sysname{} carries association state across applications, enabling progressive Sybil resistance and trustless recovery while preserving zero-knowledge credential presentation.
  The evaluation shows that \sysname{} effectively bridges the gap between privacy-preserving credentials and accountable identity, achieving identifier association and credential presentation times of 2.41s and 3.31s on consumer-grade devices.
\end{abstract}

%% file: sections/body/introduction.tex
\section{Introduction}\label{sec: introduction}

Cyberattacks on centralized identity systems have led to serious data breaches, resulting in financial losses and increased costs for enterprises and governments. Decentralized identity, integral to Web3, leverages decentralized ledgers and verifiable credentials to promise users unprecedented sovereignty and privacy~\cite{world2022decentralized, world2023verifiable}. This vision enables participation in on-chain activities like decentralized autonomous organization (DAO) voting or token airdrops with minimal data disclosure, seemingly eliminating the risks of identity tracking and data breaches common in centralized systems~\cite{wang2019decentralized}.

However, this promising vision is fundamentally undermined by three persistent and interconnected challenges that severely limit its practical application. The first challenge is the foundational threat of the \textit{Sybil attack}. The very feature that enhances privacy—the ability to create multiple identifiers—creates a fertile ground for malicious actors. A single attacker can masquerade as thousands, illegitimately draining airdrops intended for a community or seizing control of a DAO's treasury, thus corrupting the core principle of fair, decentralized governance~\cite{fan2023altruistic, nabben2021decentralized}.

The second challenge comes from the brittle nature of \textit{self-sovereign key recovery}. Compared to centralized systems, user-created keypairs are the sole gateway to digital assets. This absolute control becomes an absolute vulnerability upon key loss. The consequences are dire: an estimated \$140 billion in digital assets lie stranded in inaccessible wallets, a permanent testament to the unforgiving nature of this challenge~\cite{nyt21keyloss}.

The third challenge arises from the inherent contradiction between \textit{anonymity} and \textit{non-transferability}. Anonymous credentials are vital for privacy, shielding users from collusion and surveillance~\cite{rosenberg2023zk, doerner2023threshold}. However, this anonymity makes them transferable, creating a bustling black market for stolen credentials, evidenced by the 80 million sold on the Genesis marketplace alone~\cite{caliendo2023breaches, specops2023genesis}. Enforcing non-transferability has traditionally required sacrificing the very anonymity that makes these credentials powerful, forcing developers into a security paradigm with no good choices.

\subsection{Prior solutions and limitations}\label{subsec: introduction-prior-solutions}

Several solutions have been developed to tackle the challenges mentioned above while exhibiting certain limitations.

\spar{Sybil-resistant identification}
Sybil attacks in decentralized systems cannot be eliminated entirely—they can only be deterred by making attack costs exceed potential gains~\cite{douceur2002sybil}. In Web3 systems, offline gatherings serve as a prominent anti-Sybil strategy. Techniques such as Proof-of-Personhood (PoP)~\cite{borge2017proof} and Worldcoin's iris scanning approach~\cite{gent2023cryptocurrency} require physical presence for identification. However, this approach faces major scalability challenges, increases ordinary user costs significantly, and remains vulnerable to low-cost attacks\footnote{An attack known as \textit{Puppeteer Attack} occurs when an attacker pays someone, typically for as little as \$10, to undergo identification on its behalf~\cite{RoboTeddy2021humanity}.}. Collateralization, another traditional anti-Sybil method, creates high entry barriers while remaining susceptible to circumvention through token lending or user collusion—a vulnerability particularly evident in quadratic voting~\cite{platt2021sybil, ford2020identity, austgen2023dao, brenzikofer2023quadratic}.

One emerging strategy involves migrating legacy profiles~\cite{khalsa2022holonym} or natural unique identifiers~\cite{maram2021candid}, such as the U.S. Social Security Number (SSN). However, the inherent scope limitations of legacy systems restrict this approach's ability to serve all potential users. A related approach evaluates existing credentials to determine identifier trustworthiness~\cite{gitcoin2022streaming}. While improving usability, this approach requires reassessment during each verification process to maintain privacy protection. We envision a progressive anti-Sybil solution that can both accumulate and carry forward the identifier trustworthiness demonstrated during each verification.

\spar{Key recovery}
Social recovery has emerged as a prevalent remedial approach, enabling users to authorize a designated group of guardians to facilitate key recovery or modification in the event of key loss~\cite{he2018social, liu2020design, weyl2022decentralized}. However, this strategy encounters challenges pertaining to the necessity of trust in the guardians and the requirement for offline verifications to authenticate the user's identity. Keys can be recovered without the owners' consent if these guardians collude~\cite{camenisch2012practical, jarecki2014round}.

\spar{Credential verification with identity privacy}
The challenge of providing non-transferable yet anonymous credentials remains unsolved, forcing existing approaches to trade off between these features~\cite{kakvi2023sok}. Some methods sacrifice anonymity by requiring credential holders to reveal their identity during presentation~\cite{world2023verifiable,weyl2022decentralized}. Other approaches, focusing on anonymous credentials, abandon non-transferability and instead limit the number of presentations to prevent unauthorized use~\cite{camenisch2017practical, sonnino2019coconut, rathee2022zebra}.

\subsection{Our contributions}\label{subsec: introduction-our-contributions}

In this paper, we introduce a comprehensive decentralized identity system entitled ``\textit{\underline{B}inding and Agg\underline{r}egating \underline{A}ssociated \underline{Id}entifiers} (\sysname).'' This system is designed to enable \textit{progressive Sybil resistance} and \textit{trustless key recovery}, while maintaining \textit{non-transferability} and \textit{anonymity} of verifiable credentials.

\sysname{} establishes bindings between identifiers on the blockchain and progressively aggregates those belonging to the same holder into an associated identifier. Essentially, it strikes a balance between identity privacy and accountability. For legitimate holders, \sysname{} records their associated identifiers on the blockchain without exposing the private information of individual identifiers. It also offers a way for these users to prove ownership of individual identifiers through the associated identifier. Conversely, for malicious holders, \sysname{} stops malicious users from unfairly benefiting by forging multiple identities or transferring anonymous credentials, thanks to its binding and aggregating processes. In this way, \sysname{} effectively addresses the limitations of previous approaches and resolves the three challenges above.

\spar{Progressive Sybil resistance}
\sysname{} leverages a pivotal observation that holders often need to present multiple credentials to participate in a campaign~\cite{weyl2022decentralized}. For instance, an airdropper might require holders to provide at least three participation credentials before qualifying for a new airdrop~\cite{fan2023altruistic}.
\sysname{} achieves Sybil resistance through progressive binding of correlated identifiers. When users present multiple credentials at once, \sysname{} enforces the aggregation of corresponding identifiers into an indivisible association, which is submitted to the blockchain in a cumulative manner. Given that all identifiers within an association are assumed to belong to the same individual, any attempt at duplicate participation within that association can be effectively detected. Theoretical analysis and experiments demonstrate that, as interactions increase, associations will expand progressively while the number of manipulable credentials remains limited, thereby substantially reducing the chance of successful Sybil attacks. Notably, \sysname{} is carefully designed to handle cases where malicious holders try to deliberately spread a single identifier across multiple associations.
This persistent cross-context linkage necessitates a strategic trade-off: \sysname{} relaxes the strict global unlinkability of traditional anonymous credentials into \textit{contextual unlinkability}. By prioritizing cross-context Sybil resistance over absolute isolation when public linkability handles are reused, \sysname{} provides a practical security design for Web3 applications where Sybil resistance is critical, such as DAO governance and ecosystem-wide airdrops.

\spar{Trustless key recovery}
Traditional social recovery mechanisms depend on trust relationships, using secret sharing between custodians to reconstruct keys after loss. In contrast, \sysname{} provides \textit{trustless} key recovery through its identifier association mechanism.
\sysname{} implements two approaches for reliably identifying genuine identifier owners. Holders can demonstrate ownership of any identifier in an association by proving they have the private keys of other identifiers within that association.
Alternatively, when associated identifiers provide sufficient information entropy to meet security requirements, holders can prove ownership by demonstrating their knowledge of the hidden composition of the association without needing to memorize specific keys.

\spar{Non-transferable anonymous credential}
The success of \sysname{} hinges on the authenticity of credentials. In scenarios where identity privacy is crucial, credentials are often anonymous and vulnerable to transfer by malicious holders. To tackle this issue, \sysname{} requires holders to bind their identifiers and private keys on the blockchain. When presenting anonymous credentials, holders must prove to verifiers that the credentials' holder identifiers align with those bound on the blockchain. As only the genuine holder possesses the private key and the proof is zero-knowledge, \sysname{} can assert the credential owner without compromising identity privacy, preventing the transfer of anonymous credentials. Notably, these benefits do not require any modifications to existing credentials.

\spar{Extensions}
Beyond the contributions mentioned above, \sysname{} implements credential revocation and identifier blocklisting to enforce accountability for malicious behaviors.
\sysname{} extends tracking and penalties beyond just the offending identifier to include all associated identifiers controlled by malicious users. As a bonus feature, \sysname{} empowers credential holders to designate specific verifiers, ensuring proofs work only for designated verifiers, thereby preventing malicious proof forwarding and impersonation attacks.

\spar{Contributions}
To summarize, we design, implement, and evaluate \sysname{}, which offers significant enhancements over pure anonymous credentials and existing decentralized identity solutions. Our key contributions include:
\begin{itemize}
	\item A practical anti-Sybil mechanism named \textit{identifier association} significantly raises Sybil attack costs while maintaining low user overhead and eliminating the need for external identity profiles, offline gatherings, or collateral.
	\item A \textit{trustless key recovery} mechanism that enables holders to regain control over specific identifiers with lost keys without relying on trusted parties or authorities.
	\item \textit{Non-transferable anonymous credentials} that authenticate credential ownership while protecting identity privacy.
	\item \textit{Identifier blocklisting} and \textit{credential revocation} for enhanced accountability, along with \textit{non-forwardable proofs} to prevent replay attacks of designated verifier proofs.
	\item Security analysis and functionality evaluation, demonstrating \sysname's progressive Sybil resistance and better performance compared to baseline identity and credential schemes.
\end{itemize}

%% file: sections/body/preliminaries.tex
\section{Preliminaries}\label{sec: preliminaries}

\spar{Notation}
We denote by $\lambda\in\N$ the security parameter and by $1^\lambda$ its unary representation. Unless otherwise stated, we assume that all cryptographic algorithms in this paper take $1^\lambda$ as an input and thus omit it. If $S$ is a finite set, we denote by $x\leftarrow S$ the process of sampling $x$ according to $S$ and by $x\sample S$ a uniform one. We denote by $\negl(\lambda)$ a \textit{negligible} function with respect to $\lambda$. We denote by $[\ell]$ the non-negative integer series up to $\ell$, e.g., $\{1,2,\dots,\ell\}$. We say an algorithm is \textit{probabilistic polynomial time (p.p.t.)} if it is probabilistic and operates within polynomial time with respect to $\lambda$.

In this paper, we use two collision-resistant hash functions $H_a, H_n$ which can be modeled as random oracles. These hash functions can be constructed from a single primitive using proper domain separation. We also employ a key derivation algorithm $\pubkeygen$ whose one-wayness ensures that the public key does not divulge any knowledge about the private key.

\spar{Non-interactive zero knowledge}
Let $\relgen$ be a relation generator that takes a security parameter $\lambda$ and returns a binary relation $R$. For a pair $(x,w)$ such that $R(x,w)=1$, we call $x$ the \textit{statement} and $w$ the \textit{witness}. Given $1^\lambda$, we define the set of all relations that $\relgen$ may output as $\relgen_\lambda$.
Using the current standard notation~\cite{camenisch1997efficient}, we denote a proof for relation $R$ as:
\[
	\pi=\zkpok\{\witness{w}:R(x,\witness{w})=1\}.
\]
For clarity, we will present only the relations $R(x,\witness{w})$ in protocols of Section~\ref{sec: system}, with the witness element(s) $\witness{w}$ highlighted to enhance understanding.

\spar{Cryptographic commitment}
A commitment scheme is a pair of polynomial-time algorithms $(\cmtcommit,\cmtopen)$ which performs as:
\begin{itemize}
	\item $(c,r)\sample\cmtcommit(x)$: takes as input a message $x$, and outputs its commitment $c$ and a randomness $r$.
	\item $b\gets\cmtopen(x,c,r)$: takes as inputs a message $x$, a commitment $c$ and a randomness $r$, and outputs $b=1$ for a valid opening or $b=0$ otherwise.
\end{itemize}

\spar{Incremental Merkle tree}
Merkle tree is a vector commitment scheme that commits to a vector $\bm m=(m_1,\dots,m_\ell)$ and opens it at arbitrary index $i\in[\ell]$. An incremental Merkle tree is a constant-depth tree with placeholder leaves and pre-computed intermediate nodes. Elements are inserted sequentially by replacing the placeholders. An incremental Merkle tree could be abstracted as a tuple of polynomial-time algorithms $(\mtinsert,\mtauth)$ which performs as:
\begin{itemize}
	\item $(\bm m',r,\rho)\leftarrow\mtinsert(x,\bm m)$: takes as inputs a new leaf $x$ and $\bm m$, returning a new vector $\bm m'=(m_1,\dots,m_\ell,x)$, the updated root $r$ and a inclusion proof $\rho$ for $x$.
	\item $b\leftarrow\mtauth(r,x,\rho)$: takes as inputs a leaf $x$, a Merkle root $r$ and a proof $\rho$, returning $b=1$ for a valid tuple $(r,x,\rho)$.
\end{itemize}

\spar{Universally composable security}
In this work, we employ the Universal Composability (UC) framework for security definition and analysis~\cite{canetti2002universally}. This framework involves a protocol $\Pi$ interacting with an \textit{adversary} $\adv$ and an \textit{environment} $\env$ with initial input $z$. $\adv$ can send backdoor messages to parties, transmit outputs to $\env$, read outgoing messages from parties, and deliver messages between parties. When $\adv$ corrupts a party, it gains full access to its messages and internal states.

The security of $\Pi$ is defined by comparing its execution in the \textit{real world} to an \textit{ideal process}. The key ingredient in the ideal process is the \textit{ideal functionality} $\func$, which captures the desired specifications and security properties of the task at hand. $\func$ is a Turing machine that interacts with $\env$ and an ideal adversary (\textit{simulator}) $\simulator$ through a set of \textit{dummy parties}. When $\simulator$ is activated, it can send messages to $\func$. $\simulator$ is also responsible for delivering messages from $\func$ to parties.

We say a protocol $\Pi$ \textit{UC-realizes} the functionality $\func$ if for any adversary $\adv$, there exists a simulator $\simulator$ such that no environment $\env$ can distinguish with non-negligible probability whether it is interacting with $\adv$ and $\Pi$ or with $\simulator$ and $\func$.

%% file: sections/body/overview.tex
\section{Overview}\label{sec: overview}

In this section, we start by introducing the system and threat models of \sysname. Subsequently, we introduce the foundational concepts behind \sysname, followed by its security and privacy objectives.

\subsection{System and threat models}

\spar{System model}
\sysname{} considers a decentralized identity system composed of \textit{issuers} $\issuer$, \textit{holders} $\holder$, and \textit{verifiers} $\verifier$. To participate in \sysname, \textit{holders} create a key pair $(\pk,\sk)$ and publish the public key $\pk$ to a \sysname-specific \textit{verifiable data registry (VDR)} $\registry$ maintained by a decentralized ledger (e.g., a blockchain) for registration. The VDR then generates a DID document that encompasses $\pk$ and an identifier\footnote{All \textit{identifiers} in this paper are decentralized identifiers (DIDs) that conform to the W3C specification~\cite{world2022decentralized}, unless otherwise indicated.} $\id$ that resolves to it. Crucially, \textit{issuers} are not required to formally register within the \sysname{} system\footnote{For clarity, we record issuer identifiers and keys in \sysname's VDR in this manuscript. In practice, they may be maintained in any external registry accessible to all parties.}; they simply publish their public keys to any standard W3C-compliant registry. This allows verifiers to natively authenticate credentials without imposing adoption barriers on legacy issuers.

A \textit{verifiable credential} $\cred=(\idi,\idh,\sigma,s)$ contains a collection of claims $s$ issued by some $\issuer$ to certain $\holder$, who can then present it to some $\verifier$ for authentication, as shown in Figure~\ref{fig: overview}. In accordance with the W3C specification~\cite{world2023verifiable}, $\cred$ contains identifiers $\idi$ and $\idh$ of its issuer $\issuer$ and holder $\holder$, as well as a signature $\sigma$ generated by signing other portions of $\cred$ using $\issuer$'s private key $\ski$. A \textit{campaign} is an interaction event (e.g., token airdrop) initiated by a verifier and identified by a \textit{session ID} $\sid$. Each participant can take part \textit{only once} in a campaign.

\spar{Threat model}
Our protocol considers malicious \textit{holders} trying to bypass verification without possessing valid credentials or sharing credentials with other parties. Holders are allowed to register and control multiple identifiers. Based on rational assumptions, holders will not share private keys with others.

\textit{Issuers}, while executing the protocol honestly, maintain records of all issued credentials and may collude with verifiers to track holder activities and deduce their identities. Issuers will not collude with holders to issue fraudulent credentials.
\textit{Verifiers} are completely untrusted, as they may collude with issuers to obtain credential claims or holder identities, and may even forward holder proofs to impersonate their identities.
However, we assume verifiers are economically motivated to prevent Sybil attacks to ensure the fairness and integrity of their own campaigns.

The \textit{VDR} operates through blockchain smart contracts, with security based on standard permissionless blockchain assumptions where all states remain public and immutable. The VDR provides consistent observations to all observers at any moment. We also assume that network communications between parties remain protected from third parties and that all states except for the VDR are secure and private.

    \spar{Behavioral assumptions}
    The security of \sysname's cross-context Sybil resistance and non-transferability relies on rational behavioral assumptions driven by economic incentives in high-stakes environments. Specifically, we assume:
    \begin{itemize}[leftmargin=*]
        \item \textit{Credential scarcity:} Acquiring valuable, issuer-verified credentials requires significant capital, effort, or sustained contributions, whereas creating blank identifiers is costless.
        \item \textit{Joint presentation:} Since individual credentials rarely match complex application requirements, verifiers naturally require the joint presentation of multiple credentials, forcing attackers to acquire distinct credential sets for each Sybil identity.
        \item \textit{Incentive to reuse associations:} Credentials are bound to specific identifiers at issuance. Once associated, these identifiers are cryptographically locked. Creating new identifiers yields only credential-free blank slates. Thus, users must reuse and expand their existing associations.
        \item \textit{Incentive against key sharing:} As in any public-key infrastructure, digital asset security relies on private key secrecy. Rational users will not share their keys, as doing so equates to relinquishing control over their assets.
    \end{itemize}

\subsection{Design concepts}\label{subsec: motivation}

\begin{figure}[t]
    \centering
    \hspace{-25pt}
    \resizebox{0.92\columnwidth}{!}{\input{floats/diagrams/overview-flow}}
    \caption{The flow of credential issuance and presentation.}
    \label{fig: overview}
    \Description{A diagram showing the flow of credential issuance and presentation, highlighting the key steps and interactions.}
\end{figure}

To illustrate the core concepts of \sysname, we present a running example featuring Alice, an active participant in the Web3 ecosystem. Alice wants to join \textit{FutureDAO}, a high-value DAO.\@
Through this example, we will show how \sysname{} provides three key benefits: \textit{Sybil resistance}, \textit{anonymous but non-transferable presentation}, and \textit{trustless key recovery}.

\subsubsection{The challenge: Sybil attacks in high-stakes DAOs}

\textit{FutureDAO} governs a substantial treasury, making its voting process vulnerable to Sybil attacks. A simple \textit{one-address-one-vote} model proves inadequate since attackers could create thousands of addresses at a negligible cost. To ensure fair governance, \textit{FutureDAO} establishes a more robust policy: a member must prove it is a multifaceted contributor by presenting at least two distinct, valuable credentials. For instance, they might require a \textit{Developer Credential} issued after verification of significant code contributions and a \textit{Community Credential} awarded for sustained, active participation.

This policy's effectiveness stems from two synergistic principles. \textit{Credential scarcity} dictates that valuable credentials from reputable third parties are hard to acquire, unlike cheap identifiers~\cite{gitcoin2022streaming, khalsa2022holonym}. Simultaneously, the trend of \textit{joint presentation} requires users to prove multifaceted qualifications in high-stakes applications~\cite{world2023verifiable, khovratovich2017sovrin, naik2020uport}. This creates a prohibitively expensive dual challenge for attackers, who must acquire not just single credentials, but distinct pairs for each Sybil identity.

This exposes a limit of traditional privacy-preserving schemes. Even an attacker who legitimately obtained one credential pair could use these credentials separately across different campaigns to build reputation, violating the principle of unified identity. This specific vulnerability is what \sysname{} is engineered to address.

\subsubsection{\sysname's approach: enforcing identity association}

\begin{figure}[t]
    \centering
    \resizebox{0.97\columnwidth}{!}{\input{floats/diagrams/identifier-association}}
    \caption{An example of the identifier association process.}
    \label{fig: identifier-association}
    \Description{A diagram that indicates the basic principle and flow of identifier association in \sysname.}
\end{figure}

As shown in Figure~\ref{fig: identifier-association}, Alice holds a \textit{Developer Credential} $\cred_1$ associated with the identifier $\id_1$ and a \textit{Community Credential} $\cred_2$ associated with $\id_2$. To vote in \textit{FutureDAO}, she must present both. This is where \sysname{} comes into play.

\spar{Identifier association}
Before voting at \textit{FutureDAO} in \textit{Presentation \cirn{172}}, \sysname{} requires Alice to cryptographically associate $\id_1$ and $\id_2$ on the blockchain, forming a unified \textit{associated identifier}. She computes $\aid := Hash(id_1, id_2, nonce)$ and registers the associated identifier $\aid$ to the VDR $\registry$. This operation remains anonymous—no one knows that $\aid$ belongs to Alice—yet it creates an immutable, publicly verifiable record proving that the owner of $\id_1$ is the same as that of $\id_2$.

\spar{Cross-context Sybil resistance}
This mechanism progressively limits Sybil attacks by creating permanent linkages. Once Alice's identifiers are associated, they remain permanently bound together across the \sysname{} ecosystem. If another application, like a DeFi protocol, also requires Alice to present credentials, she must integrate these new credential identifiers into her existing $aid$. Figure~\ref{fig: identifier-association} illustrates this scenario where Alice interacts with a DeFi app \textit{InsuraFi} and presents credentials $\cred_3$ and $\cred_4$ from identifiers $\id_1$ and $\id_3$ (\textit{Presentation \cirn{173}}). Since $\id_1$ is already part of $\aid$, Alice must add $\id_3$ to her existing $\aid$ before proceeding.

Identifier association creates unprecedented cross-context Sybil resistance by preventing exploitation of information silos between campaigns. While attackers can generate unlimited identifiers, these are merely valueless \textit{blank slates}—the truly scarce resource is \textit{distinct sets of valuable credentials}.
\sysname's forced consolidation leverages this scarcity, as demonstrated in \textit{Presentation \cirn{174}}, where the attempt to reclaim airdrop with a different identifier $\id_2$ is immediately detected through established cross-context linkage. This rapidly depletes an attacker's limited pool of valuable identities across multiple campaigns, making sustained Sybil attacks exponentially more expensive.

\subsubsection{Anonymous presentation with non-transferability}

Anonymity and non-transferability present a fundamental conflict. Verifying legitimate ownership becomes difficult when credentials are anonymous, creating opportunities for unauthorized transfers.
\sysname{} resolves this by allowing Alice to prove ownership without revealing the associated identifier $\aid$ or any individual identifier $\id$.

Instead of presenting identifying information, Alice generates a zero-knowledge proof. This proof attests to \textit{FutureDAO} that she possesses the secret key corresponding to the identifier to which her credential was issued. This is possible because a cryptographic commitment, binding her identifier to her control over the secret key, was recorded on-chain during the identifier registration process. By demonstrating knowledge of this secret key in an entirely anonymous fashion, \sysname{} ensures that the ability to present the credential is non-transferable. This achieves robust non-transferability without compromising her privacy.

\subsubsection{A safety net: trustless key recovery}
Months later, Alice loses her $\id_1$ private key. In traditional systems, this would permanently erase her developer reputation and associated assets.
\sysname{} provides a safety net by shifting from \textit{atomic secrets} (single seed phrases) to \textit{structural knowledge} of her associated identity, offering two complementary recovery mechanisms.

The primary mechanism, \textit{threshold key-based recovery}, allows Alice to recover $\id_1$ by proving threshold control of other identifiers within $\aid$ (e.g., private keys of $\id_2$ and $\id_3$). This mitigates single points of failure, as one compromised key cannot compromise her entire association.

For users with complex digital footprints, \textit{entropy-based recovery} leverages the hidden composition of their associated identifiers as a high-entropy secret. Recovery becomes \textit{reconstruction} rather than \textit{memorization}, where Alice can index public on-chain records to prove knowledge of her $\aid$ structure without depending on any specific key.

These methods offer a strategic trade-off: one serves as the universally applicable mechanism with strong security, while the other provides a key-agnostic alternative for advanced users. Both approaches are \textit{trustless}, eliminating reliance on trusted third parties or centralized intermediaries.

\subsection{Functionality}

We employ an ideal functionality $\funb$ to provide a comprehensive overview of \sysname, as shown in Figure~\ref{fig: functionality-braid}.

\begin{figure}[t]
    \centering
    \input{floats/frames/function-braid}
    \caption{The ideal functionality for \sysname.}
    \label{fig: functionality-braid}
    \Description{The ideal functionality of \sysname, containing four operations: identifier generation, identifier association, credential presentation, and key recovery.}
\end{figure}

\begin{itemize}[leftmargin=*]
    \item In the \textit{identifier registration} phase, a holder $\holder$ requests to generate a new identifier $\id$, along with its keypair $(\pk,\sk)$.
    \item In the \textit{identifier association} phase, $\party$ requests to associate multiple identifiers $\id_1,\dots,\id_\ell$ belonging to it by hashing them together with a nonce $u_a$ initially set to $0$. The resulting associated identifier $\aid$ is recorded for later use, and all member $\id$'s are marked as \textit{associated}, preventing their inclusion in other associations.
          \item In the \textit{association update} phase, $\party$ requests to securely modify an existing association without revealing its structure. To concisely capture this dynamic lifecycle, $\funb$ encompasses three specific methods for this phase: \textit{Association Appending} (adding a new identifier to $\aid$), \textit{Association Merging} (combining two existing associations), and \textit{Association Refreshing} (incrementing the nonce $u_a$ to generate a new $\aid$, proactively breaking nullifier-based correlations across different campaigns).
    \item In the \textit{credential presentation} phase, a holder $\holder$ presents several credentials $\cred_1,\dots,\cred_n$ to a verifier $\verifier$ in campaign $\sid$. The validity of each credential is checked individually to ensure that it indeed meets the verification criteria $\phi$ specified by $\verifier$ and has not been revoked. To prevent reuse of an association in $\sid$, all holder identifiers $\idh$ of credentials should be associated within the same $\aid$, and the exact association state $\{\id_1,\dots,\id_\ell\}$ is recorded to ensure that this state appears only once in $\sid$.
    \item In the \textit{key recovery} phase, $\party$ requests to update the keypair of an $\id$ with a lost key by proving its knowledge about $\id$ and a certain association $\aid$.
\end{itemize}

\begin{figure}[t]
    \centering
    \input{floats/frames/function-ledger}
    \caption{The ledger functionality $\funl$.}
    \label{fig: functionality-ledger}
    \Description{The ledger functionality $\funl$ abstracting the operations of the decentralized ledger (e.g., the blockchain). It contains five operations: register, record, deprecate, retrieve, and check.}
\end{figure}

\spar{Decentralized ledger}
Our protocol requires processing and maintaining states such as participant identifiers and public keys. To capture the properties and interfaces of the decentralized ledger, we incorporate a functionality $\funl$ as shown in Figure~\ref{fig: functionality-ledger}. Throughout the protocol's operation, invocations to the decentralized ledger are modeled as interactions with $\funl$, which UC-realizes the functionalities for updating and querying on-chain states.

\subsection{Security properties}\label{subsec: security-properties}

Our ideal functionality $\funb$ clearly ensures the following security properties. As our \sysname{} protocol implements this ideal functionality, it naturally achieves these same security properties in the real world.

\begin{itemize}[leftmargin=*]
    \item \textit{Credential privacy}: $\verifier$ cannot learn anything beyond the fact that the claim $s$ in $\cred$ satisfies $\phi(s)=1$.
    \item \textit{Identifier privacy}: $\verifier$ cannot ascertain any individual identifier $\id$ that constitutes a specific associated identifier $\aid$.
    \item \textit{Contextual Unlinkability}: When $\holder$ presents the same underlying credential multiple times \textit{after refreshing the relevant association and credential handles ($n_a$ and $n_i^c$)}, $\verifier$ cannot link these presentations from the protocol transcript even if it colludes with other malicious verifiers or issuers.
    \item \textit{Sybil-resistance}: $\holder$ cannot present credentials multiple times under a campaign $\sid$, given the presented credentials belong to the same associated identifier $\aid$.
    \item \textit{Non-transferability}: $\holder$ cannot transfer its credential to another $\holder'$ and enable the latter to legally present it.
    \item \textit{Key recovery security}: a malicious $\holder'$ cannot impersonate $\holder$ and modify the public key of an identifier that doesn't belong to it.
\end{itemize}

\spar{Definition of security}
Consider a protocol $\prob$ with access to the decentralized ledger functionality $\funl$. The output of an environment $\env$ interacting with $\prob$ and an adversary $\adv$ on the security parameter $\lambda\in\N$ and input $z\in\{0,1\}^*$ is denoted as $\exec_{\prob,\adv,\env}^{\funl}(\lambda,z)$. On the other hand, the output of $\env$ interacting with $\funb$ and an ideal-world adversary $\simulator$ is denoted as $\ideal_{\funb,\simulator,\env}(\lambda,z)$. Then, we define the security of $\sysname$, whose protocol is detailed in Section~\ref{sec: system}.

\begin{definition}\label{def: protocol-security}
    Let $\lambda\in\N$ be a security parameter, and $\prob$ be a protocol under the $\funl$-hybrid model. We say $\prob$ UC-realizes $\funb$ under the $\funl$-hybrid model if for all p.p.t. adversary $\adv$ interacting with $\prob$, there exists a p.p.t. simulator $\simulator$ interacting with $\funb$, such that
    \[
        \exec_{\prob,\adv,\env}^{\funl}(\lambda,z)
        \approx
        \ideal_{\funb,\simulator,\env}(\lambda,z)
    \]
    holds for all p.p.t. environments $\env$ and for all $z\in\{0,1\}^*$.
\end{definition}

%% file: floats/diagrams/overview-flow.tex

\savebox{\badge}{
	\tikzset{
		badge upper/.style = {
			fill = kogane!80,
			draw = kogane!80,
			line width = 0pt,
		},
		badge slides/.style = {
			fill = momo,
			draw = momo,
			line width = 0pt,
		},
		blank/.style = {
			opacity = 1,
			fill = white,
			draw = white,
			line width = 0pt,
		},
	}
	\begin{tikzpicture}[x = 0.75pt, y = 0.75pt, yscale = -1, xscale = 1]
		\draw[badge upper] (326.2, 46.3) circle (6.5);
		\draw[blank] (326.2, 46.3) circle (4.5);
		\draw[badge upper] (326.2, 46.3) circle (3.8);
		\draw[badge slides]
			(322.87,52.4) .. controls (323.99,53.0) and (325.03,53.28) .. 
			(326.5,53.23) .. controls (325.92,55.51) and (325.4,57.75) .. 
			(324.85,59.77) .. controls (324.13,59.08) and (323.86,58.84) .. 
			(323.29,58.26) .. controls (322.4,58.45) and (322.04,58.52) .. 
			(321.22,58.69) .. controls (321.72,56.72) and (322.06,55.38) .. 
			(322.87,52.4) -- cycle ;
		\draw[badge slides]
			(326.38,55) .. controls (326.55,54.35) and (326.66,53.93) .. 
			(326.82,53.23) .. controls (327.68,53.1) and (328.6,52.87) .. 
			(329.55,52.32) .. controls (330,54.12) and (330.71,56.66) .. 
			(331.24,58.53) .. controls (330.28,58.32) and (329.79,58.22) .. 
			(329.17,58.12) .. controls (328.59,58.63) and (328.21,59.01) .. 
			(327.56,59.54) .. controls (327.15,57.86) and (326.74,56.47) .. 
			(326.38,55) -- cycle ;
	\end{tikzpicture}
}

\savebox{\blankfile}{
	\tikzset{
		credential outline/.style = {
        		draw = murazaki,
    	},
    	credential/.style = {
        		fill = usumurazaki,
        		drop shadow = {%
            		opacity = 0.8,
            		shadow xshift = 1.2,
            		shadow yshift = -1.2,
        		},
    	},
	}
    \begin{tikzpicture}[x = 0.75pt, y = 0.75pt, yscale = -1, xscale = 1]
        \draw[credential, credential outline] (-1,-3.8) -- (3.8, 1) -- (3.8, 15) -- (-11, 15) -- (-11, -3.8) -- cycle;
        \draw[credential outline] (3.8, 1) -- (-1, 1) -- (-1, -3.8);
    \end{tikzpicture}
}

\savebox{\filledfile}{
	\tikzset{
		items/.style = {
        		draw = white,
        		fill = white,
        		line width = 0pt,
    	},
	}
	\begin{tikzpicture}[x = 0.75pt, y = 0.75pt, yscale = -1, xscale = 1]
		\node at (0, 0) { \usebox{\blankfile} };
		\draw[items] (-4, -2.5) rectangle (7, -1.6);
		\draw[items] (-4, 0.1) rectangle (7, 1.0);
		\draw[items] (-4, 2.7) rectangle (7, 3.6);
		\draw[items] (-4, 5.3) rectangle (7, 6.2);
	\end
	{tikzpicture}
}

\savebox{\badgedfile}{
	\begin{tikzpicture}[x = 0.75pt, y = 0.75pt, yscale = -1, xscale = 1]
		\node at (0, 0) { \usebox{\blankfile} };
		\node[scale = 0.7] at (8, 6) { \usebox{\badge} };
	\end{tikzpicture}
}

\savebox{\badgedfilledfile}{
	\begin{tikzpicture}[x = 0.75pt, y = 0.75pt, yscale = -1, xscale = 1]
		\node at (0, 0) { \usebox{\filledfile} };
		\node[scale = 0.7] at (10, 6) { \usebox{\badge} };
	\end{tikzpicture}
}

\savebox{\justice}{
	\tikzset{
		issuer stage/.style = {
        		draw = kon,
        		fill = ao,
    	},
    	issuer pillar/.style = {
        		draw = kon,
        		fill = ao!30,
    	},
	}
    \begin{tikzpicture}[x = 0.75pt, y = 0.75pt, yscale = -1, xscale = 1]
        \draw[issuer stage]
            (0, 0) -- (22, 10) -- (22, 13) -- 
            (-22, 13) -- (-22, 10) -- cycle ;

        \draw[issuer pillar]
            (-17, 15) -- (-11, 15) -- (-11, 16.5) -- 
            (-12, 16.5) -- (-12, 35.5) -- (-11, 35.5) -- 
            (-11, 37) -- (-17, 37) -- (-17, 35.5) -- 
            (-16, 35.5) -- (-16, 16.5) -- (-17, 16.5) -- cycle;
        \draw[issuer pillar]
            (-9, 15) -- (-3, 15) -- (-3, 16.5) -- 
            (-4, 16.5) -- (-4, 35.5) -- (-3, 35.5) -- 
            (-3, 37) -- (-9, 37) -- (-9, 35.5) --
            (-8, 35.5) -- (-8, 16.5) -- (-9,16.5) -- cycle;
        \draw[issuer pillar]
            (3, 15) -- (9, 15) -- (9, 16.5) -- 
            (8, 16.5) -- (8, 35.5) -- (9, 35.5) -- 
            (9, 37) -- (3, 37) -- (3, 35.5) -- 
            (4, 35.5) -- (4, 16.5) -- (3, 16.5) -- cycle;
        \draw[issuer pillar]
            (11, 15) -- (17, 15) -- (17, 16.5) -- 
            (16, 16.5) -- (16, 35.5) -- (17, 35.5) -- 
            (17, 37) -- (11, 37) -- (11, 35.5) -- 
            (12, 35.5) -- (12, 16.5) -- (11, 16.5) -- cycle;

        \draw[issuer stage] (-19, 39) -- (19, 39) -- (19, 41.5) -- (-19, 41.5) -- cycle;  
        \draw[issuer stage] (-21.5, 43) -- (21.5, 43) -- (21.5, 45.5) -- (-21.5, 45.5) -- cycle;
    \end{tikzpicture}
}

\savebox{\holderalice}{
	\tikzset{
		holder alice/.style = {
        		draw = ruri,
        		fill = ruri!40,
        		line width = 0.5pt,
    	},
	}
    \begin{tikzpicture}[x = 0.75pt, y = 0.75pt, yscale = -1, xscale = 1]
        \draw[holder alice]
        (401.07,53.47) .. controls (398.41,44.46) and (402.17,39.14) .. 
        (407.23,38.3) .. controls (408.4,38.17) and (411.51,36.88) .. 
        (415.07,40.05) .. controls (420.59,40.7) and (421.36,50.95) .. 
        (418.83,53.6) .. controls (420.2,55.16) and (419.74,57.56) .. 
        (417.8,58.79) .. controls (417.41,61.39) and (416.44,62.88) .. 
        (415.46,64.3) .. controls (415.14,65.86) and (415.14,67.09) .. 
        (415.46,68.91) .. controls (418.12,73.51) and (430.05,72.86) .. 
        (431.02,82.33) .. controls (392.31,82.33) and (428.76,82.39) .. 
        (389.27,82.26) .. controls (390.17,73.19) and (402.36,73.25) .. 
        (404.57,68.91) .. controls (405.09,67.09) and (404.83,66.38) .. 
        (404.63,64.24) .. controls (403.53,62.36) and (402.43,61.26) .. 
        (401.91,58.86) .. controls (400.94,57.5) and (399.12,55.36) .. 
        (401.07,53.47) -- cycle;
    \end{tikzpicture}
}

\savebox{\holderblank}{
	\tikzset{
		holder bob/.style = {
        		draw = momo!90,
        		fill = white,
        		line width = 0.55pt,
    	},
	}
    \begin{tikzpicture}[x = 0.75pt, y = 0.75pt, yscale = -1, xscale = 1]
        \draw[holder bob]
        (401.07,53.47) .. controls (398.41,44.46) and (402.17,39.14) .. 
        (407.23,38.3) .. controls (408.4,38.17) and (411.51,36.88) .. 
        (415.07,40.05) .. controls (420.59,40.7) and (421.36,50.95) .. 
        (418.83,53.6) .. controls (420.2,55.16) and (419.74,57.56) .. 
        (417.8,58.79) .. controls (417.41,61.39) and (416.44,62.88) .. 
        (415.46,64.3) .. controls (415.14,65.86) and (415.14,67.09) .. 
        (415.46,68.91) .. controls (418.12,73.51) and (430.05,72.86) .. 
        (431.02,82.33) .. controls (392.31,82.33) and (428.76,82.39) .. 
        (389.27,82.26) .. controls (390.17,73.19) and (402.36,73.25) .. 
        (404.57,68.91) .. controls (405.09,67.09) and (404.83,66.38) .. 
        (404.63,64.24) .. controls (403.53,62.36) and (402.43,61.26) .. 
        (401.91,58.86) .. controls (400.94,57.5) and (399.12,55.36) .. 
        (401.07,53.47) -- cycle;
    \end{tikzpicture}
}

\savebox{\computer}{
    \tikzset{
        entity/.style = {
		  draw = kogane!90!black,
		  fill = kogane!50,
        },
        space/.style = {
		  draw = white,
		  line width = 0,
		  fill = white,
        },
        earth line/.style = {
		  draw = kogane!90,
		  line width = 0.7pt
        },
        stands/.style = {
          draw = kogane!90!black,
          fill = kogane,
        },
    }
    \begin{tikzpicture}[x = 0.75pt, y = 0.75pt, yscale = -1, xscale = 1]
        \draw[stands]
            (635.77,82.81) -- (637.76,76.2) -- (644.37,76.2) -- (646.35,82.81) -- cycle ;
        \draw[entity]
            (633.13,82.17) -- (648.99,82.17) -- (648.99,83.49) -- (633.13,83.49) -- cycle ;

        \draw[entity]
            (611.97,42.82) .. controls (611.97,42.27) and (612.42,41.82) .. (612.97,41.82) -- 
            (669.15,41.82) .. controls (669.7,41.82) and (670.15,42.27) .. (670.15,42.82) -- 
            (670.15,77.18) .. controls (670.15,77.73) and (669.7,78.18) .. (669.15,78.18) -- 
            (612.97,78.18) .. controls (612.42,78.18) and (611.97,77.73) .. (611.97,77.18) -- cycle;
        \draw[space, fill opacity = 0.97]
            (614.62,45.27) .. controls (614.62,44.83) and (614.98,44.47) .. (615.42,44.47) -- 
            (666.7,44.47) .. controls (667.14,44.47) and (667.5,44.83) .. (667.5,45.27) -- 
            (667.5,73.41) .. controls (667.5,73.85) and (667.14,74.21) .. (666.7,74.21) -- 
            (615.42,74.21) .. controls (614.98,74.21) and (614.62,73.85) .. (614.62,73.41) -- cycle;

        \draw[space]
            (640.73,43.36) .. controls (640.73,43.18) and (640.88,43.03) .. 
            (641.06,43.03) .. controls (641.24,43.03) and (641.39,43.18) .. 
            (641.39,43.36) .. controls (641.39,43.55) and (641.24,43.69) .. 
            (641.06,43.69) .. controls (640.88,43.69) and (640.73,43.55) .. 
            (640.73,43.36) -- cycle ;

        \draw[earth line] (632.88,60) -- (648.58,60);
        \draw[earth line] (640.73,52.15) -- (640.73,67.85);
        \draw[earth line]
            (632.88,60) .. controls (632.88,55.66) and (636.39,52.15) .. 
            (640.73,52.15) ..  controls (645.07,52.15) and (648.58,55.66) .. 
            (648.58,60) .. controls (648.58,64.34) and (645.07,67.85) .. 
            (640.73,67.85) .. controls (636.39,67.85) and (632.88,64.34) .. 
            (632.88,60) -- cycle ;
        \draw[earth line]
            (640.73,52.15) .. controls (646.57,56.21) and (646.37,64.25) ..(640.73,67.85);
        \draw[earth line]
            (640.73,52.15) .. controls (634.79,56.01) and (635.48,64.55) .. (640.73,67.85);
        \draw[earth line]
            (634.89,54.9) .. controls (639.01,57.09) and (642.45,57.09) .. (646.76,54.93);
        \draw[earth line]
            (634.79,65.1) .. controls (638.91,62.29) and (642.55,62.39) .. (646.67,65.14);
    \end{tikzpicture}
}

\savebox{\card}{
	\tikzset{
		bg/.style = {
			fill = momo!50,
			rounded corners = 3pt,
		},
		fg/.style = {
			draw = momo!90,
        		fill = white,
        		line width = 0.23pt,
        		rounded corners = 1.1pt
		}
	}
	\begin{tikzpicture}[x = 0.75pt, y = 0.75pt, yscale = -1, xscale = 1]
		\path[bg] (0, 0) rectangle (40, 25);
		\node[scale = 0.4] at (10, 12) { \usebox{\holderblank} };
		
		\draw[fg] (21.5, 4) rectangle (36.5, 7);
		\draw[fg] (21.5, 11) rectangle (36.5, 14);
		\draw[fg] (21.5, 18) rectangle (36.5, 21);
	\end{tikzpicture}
}

\savebox{\key}{%
	\tikzset{
		key body/.style = {
			fill = ruri!70!black,
		},
		blank/.style = {
			fill = white,
		},
	}%
	\begin{tikzpicture}[x = 0.75pt, y = 0.75pt, yscale = -1, xscale = 1]
		\path[key body] (121.25, 72.17) circle (3.5);
		\path[blank] (119.4, 72.07) circle (0.8);
		\path[key body]
			(133.05,70.94) -- (134.44,72.13) -- (133.58,73.23) -- (125.37,73.23) -- 
			(125.37,73.77) -- (124.29,73.77) -- (124.29,70.4) -- (125.37,70.4) -- 
			(125.37,70.94) -- (125.9,70.95) -- (126.72,71.35) -- (127.25,71.35) -- 
			(127.93,70.94) -- (128.53,71.35) -- (129,71.35) -- (129.68,70.95) -- 
			(130.28,71.35) -- (130.75,71.35) -- (131.44,70.94) -- (132.1,71.35) -- 
			(132.5,71.35) -- cycle;
		\path[blank]
			(125.37,72.56) -- (133.18,72.56) -- (133.18,72.76) -- (125.37,72.76) -- cycle;
	\end{tikzpicture}%
}

\tikzset{
    major indicator/.style = {
  		line width = 1.2pt
    },
    main indicator/.style = {
    	line width = 0.9pt,
    	rounded corners = 2pt,
    },
    associate frame/.style = {
    	fill = ruri!15,
    	draw = ruri!80,
    	line width = 1pt,
    	rounded corners = 6pt,
    },
    campaign frame/.style = {
        fill = momo!10,
        draw = beni,
        line width = 1pt,
        rounded corners = 6pt,
    },
    entity name/.style = {
    	scale = 0.75,
    	font = \itshape,
    },
}

\begin{tikzpicture}[x = 0.75pt,	y = 0.75pt,	yscale = -1, xscale = 1]

\node[scale = 0.4] at (102, -13) { \usebox{\card} };
\node[scale = 0.4] at (138, -13) { \usebox{\card} };
\node[entity name] at (103.5, -2) { $\id^{\issuer_1}$ };
\node[entity name] at (139.5, -2) { $\id^{\issuer_n}$ };
\node[entity name] at (122.5, -12) { $\cdots$ };

\node[scale = 0.9] at (0, 0) { \usebox{\justice} };
\node[scale = 0.9] at (65, 0) { \usebox{\justice} };
\node[entity name] at (2, 28) { Issuer $\issuer_1$ };
\node[entity name] at (67, 28) { Issuer $\issuer_n$ };
\node[entity name] at (34, 5) { $\cdots$ };

\node[scale = 1] at (10, 48) { \usebox{\badgedfilledfile} };
\node[scale = 1] at (75, 48) { \usebox{\badgedfilledfile} };
\node[entity name] at (15.5, 66) { $\cred_1$ };
\node[entity name] at (80.5, 66) { $\cred_n$ };

\draw[campaign frame] (-25, 85) rectangle (285, 230);

\node[scale = 0.9] at (0, 120) { \usebox{\holderalice} };
\node[scale = 0.9] at (0, 190) { \usebox{\holderalice} };
\node[entity name] at (2, 149) { Holder $\holder_1$ };
\node[entity name] at (2, 219) { Holder $\holder_m$ };
\node[entity name] at (2, 158) { $\vdots$ };

\node[scale = 0.9] at (250, 123) { \usebox{\computer} };
\node[entity name] at (252, 149) {Verifier $\verifier$};

\node[scale = 1] at (130, 100) { \usebox{\badgedfilledfile} };
\node[scale = 1] at (160, 100) { \usebox{\badgedfilledfile} };
\node[scale = 1] at (100, 170) { \usebox{\badgedfilledfile} };
\node[scale = 1] at (130, 170) { \usebox{\badgedfilledfile} };
\node[scale = 1] at (160, 170) { \usebox{\badgedfilledfile} };

\draw[associate frame] (160, -15) rectangle (280, 25);
\node[entity name, ruri] at (220, 6) { Verifiable Data Registry\,\, $\registry$ };

\node[scale = 0.4] at (237, 58) { \usebox{\card} };
\node[scale = 1, rotate = 210] at (264, 59) { \usebox{\key} };
\node[entity name] at (238, 70) { $\id^{\issuer_i}$ };
\node[entity name] at (265, 71) { $\pk^{\issuer_i}$ };

\node[entity name] at (250, 220) { Campaign $\sid$ };

\node[entity name] at (122.5, 12) {register};
\node[entity name] at (-10, 55) {issue};
\node[entity name] at (40, 108) {present};
\node[entity name] at (40, 178) {present};
\node[entity name] at (232, 38) {retrieve};

\draw[->, main indicator] (2, 34) -- (2, 98);
\draw[main indicator] (67, 34) -- (67, 75) -- (2, 75);
\draw[->, main indicator] (22, 115) -- (220, 115);
\draw[->, main indicator] (22, 185) -- (200, 185) -- (200, 125) -- (220, 125);
\draw[{-left to}, main indicator] (250, 30) -- (250, 100);
\draw[{-left to}, main indicator] (250, 100) -- (250, 30);
\draw[<->, main indicator] (90, 5) -- (155, 5);

\end{tikzpicture}

%% file: floats/diagrams/identifier-association.tex

\savebox{\badge}{%
	\tikzset{%
		badge upper/.style = {
			fill = kogane!80,
			draw = kogane!80,
			line width = 0pt,
		},
		badge slides/.style = {
			fill = momo,
			draw = momo,
			line width = 0pt,
		},
		blank/.style = {
			opacity = 1,
			fill = white,
			draw = white,
			line width = 0pt,
		},
	}%
	\begin{tikzpicture}[x = 0.75pt, y = 0.75pt, yscale = -1, xscale = 1]%
		\draw[badge upper] (326.2, 46.3) circle (6.5);
		\draw[blank] (326.2, 46.3) circle (4.5);
		\draw[badge upper] (326.2, 46.3) circle (3.8);
		\draw[badge slides]
			(322.87,52.4) .. controls (323.99,53.0) and (325.03,53.28) .. 
			(326.5,53.23) .. controls (325.92,55.51) and (325.4,57.75) .. 
			(324.85,59.77) .. controls (324.13,59.08) and (323.86,58.84) .. 
			(323.29,58.26) .. controls (322.4,58.45) and (322.04,58.52) .. 
			(321.22,58.69) .. controls (321.72,56.72) and (322.06,55.38) .. 
			(322.87,52.4) -- cycle ;
		\draw[badge slides]
			(326.38,55) .. controls (326.55,54.35) and (326.66,53.93) .. 
			(326.82,53.23) .. controls (327.68,53.1) and (328.6,52.87) .. 
			(329.55,52.32) .. controls (330,54.12) and (330.71,56.66) .. 
			(331.24,58.53) .. controls (330.28,58.32) and (329.79,58.22) .. 
			(329.17,58.12) .. controls (328.59,58.63) and (328.21,59.01) .. 
			(327.56,59.54) .. controls (327.15,57.86) and (326.74,56.47) .. 
			(326.38,55) -- cycle ;
	\end{tikzpicture}
}

\savebox{\blankfile}{%
	\tikzset{%
		credential outline/.style = {%
        		draw = murazaki,
    	},
    	credential/.style = {%
        		fill = usumurazaki,
        		drop shadow = {%
            		opacity = 0.8,
            		shadow xshift = 1.2,
            		shadow yshift = -1.2,
        		},
    	},
	}%
    \begin{tikzpicture}[x = 0.75pt, y = 0.75pt, yscale = -1, xscale = 1]%
        \draw[credential, credential outline] (-1,-3.8) -- (3.8, 1) -- (3.8, 15) -- (-11, 15) -- (-11, -3.8) -- cycle;
        \draw[credential outline] (3.8, 1) -- (-1, 1) -- (-1, -3.8);
    \end{tikzpicture}
}

\savebox{\filledfile}{%
	\tikzset{%
		items/.style = {
        		draw = white,
        		fill = white,
        		line width = 0pt,
    	},
	}%
	\begin{tikzpicture}[x = 0.75pt, y = 0.75pt, yscale = -1, xscale = 1]
		\node at (0, 0) { \usebox{\blankfile} };
		\draw[items] (-7, -2.5) rectangle (4, -1.6);
		\draw[items] (-7, 0.1) rectangle (4, 1.0);
		\draw[items] (-7, 2.7) rectangle (4, 3.6);
		\draw[items] (-7, 5.3) rectangle (4, 6.2);
	\end
	{tikzpicture}
}%

\savebox{\badgedfile}{%
	\begin{tikzpicture}[x = 0.75pt, y = 0.75pt, yscale = -1, xscale = 1]
		\node at (0, 0) { \usebox{\blankfile} };
		\node[scale = 0.7] at (8, 6) { \usebox{\badge} };
	\end{tikzpicture}
}%

\savebox{\badgedfilledfile}{%
	\begin{tikzpicture}[x = 0.75pt, y = 0.75pt, yscale = -1, xscale = 1]
		\node at (0, 0) { \usebox{\filledfile} };
		\node[scale = 0.7] at (10, 6) { \usebox{\badge} };
	\end{tikzpicture}
}%

\savebox{\justice}{%
	\tikzset{
		issuer stage/.style = {%
        		draw = kon,
        		fill = ao,
    	},
    	issuer pillar/.style = {%
        		draw = kon,
        		fill = ao!30,
    	},%
	}%
    \begin{tikzpicture}[x = 0.75pt, y = 0.75pt, yscale = -1, xscale = 1]%
        \draw[issuer stage]
            (0, 0) -- (22, 10) -- (22, 13) -- 
            (-22, 13) -- (-22, 10) -- cycle ;

        \draw[issuer pillar]
            (-17, 15) -- (-11, 15) -- (-11, 16.5) -- 
            (-12, 16.5) -- (-12, 35.5) -- (-11, 35.5) -- 
            (-11, 37) -- (-17, 37) -- (-17, 35.5) -- 
            (-16, 35.5) -- (-16, 16.5) -- (-17, 16.5) -- cycle;
        \draw[issuer pillar]
            (-9, 15) -- (-3, 15) -- (-3, 16.5) -- 
            (-4, 16.5) -- (-4, 35.5) -- (-3, 35.5) -- 
            (-3, 37) -- (-9, 37) -- (-9, 35.5) --
            (-8, 35.5) -- (-8, 16.5) -- (-9,16.5) -- cycle;
        \draw[issuer pillar]
            (3, 15) -- (9, 15) -- (9, 16.5) -- 
            (8, 16.5) -- (8, 35.5) -- (9, 35.5) -- 
            (9, 37) -- (3, 37) -- (3, 35.5) -- 
            (4, 35.5) -- (4, 16.5) -- (3, 16.5) -- cycle;
        \draw[issuer pillar]
            (11, 15) -- (17, 15) -- (17, 16.5) -- 
            (16, 16.5) -- (16, 35.5) -- (17, 35.5) -- 
            (17, 37) -- (11, 37) -- (11, 35.5) -- 
            (12, 35.5) -- (12, 16.5) -- (11, 16.5) -- cycle;

        \draw[issuer stage] (-19, 39) -- (19, 39) -- (19, 41.5) -- (-19, 41.5) -- cycle;  
        \draw[issuer stage] (-21.5, 43) -- (21.5, 43) -- (21.5, 45.5) -- (-21.5, 45.5) -- cycle;
    \end{tikzpicture}
}

\savebox{\holderalice}{%
	\tikzset{%
		holder alice/.style = {%
        		draw = kon,
        		fill = ao!70,
        		line width = 0.5pt,
    	},%
	}%
    \begin{tikzpicture}[x = 0.75pt, y = 0.75pt, yscale = -1, xscale = 1]
        \draw[holder alice]
        (401.07,53.47) .. controls (398.41,44.46) and (402.17,39.14) .. 
        (407.23,38.3) .. controls (408.4,38.17) and (411.51,36.88) .. 
        (415.07,40.05) .. controls (420.59,40.7) and (421.36,50.95) .. 
        (418.83,53.6) .. controls (420.2,55.16) and (419.74,57.56) .. 
        (417.8,58.79) .. controls (417.41,61.39) and (416.44,62.88) .. 
        (415.46,64.3) .. controls (415.14,65.86) and (415.14,67.09) .. 
        (415.46,68.91) .. controls (418.12,73.51) and (430.05,72.86) .. 
        (431.02,82.33) .. controls (392.31,82.33) and (428.76,82.39) .. 
        (389.27,82.26) .. controls (390.17,73.19) and (402.36,73.25) .. 
        (404.57,68.91) .. controls (405.09,67.09) and (404.83,66.38) .. 
        (404.63,64.24) .. controls (403.53,62.36) and (402.43,61.26) .. 
        (401.91,58.86) .. controls (400.94,57.5) and (399.12,55.36) .. 
        (401.07,53.47) -- cycle;
    \end{tikzpicture}
}

\savebox{\holderblank}{%
	\tikzset{%
		holder bob/.style = {
        		draw = momo!90,
        		fill = white,
        		line width = 0.55pt,
    	},%
	}%
    \begin{tikzpicture}[x = 0.75pt, y = 0.75pt, yscale = -1, xscale = 1]
        \draw[holder bob]
        (401.07,53.47) .. controls (398.41,44.46) and (402.17,39.14) .. 
        (407.23,38.3) .. controls (408.4,38.17) and (411.51,36.88) .. 
        (415.07,40.05) .. controls (420.59,40.7) and (421.36,50.95) .. 
        (418.83,53.6) .. controls (420.2,55.16) and (419.74,57.56) .. 
        (417.8,58.79) .. controls (417.41,61.39) and (416.44,62.88) .. 
        (415.46,64.3) .. controls (415.14,65.86) and (415.14,67.09) .. 
        (415.46,68.91) .. controls (418.12,73.51) and (430.05,72.86) .. 
        (431.02,82.33) .. controls (392.31,82.33) and (428.76,82.39) .. 
        (389.27,82.26) .. controls (390.17,73.19) and (402.36,73.25) .. 
        (404.57,68.91) .. controls (405.09,67.09) and (404.83,66.38) .. 
        (404.63,64.24) .. controls (403.53,62.36) and (402.43,61.26) .. 
        (401.91,58.86) .. controls (400.94,57.5) and (399.12,55.36) .. 
        (401.07,53.47) -- cycle;
    \end{tikzpicture}
}

\savebox{\computer}{%
    \tikzset{
        entity/.style = {
		  draw = kon,
		  fill = ao,
        },
        space/.style = {
		  draw = white,
		  line width = 0,
		  fill = white,
        },
        earth line/.style = {
		  draw = ao!60!kon,
		  line width = 0.7pt
        },
    }
    \begin{tikzpicture}[x = 0.75pt, y = 0.75pt, yscale = -1, xscale = 1]
        \draw[entity, fill = ao!90!black]
            (635.77,82.81) -- (637.76,76.2) -- (644.37,76.2) -- (646.35,82.81) -- cycle ;
        \draw[entity]
            (633.13,82.17) -- (648.99,82.17) -- (648.99,83.49) -- (633.13,83.49) -- cycle ;

        \draw[entity]
            (611.97,42.82) .. controls (611.97,42.27) and (612.42,41.82) .. (612.97,41.82) -- 
            (669.15,41.82) .. controls (669.7,41.82) and (670.15,42.27) .. (670.15,42.82) -- 
            (670.15,77.18) .. controls (670.15,77.73) and (669.7,78.18) .. (669.15,78.18) -- 
            (612.97,78.18) .. controls (612.42,78.18) and (611.97,77.73) .. (611.97,77.18) -- cycle;
        \draw[space, fill opacity = 0.97]
            (614.62,45.27) .. controls (614.62,44.83) and (614.98,44.47) .. (615.42,44.47) -- 
            (666.7,44.47) .. controls (667.14,44.47) and (667.5,44.83) .. (667.5,45.27) -- 
            (667.5,73.41) .. controls (667.5,73.85) and (667.14,74.21) .. (666.7,74.21) -- 
            (615.42,74.21) .. controls (614.98,74.21) and (614.62,73.85) .. (614.62,73.41) -- cycle;

        \draw[space]
            (640.73,43.36) .. controls (640.73,43.18) and (640.88,43.03) .. 
            (641.06,43.03) .. controls (641.24,43.03) and (641.39,43.18) .. 
            (641.39,43.36) .. controls (641.39,43.55) and (641.24,43.69) .. 
            (641.06,43.69) .. controls (640.88,43.69) and (640.73,43.55) .. 
            (640.73,43.36) -- cycle ;

        \draw[earth line] (632.88,60) -- (648.58,60);
        \draw[earth line] (640.73,52.15) -- (640.73,67.85);
        \draw[earth line]
            (632.88,60) .. controls (632.88,55.66) and (636.39,52.15) .. 
            (640.73,52.15) ..  controls (645.07,52.15) and (648.58,55.66) .. 
            (648.58,60) .. controls (648.58,64.34) and (645.07,67.85) .. 
            (640.73,67.85) .. controls (636.39,67.85) and (632.88,64.34) .. 
            (632.88,60) -- cycle ;
        \draw[earth line]
            (640.73,52.15) .. controls (646.57,56.21) and (646.37,64.25) ..(640.73,67.85);
        \draw[earth line]
            (640.73,52.15) .. controls (634.79,56.01) and (635.48,64.55) .. (640.73,67.85);
        \draw[earth line]
            (634.89,54.9) .. controls (639.01,57.09) and (642.45,57.09) .. (646.76,54.93);
        \draw[earth line]
            (634.79,65.1) .. controls (638.91,62.29) and (642.55,62.39) .. (646.67,65.14);
    \end{tikzpicture}
}

\savebox{\card}{%
	\tikzset{%
		bg/.style = {%
			fill = momo!50,
			rounded corners = 3pt,
		},
		fg/.style = {
			draw = momo,
        		fill = white,
        		line width = 0.3pt,
        		rounded corners = 1.1pt
		}%
	}%
	\begin{tikzpicture}[x = 0.75pt, y = 0.75pt, yscale = -1, xscale = 1]%
		\path[bg] (0, 0) rectangle (40, 25);
		\node[scale = 0.4] at (11, 12) { \usebox{\holderblank} };
		
		\draw[fg] (21.5, 4) rectangle (36.5, 7);
		\draw[fg] (21.5, 11) rectangle (36.5, 14);
		\draw[fg] (21.5, 18) rectangle (36.5, 21);
	\end{tikzpicture}
}

\savebox{\balloona}{%
	\tikzset{%
		fore/.style = {
			draw = white,
			fill = ao!50,
			line width = 0.6pt,
		},
		mid/.style = {
			draw = white,
			fill = beni!50,
			line width = 0.6pt,
		},
		back/.style = {
			draw = white,
			fill = ruri!50,
			line width = 0.6pt,
		},
		string/.style = {
			draw = #1,
			line width = 0.8pt,
		}
	}%
	\begin{tikzpicture}[x = 0.75pt, y = 0.75pt, yscale = -1, xscale = 1]
		\draw[back] (11, -6) circle (5pt);
		\draw[mid] (12, 5) ellipse (7pt and 8pt);
		\draw[fore] (0, 0) ellipse (9pt and 10pt);
		
		\path[fill = ao!50] (-2, 12) -- (-3, 15) -- (3, 15) -- (2, 12) -- cycle;
		\path[fill = beni!50] (10, 15) -- (9, 17) -- (15, 17) -- (14, 15) -- cycle;
		
		\draw[string = ao!50] (-0.5, 14.5) .. controls (2, 17) and (-2, 18) .. (0,20);
		\draw[string = beni!50] (11.5, 16.5) .. controls (14, 19) and (10, 20) .. (12,22);
		
		\draw[string = white] (-9.2, 0) .. controls (-9, -8) and (-3, -10.5) .. (0, -10.5);
		\draw[string = white] (19, 5) .. controls (19, 9) and (16, 12) .. (15, 12);
		\draw[string = white] (12, -10.5) .. controls (13.5, -10.5) and (15.5, -8.5) .. (15.5, -7);
	\end{tikzpicture}
}

\savebox{\balloonb}{%
	\tikzset{%
		fore/.style = {
			draw = white,
			fill = ao!50,
			line width = 0.6pt,
		},
		mid/.style = {
			draw = white,
			fill = beni!50,
			line width = 0.6pt,
		},
		back/.style = {
			draw = white,
			fill = ruri!50,
			line width = 0.6pt,
		},
		string/.style = {
			draw = #1,
			line width = 0.8pt,
		}
	}%
	\begin{tikzpicture}[x = 0.75pt, y = 0.75pt, yscale = -1, xscale = 1]
		\draw[back] (11, -6) circle (5pt);
		\draw[fore] (0, 0) ellipse (9pt and 10pt);
		\draw[mid] (12, 5) ellipse (7pt and 8pt);
        
		\path[fill = ao!50] (-2, 12) -- (-3, 15) -- (3, 15) -- (2, 12) -- cycle;
		\path[fill = beni!50] (10, 15) -- (9, 17) -- (15, 17) -- (14, 15) -- cycle;
		
		\draw[string = ao!50] (-0.5, 14.5) .. controls (2, 17) and (-2, 18) .. (0,20);
		\draw[string = beni!50] (11.5, 16.5) .. controls (14, 19) and (10, 20) .. (12,22);
		
		\draw[string = white] (-9.2, 0) .. controls (-9, -8) and (-3, -10.5) .. (0, -10.5);
		\draw[string = white] (19, 5) .. controls (19, 9) and (16, 12) .. (15, 12);
		\draw[string = white] (12, -10.5) .. controls (13.5, -10.5) and (15.5, -8.5) .. (15.5, -7);
	\end{tikzpicture}
}

\savebox{\balloonc}{%
	\tikzset{%
		fore/.style = {
			draw = white,
			fill = ao!50,
			line width = 0.6pt,
		},
		mid/.style = {
			draw = white,
			fill = beni!50,
			line width = 0.6pt,
		},
		back/.style = {
			draw = white,
			fill = ruri!50,
			line width = 0.6pt,
		},
		string/.style = {
			draw = #1,
			line width = 0.8pt,
		}
	}%
	\begin{tikzpicture}[x = 0.75pt, y = 0.75pt, yscale = -1, xscale = 1]
		\draw[mid] (12, 5) ellipse (7pt and 8pt);
		\draw[fore] (0, 0) ellipse (9pt and 10pt);
        \draw[back] (11, -6) circle (5pt);
		
		\path[fill = ao!50] (-2, 12) -- (-3, 15) -- (3, 15) -- (2, 12) -- cycle;
		\path[fill = beni!50] (10, 15) -- (9, 17) -- (15, 17) -- (14, 15) -- cycle;
		\path[fill = ruri!50] (10, 0) -- (9, 2) -- (13, 2) -- (12, 0) -- cycle;
		
		\draw[white] (9.6, 0.18) -- (9,2) -- (13,2) -- (12.4, 0.18);
		
		\draw[string = ao!50] (-0.5, 14.5) .. controls (2, 17) and (-2, 18) .. (0,20);
		\draw[string = beni!50] (11.5, 16.5) .. controls (14, 19) and (10, 20) .. (12,22);
		\draw[string = ruri!50] (11, 1.4) -- (11, 4);
		
		\draw[white] (10.2, 1.8) -- (10.2, 4.2) -- (11.8, 4.2) -- (11.8, 1.8);
		
		\draw[string = white] (-9.2, 0) .. controls (-9, -8) and (-3, -10.5) .. (0, -10.5);
		\draw[string = white] (19, 5) .. controls (19, 9) and (16, 12) .. (15, 12);
		\draw[string = white] (12, -10.5) .. controls (13.5, -10.5) and (15.5, -8.5) .. (15.5, -7);
	\end{tikzpicture}
}

\tikzset{
    major indicator/.style = {
  		line width = 2pt
    },
    main indicator/.style = {
    	line width = 0.9pt,
    },
    minor indicator/.style = {
    	line width = 0.6pt,
    },
    entity name/.style = {
    	scale = 0.65,
    	font = \itshape,
    },
    associate frame/.style = {
    	fill = ruri!15,
    	draw = ruri!80,
    	line width = 1pt,
    	rounded corners = 6pt,
    },
    aid/.style = {
    	fill = white,
    	draw = ruri,
    	line width = 0.7pt,
    },
    void/.style = {
    	dashed,
    	dash pattern = on 2pt off 1.5pt,
    },
    present/.style = {
    	entity name,
    	anchor=west,
    	align=left,
    },
}

\begin{tikzpicture}[x = 0.75pt,	y = 0.75pt,	yscale = -1, xscale = 1]

\draw[gray] (-18, 30) -- (280, 30);
\draw[gray] (-18, 98) -- (280, 98);

\node[present] at (-20, -30) {Presentation \cirn{172}: vote at FutureDAO};
\node[present] at (-20, 38) {Presentation \cirn{173}: claim airdrop at InsuraFi};
\node[present] at (-20, 106) {Presentation \cirn{174}: claim airdrop again};



\node[scale = 1.2] at (0, 0) { \usebox{\filledfile} };
\node[scale = 1.2] at (35, 0) { \usebox{\filledfile} };
\node[entity name] at (-3, -18) {$\cred_1$};
\node[entity name] at (32, -18) {$\cred_2$};

\node[scale = 0.4] at (5, 10) { \usebox{\card} };
\node[scale = 0.4] at (40, 10) { \usebox{\card} };
\node[entity name] at (5, 20) {$\id_1$};
\node[entity name] at (40, 20) {$\id_2$};

\draw[->, main indicator] (50, 0) -- (79, 0);
\draw[associate frame] (80, -21) rectangle (200, 20);
\draw[->, main indicator] (200, 0) -- (235, 0);

\node[scale = 0.5] at (110, -10) { \usebox{\card} };
\node[scale = 0.5] at (110, 10) { \usebox{\card} };
\node[entity name] at (91, -10) {$\id_1$};
\node[entity name] at (91, 10) {$\id_2$};

\draw[ruri, minor indicator, rounded corners = 2pt] (119.5, -12) -- (138, -12) -- (138, -7);
\draw[ruri, minor indicator, rounded corners = 2pt] (119.5, 12) -- (152, 12) -- (152, 7);

\draw[aid, void, fill = ruri!60] (131, -7) rectangle (145, 7);
\draw[aid, void, fill = ruri!60] (145, -7) rectangle (159, 7);
\node[entity name] at (138.5, 0.5) {$\id_1$};
\node[entity name] at (152.5, 0.5) {$\id_2$};
\node[entity name, ruri] at (160, -13) {$\aid$};

\node[scale = 0.9] at (255, 0) { \usebox{\balloona} };
\node[entity name] at (255, 22) {FutureDAO};


\node[scale = 1.2] at (0, 68) { \usebox{\filledfile} };
\node[scale = 1.2] at (35, 68) { \usebox{\filledfile} };
\node[entity name] at (-3, 50) {$\cred_3$};
\node[entity name] at (32, 50) {$\cred_4$};

\node[scale = 0.4] at (5, 78) { \usebox{\card} };
\node[scale = 0.4] at (40, 78) { \usebox{\card} };
\node[entity name] at (5, 88) {$\id_1$};
\node[entity name] at (40, 88) {$\id_3$};

\draw[->, main indicator] (50, 68) -- (79, 68);
\draw[associate frame] (80, 47) rectangle (200, 88);
\draw[->, main indicator] (200, 68) -- (235, 68);

\node[scale = 0.5] at (110, 78) { \usebox{\card} };
\node[entity name] at (91, 79) {$\id_3$};

\draw[ruri, minor indicator, rounded corners = 2pt] (119.5, 80) -- (166,80) -- (166, 75);

\draw[aid, fill = ruri!60] (131, 61) rectangle (145, 75);
\draw[aid] (145, 61) rectangle (159, 75);
\draw[aid, void, fill = ruri!60] (159, 61) rectangle (173, 75);
\node[entity name] at (138.5, 68.5) {$\id_1$};
\node[entity name] at (152.5, 68.5) {$\id_2$};
\node[entity name] at (166.5, 68.5) {$\id_3$};
\node[entity name, ruri] at (152, 55) {$\aid$};

\node[scale = 0.9] at (255, 68) { \usebox{\balloonb} };
\node[entity name] at (255, 90) {InsuraFi};

\node[scale = 1.2] at (20, 136) { \usebox{\filledfile} };
\node[entity name] at (17, 118) {$\cred_5$};

\node[scale = 0.4] at (25, 146) { \usebox{\card} };
\node[entity name] at (25, 156) {$\id_2$};

\draw[->, main indicator] (50, 136) -- (79, 136);
\draw[associate frame] (80, 115) rectangle (200, 156);
\draw[->, main indicator] (200, 136) -- (235, 136);

\draw[aid] (119, 129) rectangle (133, 143);
\draw[aid, fill = ruri!60] (133, 129) rectangle (147, 143);
\draw[aid] (147, 129) rectangle (161, 143);

\node[entity name] at (126.5, 136.5) {$\id_1$};
\node[entity name] at (140.5, 136.5) {$\id_2$};
\node[entity name] at (154.5, 136.5) {$\id_3$};

\node[entity name, ruri] at (140, 123) {$\aid$};

\node[momo, scale = 1.2] at (217, 136) {\ding{56}};

\node[scale = 0.9] at (255, 136) { \usebox{\balloonb} };
\node[entity name] at (255, 158) {InsuraFi};

\end{tikzpicture}

%% file: floats/frames/function-braid.tex
\protocol{Functionality $\funb$}{

\method{Identifier Registration:}
On receiving $(\mathtt{register},\sid)$ from a party $\party$, sample a $\sk$ and calculate $\pk:=\pubkeygen(\sk)$.
Sample an $\id$. If $\id$ has been recorded, sample one again.
Send $(\mathtt{register},\sid,\pk,\id)$ to the adversary $\simulator$.
Upon receiving $\mathtt{ok}$ from $\simulator$, record $(\party, \id, \pk, \sk)$, and send $(\mathtt{registered},\sid,\id)$ to $\party$.

\method{Identifier Association:}
On receiving $(\mathtt{associate},\sid,\{\id_i\}_\ell)$ from a party $\party$, retrieve $(\party,\id_i,\pk_i,\sk_i)$ for all $i\in[\ell]$.
Halt if some $\id_i$ is marked \textit{associated}.
Calculate $\aid:=H_a(\id_1,\dots,\id_\ell,u_a:=0)$, and send $(\mathtt{associate},\sid,\aid)$ to $\simulator$.
Upon receiving $\mathtt{ok}$ from $\simulator$, mark all $\id_i$ as \textit{associated} and record $(\party,\allowbreak \aid,\allowbreak \id_1,\dots,\id_\ell,0)$.
Finally, return $(\mathtt{associated},\sid,\aid)$ to $\party$.

\method{Association Appending:}
On receiving $(\mathtt{append},\sid,\aid,\id_{\ell+1})$ from $\party$, retrieve $(\party,\aid,\id_1,\dots,\id_\ell,u_a)$ and $(\party,\id_{\ell+1},\cdot,\cdot)$.
Halt if $\aid$ is \textit{invalidated} or $\id_{\ell+1}$ is \textit{associated}.
Calculate $\aid':=H_a(\id_1,\dots,\id_{\ell+1},u_a)$, and send $(\mathtt{append},\sid,\aid')$ to $\simulator$.
Upon receiving $\mathtt{ok}$ from $\simulator$, mark $\aid$ as \textit{invalidated}, mark $\id_{\ell+1}$ as \textit{associated}, and record $(\party,\aid',\id_1,\dots,\id_{\ell+1},u_a)$.
Finally, return $(\mathtt{appended},\sid,\aid')$ to $\party$.

\method{Association Merging:}
On receiving $(\mathtt{merge},\sid,\aid_1,\aid_2)$ from $\party$, retrieve $(\party,\aid_1,\id_1,\dots,\id_t,u_{a1})$ and $(\party,\aid_2,\id_{t+1},\dots,\id_\ell,u_{a2})$.
Halt if $\aid_1$ or $\aid_2$ is \textit{invalidated}.
Set $u_a := \max(u_{a1}, u_{a2})$.
Calculate $\aid':=H_a(\id_1,\dots,\id_\ell,u_a)$, and send $(\mathtt{merge},\sid,\aid')$ to $\simulator$.
Upon receiving $\mathtt{ok}$ from $\simulator$, mark $\aid_1$ and $\aid_2$ as \textit{invalidated}, and record $(\party,\aid',\id_1,\dots,\id_\ell,u_a)$.
Finally, return $(\mathtt{merged},\sid,\aid')$ to $\party$.

\method{Association Refreshing:}
On receiving $(\mathtt{refresh},\sid,\aid)$ from a party $\party$, retrieve $(\party,\aid,\id_1,\dots,\id_\ell,u_a)$.
Halt if $\aid$ is \textit{invalidated}.
Set $u'_a := u_a + 1$.
Calculate $\aid':=H_a(\id_1,\dots,\id_\ell,u'_a)$, and send $(\mathtt{refresh},\sid,\aid')$ to $\simulator$.
Upon receiving $\mathtt{ok}$ from $\simulator$, mark $\aid$ as \textit{invalidated}, and record $(\party,\aid',\id_1,\dots,\id_\ell,u'_a)$.
Finally, return $(\mathtt{refreshed},\sid,\aid')$ to $\party$.

\method{Credential Presentation:}
On receiving $(\mathtt{campaign},\sid,\phi)$ from $\verifier$, wait for input $(\mathtt{present},\sid,\aid,\{\cred_i\}_n)$ from $\holder$.
For all $i\in[n]$, parse $\cred_i:=(\idi_i, \sigma_i,\allowbreak \idh_i, s_i)$, and retrieve $(\holder,\idh_i,\cdot,\skh_i)$.
Then, check:
\begin{itemize}[leftmargin=*,parsep=0.5pt]
    \item If $(\holder,\aid,\id_1,\dots,\id_\ell,u_a)$ is recorded and not \textit{invalidated}.
    \item If $\idh_i\in\{\id_1,\dots,\id_\ell\}$.
    \item If $(\mathtt{presented},\sid,\{\id_1,\dots,\id_\ell\})$ is not recorded.
    \item If $\cred_i$ is valid, recorded, and non-\textit{revoked}, and $\phi(s_i)=1$ holds.
    \item If $(\issuer,\idi_i,\pki_i,\cdot)$ is recorded and $\versign\big(\pki_i,\sigma_i,(\idi_i,\idh_i,s_i)\big)=1$.
\end{itemize}
Halt if any check fails.
Else, send $(\mathtt{present},\sid,\{\idi_i\}_n)$ to $\simulator$.
On receiving $\mathtt{ok}$ from $\simulator$, record $(\mathtt{presented},\sid,\{\id_1,\dots,\id_\ell\})$ and send $(\mathtt{verified},\sid,\holder)$ to $\verifier$.

\method{Key Recovery:}
On receiving $(\mathtt{recover},\sid,\aid,\id,\{\id'\}_t,\{\sk'\}_t)$ from a party $\party$, check:
\begin{itemize}[leftmargin=*,parsep=0.5pt]
    \item If $(\party,\aid,\id_1,\dots,\id_\ell,\cdot)$ is recorded and not \textit{invalidated}.
    \item If $\id\in\{\id_1,\dots,\id_\ell\}$ and $\{\id'\}_t\subset\{\id_1,\dots,\id_\ell\}$.
    \item If $(\party,\id,\pk,\sk)$ is recorded.
    \item If $(\party,\id'_j,\cdot,\sk'_j)$ is recorded and \textit{associated} for all $j\in[t]$.
\end{itemize}
Halt if any check fails.
Else, sample a $\sk''$ and calculate $\pk'':=\pubkeygen(\sk'')$.
Send $(\mathtt{recover},\sid,\id,\pk'')$ to $\simulator$.
On receiving $\mathtt{ok}$ from $\simulator$, replace the recorded $(\party,\id,\pk,\sk)$ with $(\party,\id,\pk'',\sk'')$ and return $(\mathtt{recovered},\sid)$ to $\party$.
}

%% file: floats/frames/function-ledger.tex
\protocol{Decentralized ledger $\funl$}{
The ledger functionality maintains a Merkle tree $T$, a nullifier list $N$, and an identifier-public key map $M$. It interacts with an ideal-world adversary $\simulator$.

\method{Register:}
On receiving $(\mathtt{register},\sid,k,v)$ from a party $\party$, send $(\mathtt{register},\allowbreak \sid, k,v)$ to $\simulator$.
Upon receiving $\mathtt{ok}$ from $\simulator$, if there is already a record $(k,\cdot)$ in $M$, update it to $(k,v)$; else, record $(k,v)$ in $M$. Send $(\mathtt{registered},\sid)$ to $\party$.

\method{Record:}
On receiving $(\mathtt{record},\sid,v)$ from a party $\party$, calculate $(T,r,\allowbreak\rho)\gets\mtinsert(v,T)$ and send $(\mathtt{record},\sid,r,\rho,v)$ to $\simulator$.
Upon receiving $\mathtt{ok}$ from $\simulator$, record $r$ and send $(\mathtt{recorded},\sid,r,\rho)$ to $\party$.

\method{Deprecate:}
On receiving $(\mathtt{invalidate},\sid,v)$ from a party $\party$, send $(\mathtt{invalidate},\allowbreak \sid, v)$ to $\simulator$.
Upon receiving $\mathtt{ok}$ from $\simulator$, mark $v$ as \textit{invalidated} by appending it to $N$, and send $(\mathtt{invalidated},\sid)$ to $\party$.

\method{Retrieve:}
On receiving $(\mathtt{retrieve},\sid,k)$ from a party $\party$, send $(\mathtt{retrieve},\sid,\allowbreak k)$ to $\simulator$.
Upon receiving $\mathtt{ok}$ from $\simulator$, if $(k,v)$ is recorded in $M$, send $(\mathtt{retrieved},\sid,v)$ to $\party$, else send $(\mathtt{unrecorded},\sid)$.

\method{Check:}
On receiving $(\mathtt{check},\sid,v)$ from a party $\party$, send $(\mathtt{check},\allowbreak \sid,v)$ to $\simulator$.
Upon receiving \(\mathtt{ok}\) from \(\simulator\), if \(v\in N\), send \((\mathtt{invalid},\sid)\) to \(\party\). Otherwise, if \(v\) is recorded, send \((\mathtt{checked},\sid)\) to \(\party\); else send \((\mathtt{unrecorded},\sid)\).
}

%% file: sections/body/protocol.tex
\section{Main construction}\label{sec: system}

In this section, we introduce the \sysname{} construction and protocol: identifier association and Sybil-resistant credential presentation, trustless key recovery, and the resulting security properties. Figure~\ref{fig: protocol-full} formalizes the core protocol. Appendix~\ref{sec: extensions} gives the association-update and credential-management protocols and extensions. Appendices~\ref{app: definition}--\ref{app: security-uc} provide formal definitions and proofs.

\subsection{Identifier registration and association}\label{subsec: identifier-association}

\sysname{} provides a mechanism called \textit{identifier association} that mandates $\holder$ to provide a \textit{proof of association} before presenting to $\verifier$ multiple credentials under different identifiers.

We note that $\registry$ maintains an incremental Merkle tree (with leaf nodes of tags $h$ or $\aid$) and a nullifier list $N$. All these data structures are publicly accessible on the blockchain.
In a highly concurrent blockchain environment, the global tree state updates frequently. Since local zero-knowledge proof generation takes time (e.g., several seconds), the specific Merkle root a user fetches to construct a proof might become outdated by the time their transaction is mined. To address this race condition, $\registry$ caches the 100 most recent Merkle roots. This sliding window allows the contract to accept proofs constructed against any recently valid historical state, preventing gas-wasting transaction reverts and synchronization failures.

\spar{Identifier registration}
$\holder$ can arbitrarily register new identifiers. To do this, $\holder$ first generates a key pair by sampling $\sk\sample\Zp$ and computing $\pk:=\pubkeygen(\sk)$, then sends $\pk$ to $\registry$. $\registry$ samples $\id\sample\Zp$ and persists the pair $(\id,\pk)$, then returns $\id$ to $\holder$.

Afterwards, $\holder$ calculates a tag $h:=H_a(\id,\sk)$ and proves in zero-knowledge that:
\begin{enumerate}[parsep=0.5pt]
    \item[\cirn{182}] $\pk$ is the public key of $\sk$, and
    \item[\cirn{183}] $h$ derives from $\id$ and $\sk$.
\end{enumerate}
Concretely, this emits a proof $\pi_g$ for relation $R_g$ below:
\[
    \pk=\pubkeygen(\witness{\sk})
    \;\land\;
    h=H_a(\id,\witness{\sk}).
\]
$\holder$ sends $(\id,h,\pi_g)$ to $\registry$, who verifies $\pi_g$ and records $h$ in the incremental Merkle tree maintained by itself.

\spar{Identifier association}
$\holder$ could associate multiple identifiers $\id_1, \dots, \id_\ell$ belonging to it into an $\aid$. Registration and association are asynchronous, which is designed to prevent any party from inferring the components of $\aid$ by monitoring the order in which identifiers are registered and associated.

To associate identifiers, $\holder$ generates a nonce $u_a:=0$.
We adopt the convention that the identifier vector $\id_1,\dots,\id_\ell$ is always canonically (lexicographically) sorted before hashing to ensure deterministic aggregation.
Then, $\holder$ computes $\aid:=H_a(\id_1,\dots,\id_\ell,u_a)$.
The nonce $u_a$ supports unlinkability by allowing a holder to refresh an association and rotate its public handles.
To justify the association, $\holder$ proves in zero-knowledge that:
\begin{enumerate}[parsep=0.5pt]
    \item[\cirn{182}] $\aid$ is the aggregation of $\id_1,\dots,\id_\ell$ with a nonce $u_a$,
    \item[\cirn{183}] For each $i\in[\ell]$, $h_i$ is recorded in the ledger,
    \item[\cirn{184}] $h_i$ derives from $\id_i$ and its private key $\sk_i$, and
    \item[\cirn{185}] $n_i$ also derives from $\id_i$ and $\sk_i$.
\end{enumerate}
This emits a proof $\pi_a$ for relation $R_a$:
\[
    \begin{aligned}
         & \aid= H_a(\witness{\id_1,\dots,\id_\ell},u_a)
        \land
        \mtauth(r_i,\witness{h_i},\witness{\rho_i})=1    \\
         & \land\;
        \witness{h_i}=H_a(\witness{\id_i},\witness{\sk_i})
        \;\land\;
        n_i=H_n(\witness{\id_i},\witness{\sk_i}).
    \end{aligned}
\]
To clarify, \cirn{183} and \cirn{184} ensure that $\id_i$ was indeed obtained through registration and belongs to $\holder$ who possesses $\sk_i$, while \cirn{185} establishes bindings for $\id_i$ and emits a nullifier $n_i$ to $\registry$ to implicitly mark identifiers that have been \textit{associated}.

$\holder$ sends $(\aid,\{r_i\}_\ell,\{n_i\}_\ell, \pi_a)$ to $\registry$, who verifies $\pi_a$ and confirms that $n_i$'s are not recorded in $N$. If this is the case, $\registry$ records $\aid$ and invalidates all $n_i$'s to $N$, indicating that all $\id_i$'s are \textit{associated}.

\spar{Update by appending}
\sysname{} facilitates updating previously submitted associated identifiers by enabling $\holder$ to append new identifiers to an existing association.

Concretely, \sysname{} allows $\holder$ to append a new $\id_{\ell+1}$ to an $\aid$. To this end, $\holder$ computes an updated $\aid':=H_a(\id_1,\dots,\id_\ell, \id_{\ell+1}, u_a)$, a nullifier $n_a:=H_n(\id_1,\dots,\allowbreak \id_\ell, u_a)$ to invalidate the old $\aid$, and another nullifier $n_{\ell+1}:=H_n(\id_{\ell+1},\allowbreak \sk_{\ell+1})$ to invalidate $\id_{\ell+1}$. After this, $\holder$ proves in zero-knowledge that:
\begin{enumerate}[parsep=0.5pt]
    \item[\cirn{182}] $\aid$ is the association of $\id_1,\dots,\id_\ell$ with the nonce $u_a$,
    \item[\cirn{183}] $\aid'$ is the association of $\id_1,\dots,\id_\ell,\id_{\ell+1}$ with $u_a$,
    \item[\cirn{184}] $\aid$ is recorded by the ledger and is not \textit{blocklisted}, and
    \item[\cirn{185}] $\id_{\ell+1}$ is recorded but not \textit{associated}.
\end{enumerate}
This emits a proof $\pi_p$ for relation $R_p$:
\[
    \begin{aligned}
         &
        \witness{\aid} = H_a(\witness{\id_1, \dots, \id_\ell}, u_a)
        \land
        \aid' = H_a(\witness{\id_1,\dots, \id_{\ell+1}}, u_a)                \\
         & \land\;
        \mtauth(r_a, \witness{\aid}, \witness{\rho_a}) = 1
        \;\land\;
        n_a = H_n(\witness{\id_1, \dots, \id_\ell}, u_a)                     \\
         & \land\;
        \mtauth(r_{\ell+1},\witness{h_{\ell+1}},\witness{\rho_{\ell+1}}) = 1 \\
         & \land\;
        \witness{h_{\ell+1}}=H_a(\witness{\id_{\ell+1}}, \witness{\sk_{\ell+1}})
        \;\land\;
        n_{\ell+1}=H_n(\witness{\id_{\ell+1}}, \witness{\sk_{\ell+1}}).
    \end{aligned}
\]
$\holder$ sends $(r_a,r_{\ell+1},n_a,n_{\ell+1},\aid',\pi_p)$ to $\registry$, who verifies $\pi_p$ and confirms that $r_a, r_{\ell+1}$ are recorded tree roots and $n_a,n_{\ell+1}$ not recorded in $N$. If this is the case, $\registry$ invalidates $n_{\ell+1}$ and $n_a$, indicating that $\id_{\ell+1}$ is \textit{associated}, and the old version of $\aid$ has been \textit{deprecated}.

\spar{Update by merging}
\sysname{} also enables $\holder$ to merge two association $\aid_1$ and $\aid_2$ into one. To this end, $\holder$ computes $\aid'$ and two nullifiers $n_{a1}$ and $n_{a2}$ with respect to $\aid_1$ and $\aid_2$. Then, $\holder$ proves in zero-knowledge that:
\begin{enumerate}[parsep=0.5pt]
    \item[\cirn{182}] $\aid_1$ derives from $\id_1,\dots,\id_t$ with the nonce $u_{a1}$,
    \item[\cirn{183}] $\aid_2$ derives from $\id_{t+1},\dots,\id_\ell$ with the nonce $u_{a2}$,
    \item[\cirn{184}] $\aid_1$ and $\aid_2$ are recorded and not \textit{blocklisted}, and
    \item[\cirn{185}] $\aid$ derives from all $\id_1,\dots,\id_\ell$ with a nonce $u_a'$, which is the maximum of $u_{a1}$ and $u_{a2}$.
\end{enumerate}
This emits a proof $\pi_m$ for relation $R_m$:
\[
    \begin{aligned}
         &
        \witness{\aid_1} = H_a(\witness{\id_1,\dots,\id_t},u_{a1})
        \land
        \mtauth(r_{a1},\witness{\aid_1},\witness{\rho_{a1}}) = 1
        \;\land                                                \\
         &
        \witness{\aid_2} = H_a(\witness{\id_{t+1},\dots,\id_\ell},u_{a2})
        \land
        \mtauth(r_{a2},\witness{\aid_2},\witness{\rho_{a2}}) = 1
        \;\land                                                \\
         &
        n_{a1}=H_n(\witness{\id_1,\dots,\id_t},u_{a1})
        \land
        n_{a2}=H_n(\witness{\id_{t+1},\dots, \id_\ell},u_{a2}) \\
         & \land\;
        \aid' = H_a(\witness{\id_1,\dots,\id_\ell},u_a')
        \;\land\;
        u_a'=\max(u_{a1},u_{a2}).
    \end{aligned}
\]
$\holder$ sends $(r_{a1},r_{a2}, n_{a1}, n_{a2}, u_{a1}, u_{a2}, \aid',\pi_m)$ to $\registry$, who verifies $\pi_m$ and confirms that $r_{a1},r_{a2}$ are recorded tree roots and $n_{a1},n_{a2}$ not invalidated in $N$. If this is the case, $\registry$ invalidates $n_{a1}$ and $n_{a2}$, indicating that $\aid_1$ and $\aid_2$ are both \textit{deprecated}.

\subsection{Credential presentation}\label{subsec: credential-presentation}

Associated identifiers allow verifiers to require robust anti-Sybil proofs from credential holders. Specifically, verifiers can demand pre-association of all identifiers linked to presented credentials. This enables \sysname{} to instantly identify repeated participation in the same campaign.

\spar{Anti-Sybil credential presentation}
In a Sybil-sensitive campaign (e.g., an airdrop), the verifier $\verifier$ samples a session (campaign-wise) identifier $\sid\sample\Zp$ and broadcasts it to all potential participants, along with a public predicate $\phi$ that indicates the verification criterion for accepting credentials.

To participate in $\sid$ with credentials $\cred_1,\dots,\cred_n$, the holder $\holder$ justifies that none of these credentials has been presented in campaign $\sid$. Concretely, $\holder$ proves in zero-knowledge that:
\begin{enumerate}[parsep=0.5pt]
    \item[\cirn{182}] For each $i\in[n]$, $\cred_i$ has $\idh_i$ as its holder identifier,
    \item[\cirn{183}] For each $i\in[n]$, $\idh_i$ is a component of $\aid$,
    \item[\cirn{184}] $\aid$ is recorded by the ledger and is not \textit{blocklisted}, and
    \item[\cirn{185}] $\aid$ has not been \textit{presented} in $\sid$ before.
\end{enumerate}

\spar{Credential validity and authenticity}
\sysname{} supports selective disclosure of credentials, where $\holder$ demonstrates that the credential $\cred_i$ presented satisfies the criterion $\phi$. For each $i\in[n]$, $\holder$ proves in zero-knowledge that:
\begin{enumerate}[parsep=0.5pt]
    \item[\cirn{186}] The claim $s_i$ in $\cred_i$ satisfies $\phi(s_i)=1$,
    \item[\cirn{187}] $\cred_i$ is issued and signed by $\issuer_i$, and
    \item[\cirn{188}] $\cred_i$ is not \textit{revoked}.
    \item[\cirn{189}] $\cred_i$ is committed and recorded by $\issuer_i$.
\end{enumerate}

\spar{Non-transferable ownership}
\sysname{} ensures anonymity during the presentation while preventing the malicious transfer of credentials. It resolves this inherent dilemma by leveraging the above-established identity infrastructure. The key lies in ensuring that $\holder$ possesses the private key $\skh_i$ corresponding to $\idh_i$. For each $i\in[n]$, $\holder$ proves in zero-knowledge that:
\begin{enumerate}[parsep=0.5pt]
    \item[\cirn{190}] A tag $h_i$ is recorded by the registry $\registry$, and
    \item[\cirn{191}] $h_i$ derives from $\idh_i$ and $\skh_i$, where $\idh_i$ is exactly the holder identifier of $\cred_i$.
\end{enumerate}

\spar{Combining the constraints}
As described above, $\holder$ needs to justify the satisfaction of 10 constraints across the above three parts.
To clarify, for \cirn{184} and \cirn{185}, $\holder$ computes a global nullifier $n_a$ for $\aid$ and a campaign-wise nullifier $n_e$ with respect to $\aid$ in campaign $\sid$.
To justify \cirn{188}, $\holder$ hashes $\cred_i$ into $n^c_i$ along with a randomness $u^c_i$ sampled by $\holder$ during credential issuance.

Finally, $\holder$ generates a proof $\pi_c$ for relation $R_c$ composed of constraints \cirn{182}--\cirn{191}:
\[
    \begin{aligned}
         &
        \witness{\cred_i}=(\idi_i,\sigma_i,\witness{\idh_i},\witness{s_i})
        \;\land\;
        \witness{\aid}=H_a(\witness{\id_1,\dots,\id_\ell},\witness{u_a})            \\
         & \land\;
        \witness{\idh_i}\in\witness{\{\id_1,\dots,\id_\ell\}}
        \;\land\;
        n_a=H_n(\witness{\id_1,\dots,\id_\ell},\witness{u_a})                       \\
         & \land\;
        \mtauth(r_a, \witness{\aid}, \witness{\rho_a})=1
        \;\land\;
        n_e = H_n(\witness{\id_1,\dots,\id_\ell}, \sid)                             \\
         & \land\;
        \phi(\witness{s_i})=1
        \;\land\;
        \versign\big(\pki_i,\sigma_i,(\idi_i,\witness{\idh_i},\witness{s_i})\big)=1 \\
         & \land\;
        \witness{c^c_i}=H_a(\witness{\cred_i},\witness{u^c_i})
        \;\land\;
        n^c_i=H_n(\witness{\cred_i},\witness{u^c_i})                                \\
         & \land\;
        \witness{h_i}=H_a(\witness{\idh_i},\witness{\skh_i})
        \;\land\;
        \cmtopen(\witness{\skh_i},c^s_i,u^s_i)=1                                    \\
         & \land\;
        \mtauth(r_i,\witness{h_i},\witness{\rho_i})=1
        \;\land\;
        \mtauth(r^c_i, \witness{c^c_i}, \witness{\rho^c_i})=1.
    \end{aligned}
\]
where $\versign$ is the signature verification algorithm.

$\holder$ sends $(\{c^s_i\}_n, \{u^s_i\}_n, \{n^c_i\}_n, \{\idi_i\}_n, \{r_i\}_n, \{r^c_i\}_n, r_a, n_a, n_e, \pi_c)$ to $\verifier$, who resolves $\pki_i$ from the VDR using $\idi_i$.
Then, $\verifier$ confirms that $r_a$ and all $r_i$ are recorded tree roots, and that $n_a$ is not invalidated.
Also, $\verifier$ queries the credential management contract of $\issuer_i$ with $r_i^c$ and $n^c_i$ to confirm that all $\cred_i$ are issuer-recorded and not \textit{revoked}. This asynchronous, ledger-based check ensures that issuers do not need to remain online during credential presentations.


Finally, $\verifier$ verifies $\pi_c$, and confirms that $(\sid,n_e)$ is not recorded.
If all checks pass, $\verifier$ records $(\sid,n_e)$, indicating that $\aid$ has been presented in $\sid$, and any further presentation in $\sid$ using the same $\aid$ will be detected by $\verifier$ and rejected.
This also implies that if $\holder$ presents multiple credentials in a campaign, it must have already formed an $\aid$ between the holder identifiers $\id_i$ of these credentials.

    \spar{On-chain privacy boundaries and sequential ordering}
    To explicitly clarify the privacy guarantees and the underlying anti-Sybil mechanics during credential presentation, we delineate the strict boundary between public and private data:
    \begin{itemize}[leftmargin=*]
        \item \textit{Data published on-chain:} The protocol-level public values revealed to \(\verifier\) are the verification transcript
              \(
              (\{c^s_i\}_n,\{u^s_i\}_n,\allowbreak\{n^c_i\}_n,\allowbreak\{\idi_i\}_n,\{r_i\}_n,\{r_i^c\}_n,r_a,n_a,n_e,\pi_c).
              \)
              In smart contracts, this transcript may be visible as calldata, whereas the presentation step permanently records only \((\sid,n_e)\).               The transcript discloses issuer identifiers, roots, commitments, commitment randomness, nullifiers \((n_e,n_a,\{n_i^c\})\), and the proof \(\pi_c\). The nullifiers and commitments appear pseudo-random, while issuer identifiers and roots serve as public verification metadata.
              

        \item \textit{Data hidden in the presentation proof:} The holder identifiers \(\idh_i\), private keys \(\skh_i\), undisclosed claims and signatures in \(\cred_i\), Merkle paths \((\rho_a,\rho_i,\rho_i^c)\), and the association witness \((\aid,\{\id_j\},\allowbreak u_a)\) are encapsulated as private circuit witnesses. Although \(\aid\) is a public leaf inserted during association, \(\pi_c\) hides which recorded \(\aid\) leaf is used and which identifiers compose it.
    \end{itemize}
    Crucially, this enforces a strict \textit{sequential ordering invariant}. The constraint $\mtauth(r_a, {\aid}, {\rho_a})=1$ dictates that $\holder$ must finalize the association $\aid$ on the ledger \textit{before} generating the presentation proof. Utilizing the ledger as a chronological anchor binds each presentation to a pre-existing association state and prevents ephemeral, unrecorded associations from bypassing the Sybil check.

\spar{Updating after verification}
A notable attack vector involves $\holder$ participating in campaign $\sid$ with an anti-Sybil proof using $\aid$, then updating to $\aid'$ after the interaction ends to participate again. This creates a different but valid campaign-wise nullifier, circumventing \sysname's anti-Sybil protection.

To counter this, $\verifier$ can defer verification of proof $\pi_c$ until campaign $\sid$ concludes, ensuring that any updated nullifier $n_a$ already appears in $\registry$'s nullifier set $N$. This works for airdrops and DAO voting. For immediate authentication, Appendix~\ref{subapp: instant-verification} gives an extension that requires the association to predate the campaign and proves that its Merkle-tree index lies below the campaign-start threshold.

\spar{Nullifier-based correlation}
Multiple presentations of the same $\aid$ and $\cred$ repeat nullifiers $n_a$ and $n_i^c$, creating a \textit{linkability risk}: a curious $\verifier$ can correlate claims without learning the holder's identity. Holders mitigate this risk by refreshing the association nonce $u_a$ and credential nonce $u_i^c$, thereby rotating the corresponding public handles; Appendices~\ref{subsec: identifier-accountability} and~\ref{subsec: credential-revocation} formalize both procedures.

\spar{Accountability and proof integrity}
Beyond Sybil resistance, \sysname{} provides accountability through \textit{credential revocation} and \textit{identifier blocklisting}. The identifier association mechanism delivers superior accountability compared to previous systems by extending penalties from a single identifier to its entire association.
\sysname{} also prevents \textit{proof-forwarding} through designated-verifier proofs. Appendix~\ref{sec: extensions} specifies these constructions together with credential revocation and identifier blocklisting.

\subsection{Key recovery}\label{subsec: key-recovery}

Identifier association serves dual purposes: countering Sybil attacks and providing robust key recovery. While traditional key management depends on a \textit{single atomic secret} (whose loss means irreversible control loss), \sysname{} shifts to \textit{provable structural knowledge}: demonstrating control over components that make up a user's aggregated identity $\aid$.

This reframes recovery from \textit{`Do you remember a secret?'} to \textit{`Can you prove you still control parts of your identity?'} \sysname{} offers two trustless recovery modes balancing security, usability, and identity complexity.

\spar{Threshold key-based recovery}
This is \sysname's default recovery mechanism, providing strong security against partial key compromise. The core principle: to recover a lost key for one identifier, $\holder$ must prove control over a \textit{threshold $t$} of other identifiers within the same association.

When $\holder$ loses key $\sk$ for $\id$, it can select $t$ distinct identifiers $\{\id'_1, \dots, \id'_t\}$ from the same $\aid$, and demonstrate to $\registry$ its knowledge of corresponding keys $\{\sk'_1, \dots, \sk'_t\}$. To recover, $\holder$ samples a new private key $\sk'' \sample \Zp$ and computes $\pk'' := \pubkeygen(\sk'')$, along with a tag $h'' := H_a(\id, \sk'')$ and a nullifier $n'' := H_n(\id, \sk'')$. $\holder$ proves in zero-knowledge that:
\begin{enumerate}[parsep=0.5pt]
    \item[\cirn{182}] $\aid$ derives from $\id_1,\dots,\id_\ell$ with the nonce $u_a$,
    \item[\cirn{183}] $\aid$ is recorded by the ledger and not \textit{blocklisted},
    \item[\cirn{184}] $\pk''$ is the corresponding public key of $\sk''$,
    \item[\cirn{185}] Both $h''$ and $n''$ derive from $\id$ and $\sk''$,
    \item[\cirn{186}] $\id$ and $\{\id'_1, \dots, \id'_t\}$ are all components of $\aid$, and
    \item[\cirn{187}] For all $\id'$ in $\{\id'_1, \dots, \id'_t\}$, $h'$ derives from $\id'$ and its $\sk'$.
\end{enumerate}
This emits a proof $\pi_k$ for relation $R_k$:
\[
    \begin{aligned}
         &
        \witness{\aid}=H_a(\witness{\id_1,\dots,\id_\ell},\witness{u_a})
        \;\land\;
        \mtauth(r_a,\witness{\aid},\witness{\rho_a}) = 1 \\
         & \land\;
        n_a = H_n(\witness{\id_1,\dots,\id_\ell,u_a})
        \;\land\;
        \id\in\witness{\{\id_1,\dots,\id_\ell\}}         \\
         & \land\;
        \witness{\{\id'_1, \dots, \id'_t\}}\subset\witness{\{\id_1,\dots,\id_\ell\}}
        \;\land\;
        \pk''=\pubkeygen(\witness{\sk''})                \\
         & \land\;
        \forall j \in [t]:
        \left(
        \witness{h'_j}=H_a(\witness{\id'_j}, \witness{\sk'_j})
        \;\land\;
        \mtauth(r',\witness{h'_j},\witness{\rho'_j}) = 1
        \right)                                          \\
         & \land\;
        h''=H_a(\id,\witness{\sk''})
        \;\land\;
        n''=H_n(\id,\witness{\sk''}).
    \end{aligned}
\]
$\holder$ sends $(r',r_a,n_a,n'',h'',\id,\pk'',\pi_k)$ to $\registry$. $\registry$ verifies $\pi_k$ and checks the on-chain state (that $n_a$ and $n''$ are not nullified, and $r$ and $r_a$ are recorded roots). If valid, $\registry$ records $h''$ and $(\id,\pk'')$, and invalidates $n''$ to $N$.


\spar{Entropy-based recovery}
For $\holder$ with a large and complex digital identity, \sysname{} offers a key-agnostic recovery option. When an $\aid$ contains a sufficiently large number of identifiers, the exact composition of $\aid$, represented as the canonically ordered identifier vector used in the association, becomes a high-entropy secret known only to $\holder$.
In this mode, recovery shifts from proving key control to demonstrating knowledge of this structure. Users can retrieve public blockchain records to reconstruct this proof, eliminating the need to memorize complex details.

Concretely, to recover the key for $\id \in \{\id_1, \dots, \id_\ell\}$, $\holder$ proves in zero-knowledge that:
\begin{enumerate}[parsep=0.5pt]
    \item[\cirn{182}] $\aid$ derives from $\id_1,\dots,\id_\ell$ with the nonce $u_a$,
    \item[\cirn{183}] $\aid$ is recorded by the ledger and not \textit{blocklisted},
    \item[\cirn{184}] $\id$ is a component of $\aid$, i.e., $\id \in \{\id_1, \dots, \id_\ell\}$,
    \item[\cirn{185}] $\pk''$ is the corresponding public key of $\sk''$, and
    \item[\cirn{186}] Both $h''$ and $n''$ derive from $\id$ and $\sk''$.
\end{enumerate}
This emits a proof $\pi_k'$ for relation $R_k'$:
\[
    \begin{aligned}
         &
        \witness{\aid}=H_a(\witness{\id_1,\dots,\id_\ell},\witness{u_a})
        \;\land\;
        \mtauth(r_a,\witness{\aid},\witness{\rho_a}) = 1 \\
         & \land\;
        n_a = H_n(\witness{\id_1,\dots,\id_\ell,u_a})
        \;\land\;
        \id\in\witness{\{\id_1,\dots,\id_\ell\}}         \\
         & \land\;
        \pk''=\pubkeygen(\witness{\sk''})                \\
         & \land\;
        h''=H_a(\id,\witness{\sk''})
        \;\land\;
        n''=H_n(\id,\witness{\sk''}).
    \end{aligned}
\]
For these operations to maintain adequate security, $\holder$ must ensure that $\aid$ contains more than $q$ identifiers, where $n$ is the total number of leaves in Merkle tree $T$. With $n\gg q$, the probability of an adversary satisfying relation $R_k'$ is approximately $p\approx1/n^q$. Security requires that $p<1/2^\lambda$, comparable to the probability of guessing a private key for an identifier.

\begin{figure*}[t]
    \centering
    \input{floats/frames/protocol-main}
    \caption{\sysname{} main protocol.}
    \label{fig: protocol-full}
    \Description{The full protocol of \sysname{}, containing four operations.}
\end{figure*}

\subsection{Security}\label{subsec: security}

In the subsequent analysis, we examine the security properties of protocol $\prob$ by motivating the underlying mechanisms that enable it to fulfill the security requirements delineated in Section~\ref{subsec: security-properties}.

\spar{Credential and identifier privacy}
Privacy is ensured by limiting the verifier $\verifier$'s knowledge to only the satisfaction of $\phi(s)=1$ for $s$ in $\cred$. Concretely, $\verifier$ cannot obtain extra information about $\cred$ or identify individual $\id$ in an $\aid$. This is guaranteed by the zero-knowledge property of $\pi_c$, which ensures that $\verifier$ gains no knowledge about the witness beyond what is stated in $R_c$. For identifier privacy, we consider the following two scenarios:
\begin{itemize}
    \item During \textit{identifier association}, the nullifier $n_i$ represents the sole potential source of information facilitating the deduction of $\aid$'s components, while the one-wayness of $H_n$ ensures that $\id_i$ cannot be reversed from $n_i$.
    \item During \textit{credential presentation}, $\verifier$ is computationally incapable of reversing any $\id$ constituting $\aid$ or $n_a$, given the collision-resistance of $H_a$ and $H_n$.
\end{itemize}

\spar{Sybil-resistance}
$\holder$ cannot submit $\cred$'s under the same $\aid$ within multiple interactions of $\sid$. To illustrate this, assume that $\holder$ presents $\cred_1$ in one interaction and $\cred_2$ in another, but the holder identifiers $\idh_1$ and $\idh_2$ of both credentials have been incorporated into the $\aid$. To accomplish this, $\holder$ would need to construct a $\pi_c$ in the second interaction using a $n_e'$ different from the $n_e$ in the former interaction. This is computationally infeasible unless the holder compromises the soundness of $\pi_c$. Formally, the following theorem bounds the attack probability; its proof appears in Appendix~\ref{app: sybil-resistance}.

\begin{theorem}\label{thm: sybil-resistance}
    Consider a holder $\holder$ who participates in $q$ consecutive campaigns $\sid_1,\dots,\sid_q$, each of which requires $\holder$ to present at least $k$ credentials meeting specific predicate requirements. Let $W$ be the event that $\holder$ successfully launches a Sybil attack at the $q$-th campaign with a valid proof to the verifier $\verifier$. Then, we have:
    \begin{equation*}
        \Pr[W]
        \leq
        \binom{s(q)}{k}
        \cdot
        \left(
        \frac{\ell(q)-1}{\ell(q)^{2k-1}}
        \right)
        +
        \frac{q^2\cdot\ell(q)^2}{2^\lambda},
    \end{equation*}
    where $s(q)$ is the number of credentials satisfying the predicates at the $q$-th campaign, and $\ell(q)$ is the number of unique identifiers possessed by $\holder$ at the $q$-th campaign. We assume $s(q)=\poly(\lambda)$ and $\ell(q)=\poly(\lambda)$.
\end{theorem}

\spar{Key recovery}
The capability to update a keypair associated with a specific $\id$ is restricted exclusively to its legitimate owner. This constraint is rooted in the fact that only the rightful owner possesses knowledge of the $\aid$ into which the $\id$ has been incorporated, as well as the private key $\sk'$ corresponding to another $\id'$ within that $\aid$. The probability of any unauthorized entity generating a valid proof $\pi_k$ and successfully convincing $\registry$ is negligible.


\spar{Unlinkability}
\sysname{} provides contextual unlinkability rather than global unlinkability. The proof \(\pi_c\) hides holder identifiers, undisclosed claims, Merkle paths, and the selected association witness. However, public nullifiers remain equality handles: reusing the same association state repeats \(n_a\), and reusing the same credential state repeats \(n_i^c\). Thus, presentations are unlinkable only after the relevant handles have changed. The public roots used for membership checks are registry states and do not by themselves reveal which credential or association was used.

\spar{Non-transferability}
$\holder$ without the private key $\skh$ corresponding to the holder identifier $\idh$ of the credential $\cred$ cannot legitimately present it. The relation $R_c$ incorporates a tag $h$ derived from $\skh$ into the ledger's Merkle tree. Given the soundness of $\pi_c$ and collision resistance of the underlying hash function of the Merkle construct, the probability of constructing a valid $\pi_c$ without $\skh$ is negligible.

\spar{Formal security definition}
To formally define the security of protocol $\prob$, we use the UC security definition introduced in Section~\ref{sec: overview}.

\begin{theorem}\label{thm: uc-security}
    A multi-party protocol $\prob$ exists that UC-realizes the ideal functionality $\funb$ under the $\funl$-hybrid model.
\end{theorem}
\noindent
Appendix~\ref{app: security-uc} proves Theorem~\ref{thm: uc-security} by constructing simulators that show the execution of protocol $\prob$ in the $\funl$-hybrid model is indistinguishable from the execution of $\funb$ under the stated corruption cases.

%% file: floats/frames/protocol-main.tex
\protocol{Protocol $\prob$} {
This protocol runs between the issuer $\issuer$, the credential holder $\holder$, the verifier $\verifier$, and the verifiable registry $\registry$.
Parties in this protocol have access to the decentralized ledger functionality $\funl$.
In this functionality, we use $\party$ to refer to any possible entity other than $\registry$.

\method{Identifier Registration:}
On receiving $(\mathtt{register},\sid)$, party $\party$ performs as follows:
\begin{enumerate}[parsep=0.5pt]
    \item $\party$ samples $\sk\sample\Zp$, calculates $\pk:=\pubkeygen(\sk)$, and then sends $\pk$ to $\registry$.
    \item $\registry$ samples $\id\sample\Zp$, and sends $(\mathtt{register},\sid,\id,\pk)$ to $\funl$. On receiving $(\mathtt{registered},\sid)$ from $\funl$, $\registry$ returns $\id$ to $\party$.
    \item $\party$ records $(\id,\pk,\sk)$, computes $h:=H_a(\id,\sk)$, and generates a $\pi_g$ for relation $R_g$. $\party$ sends $(h,\pi_g)$ to $\registry$.
    \item $\registry$ verifies $\pi_g$ and aborts if verification fails. Else, $\registry$ sends $(\mathtt{record},\sid,h)$ to $\funl$. On receiving $(\mathtt{recorded},\sid,r,\rho)$ from $\funl$, $\registry$ sends $(r,\rho)$ to $\party$.
    \item $\party$ records $(\id,h,r,\rho)$, and then outputs $(\mathtt{registered},\sid,\id)$.
\end{enumerate}

\method{Identifier Association:}
On receiving $(\mathtt{associate},\sid,\{\id_i\}_\ell)$, party $\party$ performs as follows:
\begin{enumerate}[parsep=0.5pt]
    \item $\party$ checks if $(\id_i,\pk_i,\sk_i)$ and $(\id_i,h_i,r_i,\rho_i)$ are recorded for all $i\in[\ell]$. If not, $\party$ aborts.
    Else, $\party$ computes $\aid:=H_a(\id_1,\dots,\id_\ell,u_a:=0)$ and $n_i:=H_n(\id_i,\sk_i)$ for all $i$. $\party$ generates a proof $\pi_a$ for relation $R_a$. Finally, $\party$ sends $(\aid,\{r_i\}_\ell,\{n_i\}_\ell,u_a,\pi_a)$ to $\registry$.
    \item $\registry$ verifies $\pi_a$ and aborts if verification fails. For all $i\in[\ell]$, $\registry$ sends $(\mathtt{check},\sid,r_i)$ to $\funl$. If $\funl$ returns $(\mathtt{unrecorded},\sid)$ for any $i$, $\registry$ aborts.
    \item For all $i\in[\ell]$, $\registry$ sends $(\mathtt{check},\sid,n_i)$ to $\funl$. If $\funl$ returns $(\mathtt{invalid},\sid)$ for any $i$, $\registry$ aborts.
    \item $\registry$ sends $(\mathtt{record},\sid,\aid)$ to $\funl$. On receiving $(\mathtt{recorded},\sid,r_a,\rho_a)$ from $\funl$, $\registry$ sends $(r_a,\rho_a)$ to $\party$, and sends $(\mathtt{invalidate},\sid,n_i)$ for all $i\in[\ell]$ to $\funl$.
    \item $\party$ records $(\aid,r_a,\rho_a)$ and $(\aid,\id_1,\dots,\id_\ell,u_a)$, and then outputs $(\mathtt{associated},\sid,\aid)$.
\end{enumerate}

\method{Credential Presentation:}
On receiving $(\mathtt{campaign},\sid,\phi)$ and $(\mathtt{present},\sid,\aid,\{\cred_i\}_n)$, $\holder$ performs as follows:
\begin{enumerate}[parsep=0.5pt]
    \item For all $i\in[n]$, $\holder$ parses $\cred_i:=(\idi_i,\sigma_i,\idh_i,s_i)$. Then, $\holder$ sends $(\mathtt{retrieve},\sid,\idi_i)$ to $\funl$, receiving $(\mathtt{retrieved},\sid,\pki_i)$.
    \item $\holder$ retrieves $(\aid,\id_1,\dots,\id_\ell,u_a)$ and $(\aid,r_a,\rho_a)$ from its record, along with $(\cred_i,u^c_i,c^c_i,r^c_i,\rho^c_i)$, $(\idh_i,\cdot,\skh_i)$ and $(\idh_i,h_i,r_i,\rho_i)$ for all $i\in[n]$.
    Then, $\holder$ computes $n_a:= H_n(\id_1,\dots,\id_\ell,u_a)$ and $n_e:= H_n(\id_1,\dots,\id_\ell,\sid)$. For all $i\in[n]$, $\holder$ computes $(c^s_i,u^s_i):=\cmtcommit(\skh_i)$ and $n^c_i:=H_n(\cred_i,u^c_i)$.
    Finally, $\holder$ generates a proof $\pi_c$ for relation $R_c$, and sends $(\{c^s_i\}_n,\{u^s_i\}_n,\{n^c_i\}_n,\{\idi_i\}_n,\{r_i\}_n,\{r_i^c\}_n,r_a,n_a,n_e,\pi_c)$ to $\verifier$.
    \item $\verifier$ sends $(\mathtt{check},\sid,r_a)$ and $(\mathtt{check},\sid,n_a)$ to $\funl$, along with $(\mathtt{check},\sid,r_i)$ and $(\mathtt{retrieve}, \sid, \idi_i)$ for all $i\in[n]$. It also checks, via the contract of \(\idi_i\), that \(r_i^c\) is a valid credential-tree root and that \(n_i^c\) is not in the revocation list.
    On receiving $\mathtt{checked}$ for $r_a$ and all $r_i$, $\mathtt{unrecorded}$ for $n_a$, successful issuer-contract checks for all \((r_i^c,n_i^c)\), and $(\mathtt{retrieved},\sid,\pki_i)$ for all $\idi_i$, $\verifier$ checks if $(\sid,n_e)$ is recorded and rejects $\holder$ if this is the case.
    Finally, $\verifier$ verifies $\pi_c$ and aborts if verification fails.
    Else, $\verifier$ records $(\sid,n_e)$, and outputs $(\mathtt{verified},\sid,\holder)$.
\end{enumerate}

\method{Key Recovery:}
On receiving $(\mathtt{recover},\sid,\aid,\id,\{\id'\}_t,\{\sk'\}_t)$, party $\party$ performs as follows:
\begin{enumerate}[parsep=0.5pt]
    \item $\party$ retrieves $(\aid,\id_1,\dots,\id_\ell,u_a)$, $(\aid,r_a,\rho_a)$ and $(\id'_j,h'_j,r',\rho'_j)$ ($\forall j\in[t]$) from its record, and computes $n_a:= H_n(\id_1,\dots,\id_\ell,u_a)$.
    Then, $\party$ samples a new private key $\sk''\sample\Zp$, and computes $\pk'':=\pubkeygen(\sk'')$, $h'':= H_a(\id,\sk'')$ and $n'':= H_n(\id,\sk'')$.
    Finally, $\party$ generates a proof $\pi_k$ for the relation $R_k$, and sends $(r',r_a,n_a,n'',h'',\id,\pk'',\pi_k)$ to $\registry$.
    \item $\registry$ sends $(\mathtt{check},\sid,r_a)$, $(\mathtt{check},\sid,r')$, $(\mathtt{check},\sid,n_a)$ and $(\mathtt{check},\sid,n'')$ to $\funl$.
    On receiving $\mathtt{checked}$ for $r_a, r'$ and $\mathtt{unrecorded}$ for $n_a, n''$, $\registry$ verifies $\pi_k$, and aborts if verification fails.
    Then, $\registry$ sends $(\mathtt{register},\sid,\id,\pk'')$, $(\mathtt{record},\sid,h'')$ and $(\mathtt{invalidate},\sid,n'')$ to $\funl$.
    On receiving $(\mathtt{registered},\sid)$, $(\mathtt{recorded},\sid,r'',\rho'')$ and $(\mathtt{invalidated},\sid)$ from $\funl$, $\registry$ sends $(r'',\rho'')$ to $\party$.
    \item $\party$ records $(\id,h'',r'',\rho'')$, and outputs $(\mathtt{recovered},\sid)$.
\end{enumerate}
}

%% file: sections/body/evaluation.tex
\section{Implementation and evaluation}\label{sec: evaluation}

We evaluate \sysname{} through micro-benchmarks of its core operations, end-to-end latency from proof generation to confirmation, on-chain throughput and bottlenecks, and comparisons with two systems that provide weaker security guarantees.

\spar{Experimental setup}
Holder experiments run on an 8-core M2 Mac mini with 8GB RAM and an 80Mbps/15Mbps downstream/upstream connection. Verifier experiments run on an Intel Ultra 7-265K desktop with 32GB RAM and a 1Gbps connection. These configurations represent a consumer device and a higher-performance workstation, respectively.

\spar{Implementation details}
We implement the arithmetic circuit in \sysname\footnote{Repository: \artifacturl} and construct zero-knowledge proofs using the Plonk implementation provided by gnark~\cite{gabizon2019plonk, gnarkv0.9, consensys2021gnark}, which uses BLS12-381 as the underlying elliptic curve and the KZG scheme for polynomial commitment~\cite{kate2010constant}.
\sysname{} utilizes Poseidon as the commitment primitive, which is an SNARK-friendly sponge construction~\cite{grassi2021poseidon}. Specifically, we use a $t=12$ Poseidon-128 with $R_F=8$ and $R_P=56$, referring to the instantiation of Plonky2~\cite{mir2021plonky}. The height of the Merkle tree $T$ maintained by the registry $\registry$ is set to 32, which is consistent with the configuration in Tornado Cash~\cite{pertsev2019tornado, yang2022reducing}.

\subsection{Micro-benchmarks of core operations}

We first analyze the performance of individual cryptographic operations in \sysname.

\subsubsection{Identifier association and key recovery}

We evaluate proof generation time for identifier association operations, including association, appending, merging, and refreshing. Figure~\ref{plot: association-recovery} shows how proof time scales with the number of identifiers incorporated by the association.

\begin{figure}[t]
	\centering
	\begin{minipage}{0.47\columnwidth}
		\hspace{-8pt}
		\resizebox{1.05\columnwidth}{!}{\input{floats/plots/association-recovery}}
		\caption{Proving time for association and key recovery.}
		\label{plot: association-recovery}
	\end{minipage}
	\hspace{8pt}
	\begin{minipage}{0.47\columnwidth}
		\hspace{-8pt}
		\resizebox{1.05\columnwidth}{!}{\input{floats/plots/credential-presentation}}
		\caption{Proving time for credential presentation.}
		\label{plot: presentation-verification}
	\end{minipage}
\Description{Two line charts of holder proving time. The left plots proving time against 1--20 identifiers for association, appending, merging, refreshing, and key recovery; association grows from 0.13 to 2.41 seconds and is the steepest series. The right plots proving time against 1--10 credentials containing 2, 5, or 10 claims; time increases with both credential and claim count.}
\end{figure}

\spar{Identifier generation and aggregation}
Identifier generation averages 12,442 constraints, comprising one hash calculation and one key-derivation evaluation, and takes 39ms to prove. Figure~\ref{plot: association-recovery} shows linear association scaling: 10 identifiers produce 403,502 constraints and require 1.19s, while 20 produce 0.8M constraints and require 2.41s. Each additional identifier adds about 40,350 constraints and 120ms of proving time, or approximately 160ms including identifier generation. Thus, association remains a seconds-scale holder operation on the tested consumer-grade device.

\spar{Association update}
Appending cost grows with the existing association because the circuit verifies the correct construction of $\aid$. Appending to a 20-identifier association produces 120,770 constraints and takes approximately 410ms. For the same final identifier count, merging is costlier because the circuit evaluates the $\aid$ construction three times: merging two 10-identifier associations into one with 20 identifiers produces 138,088 constraints and takes 440ms. Refreshing is lighter; incrementing $u_a$ for a 20-identifier association proves an implicit correlation between the two states, producing 90,424 constraints and taking 280ms.

\spar{Key recovery}
Key recovery resembles refreshing because both prove the construction of $\aid$, its inclusion in $T$, and constant-size assertions. Recovery also requires enough associated identifiers for cryptographic entropy. With 20 identifiers, the circuit has 86,768 constraints and takes 0.26s to prove.

\spar{Gas costs and on-chain footprint}
We measured on-chain costs on a local Hardhat network emulating Ethereum mainnet. For $\ell=20$, Table~\ref{tab: gas-consumption} reports about 690K gas for association, or \$0.61 at 0.20 Gwei using the experiment's conversion; this is a one-time identity-base cost. Refreshing and key recovery cost about 336K and 313K gas, respectively (approximately \$0.30 and \$0.28 under the same experiment conversion). Each nullifier occupies 32 bytes, and retaining the last 100 Merkle roots uses 3.2 KB.

\begin{table}[t]
	\centering
	\caption{Gas consumption of interacting with the registry.}
	\label{tab: gas-consumption}
	\resizebox{0.63\columnwidth}{!}{%
		\input{floats/tables/contract}
	}
\end{table}

\subsubsection{Credential presentation and verification}

For credential presentation, we evaluate the zero-knowledge proofs used in selective disclosure, examining their theoretical complexity and computational costs.

\spar{Theoretical complexity}
Regarding credential verification, we compare the theoretical complexity of \sysname{} with several recent milestone schemes, as illustrated in Table~\ref{tab: proving-time-constraints}. Coconut and \texttt{zk-creds} have proof sizes and verification complexity that scale linearly with the number of claims $n$. Adding features like credential revocation further increases this verification complexity. While ZEBRA achieves constant-size proofs and verification complexity, its verification requires 4 pairing operations—significantly more time-consuming than \sysname's 2 pairings.
\sysname{} integrates revocation natively without adding extra pairing operations to proof verification; the revocation-supporting relation only adds a credential-tree membership check inside the proved statement.

\begin{table}[t]
	\centering
	\caption{Comparison of complexity with established schemes.}
	\label{tab: proving-time-constraints}
	\resizebox{0.88\columnwidth}{!}{%
		\input{floats/tables/theory-complexity}
	}
\end{table}

\spar{Holder overhead}
During credential presentation, the holder calculates a proof for all $\cred_i$ in relation to $R_c$. The corresponding arithmetic circuit performs two main computations: the evaluation of $\phi$ against $s$ and signature verification, with the latter comprising about 95,000 constraints. Hash-based computations add another 71,000 constraints. The total number of constraints varies based on the number of elements in $s$ and the complexity of $\phi$. Figure~\ref{plot: presentation-verification} illustrates the computational overhead for holders generating proofs when presenting multiple credentials at once. The overhead varies with the number of claims per credential, with 5 credentials (5 claims each) taking about 1.14s.

\spar{Verifier overhead}
The time consumed for credential verification in \sysname{} remains constant regardless of credential complexity. This is evidenced by the verification process requiring only 2 pairing operations on BLS12-381 and 12 multiplications on $\Gpa$, as outlined in Table~\ref{tab: proving-time-constraints}. However, the overall verifier time does increase slightly with credential size due to statement processing overhead. For a presentation of 5 credentials containing 5 claims each, the total verification time is approximately 21ms.

\subsection{End-to-end system performance}

While micro-benchmarks are informative, end-to-end latency determines the real-world user experience. We measured the total time from the moment a holder initiates an action to its final confirmation on the blockchain.

\spar{Baseline systems}
We compare against two weaker designs:
\begin{itemize}
	\item \textbf{Baseline-Anon} proves ownership of one anonymous credential but provides no Sybil resistance; it isolates the overhead of \sysname's anti-Sybil mechanisms.
	\item \textbf{Baseline-SID} derives $H(\id \Vert \sid)$ as a nullifier to prevent duplicate participation within one campaign but not across campaigns; it isolates the cost of \sysname's progressive linkage.
\end{itemize}

\spar{Wall-clock latency}
Figure~\ref{plot: e2e} reports credential-presentation latency for an association of 20 identifiers. At five credentials, \sysname{} takes about 14.3s: 1.2s for proof generation, 0.8s for submission through a public RPC endpoint, 12s on average for Ethereum propagation and block inclusion, and 0.3s for on-chain verification. Baseline-SID takes about 13.4s and Baseline-Anon about 12.9s; the difference reflects their smaller proofs and, for Baseline-Anon, absence of nullifier checks.

\spar{Communication overhead}
For three credentials and an association of 10 identifiers, \sysname's proof and public inputs total 1,271 bytes. Baseline-SID requires 890 bytes for its nullifier and proof, while Baseline-Anon requires 780 bytes for credential-validity proof alone.

\spar{Scalability and system throughput}
We benchmarked association and key-recovery throughput on Ethereum and Arbitrum~\cite{arbitrum2023site, tracker2023gas}. Figure~\ref{plot: throughput} reports roughly a 5--8$\times$ Layer-2 increase across the plotted points; key recovery outperforms the more state-intensive association operation on both platforms. Throughput falls as association size grows, indicating that sequential on-chain Merkle-tree and nullifier updates, rather than users' parallel proof generation enabled by caching the last 100 roots, limit settlement throughput.

Tornado Cash provides a deployment precedent for privacy-preserving protocols~\cite{pertsev2019tornado}, but does not by itself establish Layer-2 feasibility. Each \sysname{} interaction has a constant-size state footprint, such as a 32-byte nullifier, and each user pays the corresponding gas fee. Together with the measured Layer-2 throughput, these properties suggest that settlement overhead can remain economically manageable at scale under this fee model.

\begin{figure}[t]
	\centering
	\begin{minipage}{0.47\columnwidth}
		\hspace{-8pt}
		\resizebox{1.05\columnwidth}{!}{\input{floats/plots/wall-clock}}
		\caption{The E2E latency for an $\aid$ with 20 identifiers.}
		\label{plot: e2e}
	\end{minipage}
	\hspace{8pt}
	\begin{minipage}{0.47\columnwidth}
		\hspace{-8pt}
		\resizebox{1.05\columnwidth}{!}{\input{floats/plots/throughput}}
		\caption{On-chain throughput on different platforms.}
		\label{plot: throughput}
	\end{minipage}
\Description{Two line charts of end-to-end performance. The left compares latency for Braid, Baseline-SID, and Baseline-Anon across 1--10 credentials; all rise with credential count, with Braid highest and Baseline-Anon lowest. The right compares association and key-recovery throughput on Ethereum and Arbitrum for 2--25 identifiers; throughput falls as associations grow, while Arbitrum and key recovery attain higher rates.}
\end{figure}

\begin{figure}[t]
	\centering
	\begin{minipage}{0.47\columnwidth}
		\hspace{-8pt}
		\resizebox{1.05\columnwidth}{!}{\input{floats/plots/sybil-m10}}
		\caption{Sybil attack success rate with initial $\ell=10$.}
		\label{plot: m10}
	\end{minipage}
	\hspace{8pt}
	\begin{minipage}{0.47\columnwidth}
		\hspace{-8pt}
		\resizebox{1.05\columnwidth}{!}{\input{floats/plots/sybil-m100}}
		\caption{Sybil attack success rate with initial $\ell=100$.}
		\label{plot: m100}
	\end{minipage}
\Description{Two log-scale charts compare theoretical solid curves and simulated dashed curves for Sybil-attack success across 1--20 campaigns with $k=3,4,5$ credentials per presentation. The left starts with $\ell=10$ identifiers and the right with $\ell=100$; every curve declines with additional campaigns, and larger $k$ and $\ell$ yield lower success rates.}
\end{figure}

\spar{Sybil resistance analysis}
Figures~\ref{plot: m10} and~\ref{plot: m100} compare theoretical and simulated Sybil-attack success probabilities across campaigns for a malicious holder $\holder$ starting with associated sets of $10$ and $100$ identifiers. Verifiers require $k=3,4,5$ qualifying credentials per presentation. Both analysis and simulation assume that $s(q)$ grows linearly with $q$, modeling credentials that $\holder$ obtains dynamically from issuers during presentation.

In every evaluated case, success probability falls rapidly with each additional campaign and approaches negligible levels in the plotted ranges. Increasing $k$ also makes it more likely that the presented credentials belong to different identifiers, forcing earlier association and further reducing attack success.

A deployment-relevant regime may have far more unique identifiers $\ell(q)$ than valid credentials $s(q)$: raw identifiers can be generated cheaply, whereas issuer-verified credentials may require time and capital. For illustration, an attacker might acquire $s=10$ credentials but create $\ell=1000$ bot identifiers. With the evaluated baseline $k=3$ credentials per presentation, the analytic bound is approximately $1.2 \times 10^{-10}$. Under these assumptions, credential scarcity substantially reduces both the success probability and economic attractiveness of large-scale Sybil attacks.

%% file: floats/plots/association-recovery.tex
\pgfplotsset{
    scale only axis,
    every axis y label/.style = {
        at = {(-0.13, 0.5)},
        rotate = 90,
        anchor = center,
        font=\LARGE,
    },
    every axis x label/.style = {
        at = {(0.5,-0.17)},
        anchor = center,
        font=\LARGE,
    },
    tick label style = {
    	scale = 1.4,
        yshift=-0.1ex,
    },
    axis background/.style={
        fill = white,
    },
    grid=both,
    major grid style={
        opacity=.5
    },
    legend style={
        legend cell align = left,
        align = left,
        draw = white!15!black,
        font=\Large,
    }
}

\begin{tikzpicture}

\begin{axis}[%
    width=3.3in,
    height=2.4in,
    xmin=0,
    xmax=20,
    ymin=0,
    ymax=2.5,
    xlabel={\# Identifiers},
    ylabel={Proving time/\textit{s}},
    legend pos=north west,
]

\addplot [color=ruri, line width=1pt, mark size=2pt, mark=*, mark options={solid, ruri, fill=white}, smooth]
  table[row sep=crcr]{%
        1	0.13\\
        2	0.25\\
        3	0.36\\
        4	0.49\\
        5	0.59\\
        6	0.72\\
        7	0.83\\
        8	0.95\\
        9	1.07\\
        10	1.19\\
        11	1.31\\
        12	1.42\\
        13	1.55\\
        14	1.66\\
        15	1.78\\
        16	1.88\\
        17	2.03\\
        18	2.15\\
        19	2.24\\
        20	2.41\\
};
\addlegendentry{\textit{association}}

\addplot [color=momo, line width=1pt, mark size=1.8pt, mark=*, mark options={solid, momo}, smooth]
  table[row sep=crcr]{%
        1	0.26\\
        2	0.27\\
        3	0.28\\
        4	0.29\\
        5	0.30\\
        6	0.30\\
        7	0.31\\
        8	0.32\\
        9	0.32\\
        10	0.34\\
        11	0.34\\
        12	0.34\\
        13	0.35\\
        14	0.36\\
        15	0.36\\
        16	0.37\\
        17	0.38\\
        18	0.39\\
        19	0.39\\
        20	0.41\\
};
\addlegendentry{\textit{appending}}

\addplot [color=kogane!95!black, line width=1pt, mark size=2.6pt, mark=triangle*, mark options={solid, kogane!95!black, fill=white}, smooth]
  table[row sep=crcr]{%
        2	0.29\\
        3	0.29\\
        4	0.30\\
        5	0.31\\
        6	0.31\\
        7	0.33\\
        8	0.33\\
        9	0.33\\
        10	0.34\\
        11	0.35\\
        12	0.35\\
        13	0.37\\
        14	0.37\\
        15	0.38\\
        16	0.39\\
        17	0.39\\
        18	0.40\\
        19	0.41\\
        20	0.44\\
};
\addlegendentry{\textit{merging}}

\addplot [color=ao, line width=1pt, mark size=2.4pt, mark=diamond*, mark options={solid, ao, fill=white}, smooth]
  table[row sep=crcr]{%
        1	0.14\\
        2	0.14\\
        3	0.15\\
        4	0.15\\
        5	0.16\\
        6	0.17\\
        7	0.18\\
        8	0.18\\
        9	0.19\\
        10	0.20\\
        11	0.20\\
        12	0.22\\
        13	0.22\\
        14	0.23\\
        15	0.23\\
        16	0.24\\
        17	0.25\\
        18	0.26\\
        19	0.26\\
        20	0.28\\
};
\addlegendentry{\textit{refreshing}}

\addplot [color=mizu, line width=1pt, mark size=1.6pt, mark=square*, mark options={solid, mizu, fill=white}, smooth]
  table[row sep=crcr]{%
        1	0.17\\
        2	0.18\\
        3	0.18\\
        4	0.18\\
        5	0.19\\
        6	0.19\\
        7	0.20\\
        8	0.20\\
        9	0.21\\
        10	0.21\\
        11	0.21\\
        12	0.22\\
        13	0.23\\
        14	0.23\\
        15	0.23\\
        16	0.24\\
        17	0.24\\
        18	0.25\\
        19	0.25\\
        20	0.26\\
};
\addlegendentry{\textit{key recovery}}

\end{axis}
\end{tikzpicture}

%% file: floats/plots/credential-presentation.tex
\pgfplotsset{
    scale only axis,
    every axis y label/.style = {
        at = {(-0.1, 0.5)},
        rotate = 90,
        anchor = center,
        font=\LARGE,
    },
    every axis x label/.style = {
        at = {(0.5,-0.15)},
        anchor = center,
        font=\LARGE,
    },
    tick label style = {
    	scale = 1.4,
        yshift=-0.1ex,
    },
    axis background/.style={
        fill = white,
    },
    grid=both,
    major grid style={
        opacity=.5
    },
    legend style={
        legend cell align = left,
        align = left,
        draw = white!15!black,
        font=\Large,
    }
}

\begin{tikzpicture}

\begin{axis}[%
    width=3.3in,
    height=2.4in,
    xmin=0,
    xmax=10,
    ymin=0,
    ymax=5,
    xlabel={\# Credentials},
    ylabel={Proving time/\textit{s}},
    legend pos=north west,
]

\addplot [color=ruri, line width=1pt, mark size=2pt, mark=*, mark options={solid, ruri, fill=white}, smooth]
  table[row sep=crcr]{%
        1	0.547\\
        2	0.582\\
        3	0.628\\
        4	0.686\\
        5	0.744\\
        6	0.810\\
        7	0.897\\
        8	0.996\\
        9	1.126\\
        10	1.355\\
};
\addlegendentry{\textit{2 claims}}

\addplot [color=momo, line width=1pt, mark size=1.8pt, mark=*, mark options={solid, momo}, smooth]
  table[row sep=crcr]{%
        1	0.691\\
        2	0.810\\
        3	0.901\\
        4	1.028\\
        5	1.146\\
        6	1.286\\
        7	1.452\\
        8	1.682\\
        9	2.013\\
        10	2.402\\
};
\addlegendentry{\textit{5 claims}}

\addplot [color=kogane!95!black, line width=1pt, mark size=2.6pt, mark=triangle*, mark options={solid, kogane!95!black, fill=white}, smooth]
  table[row sep=crcr]{%
        1	0.959\\
        2	1.189\\
        3	1.426\\
        4	1.739\\
        5	2.119\\
        6	2.508\\
        7	2.952\\
        8	3.524\\
        9	4.101\\
        10	4.898\\
};
\addlegendentry{\textit{10 claims}}

\end{axis}
\end{tikzpicture}

%% file: floats/tables/contract.tex
\small
\begin{tabular}{lc}
	\toprule
		\multicolumn{1}{c}{\textbf{Operation}} &
		\textbf{Gas Consumption} \\
	\midrule
		\textit{Identifier generation}
			& 313,656 \\
		\textit{Identifier association}
			& 689,739 \\
		\textit{Association appending}
			& 315,884 \\
		\textit{Association merging}
			& 335,421 \\
		\textit{Association refreshing}
			& 336,183 \\
		\textit{Key recovery}
			& 312,988 \\
	\bottomrule
\end{tabular}

%% file: floats/tables/theory-complexity.tex
\begin{tabular}{ccc}
	\toprule	
	\textbf{Scheme} & \textbf{Proof Size} & \textbf{Verifier Complexity} \\
	\midrule
	\sysname
		& 7 $\Gpa$, 7 $\Fp$
		& 12 $\Gpa$-exp, 2 $\Pair$ \\[7pt]
	ZEBRA \cite{rathee2022zebra}
		& 2 $\Gpa$, 1 $\Gpb$, 10 $\Fp$
		& 3 $\Gpa$-exp, 3 $\Gpa$-op, 4 $\Pair$ \\[7pt]
	Coconut \cite{sonnino2019coconut}
		& 1 $\Gpa$, 3 $\Gpb$, (n+2) $\Fp$ & 
		\begin{tabular}[c]{@{}c@{}}
			(n+3) $\Gpa$-exp, (n+3) $\Gpa$-op,\\
			2 $\Gpb$-exp, 2 $\Gpb$-op, 2 $\Pair$
		\end{tabular} \\[10pt]
	\texttt{zk-creds} \cite{rosenberg2023zk}
		& 2 $\Gpa$, 1 $\Gpb$
		& (n+1) $\Gpa$-exp, 4 $\Pair$\\
	\bottomrule
\end{tabular}

%% file: floats/plots/wall-clock.tex
\pgfplotsset{
    scale only axis,
    every axis y label/.style = {
        at = {(-0.13, 0.5)},
        rotate = 90,
        anchor = center,
        font=\LARGE,
    },
    every axis x label/.style = {
        at = {(0.5,-0.15)},
        anchor = center,
        font=\LARGE,
    },
    tick label style = {
    	scale = 1.4,
        yshift=-0.1ex,
    },
    axis background/.style={
        fill = white,
    },
    grid=both,
    major grid style={
        opacity=.5
    },
    legend style={
        legend cell align = left,
        align = left,
        draw = white!15!black,
        font=\Large,
    }
}

\begin{tikzpicture}

\begin{axis}[%
    width=3.3in,
    height=2.4in,
    xmin=0,
    xmax=10,
    ymin=12,
    ymax=16,
    xlabel={\# Credentials},
    ylabel={Latency/\textit{s}},
    legend pos=north west,
]

\addplot [color=momo, line width=1pt, mark size=1.8pt, mark=*, mark options={solid, momo}, smooth]
  table[row sep=crcr]{%
        1	13.835\\
        2	13.922\\
        3	14.013\\
        4	14.134\\
        5	14.259\\
        6	14.391\\
        7	14.563\\
        8	14.791\\
        9	15.122\\
        10	15.514\\
};
\addlegendentry{\sysname}

\addplot [color=ruri, line width=1pt, mark size=2pt, mark=*, mark options={solid, ruri, fill=white}, smooth]
  table[row sep=crcr]{%
        1	13.142\\
        2	13.185\\
        3	13.223\\
        4	13.281\\
        5	13.346\\
        6	13.437\\
        7	13.531\\
        8	13.728\\
        9	13.971\\
        10	14.355\\
};
\addlegendentry{\textit{Baseline-SID}}

\addplot [color=kogane!95!black, line width=1pt, mark size=2.6pt, mark=triangle*, mark options={solid, kogane!95!black, fill=white}, smooth]
  table[row sep=crcr]{%
        1	12.743\\
        2	12.791\\
        3	12.821\\
        4	12.848\\
        5	12.890\\
        6	12.936\\
        7	13.032\\
        8	13.130\\
        9	13.312\\
        10	13.608\\
};
\addlegendentry{\textit{Baseline-Anon}}

\end{axis}
\end{tikzpicture}

%% file: floats/plots/throughput.tex
\pgfplotsset{
    scale only axis,
    every axis y label/.style = {
        at = {(-0.13, 0.5)},
        rotate = 90,
        anchor = center,
        font=\LARGE,
    },
    every axis x label/.style = {
        at = {(0.5,-0.15)},
        anchor = center,
        font=\LARGE,
    },
    tick label style = {
    	scale = 1.4,
        yshift=-0.1ex,
    },
    axis background/.style={
        fill = white,
    },
    grid=both,
    major grid style={
        opacity=.5
    },
    legend style={
        legend cell align = left,
        align = left,
        draw = white!15!black,
        font=\normalsize,
    }
}

\begin{tikzpicture}

\begin{axis}[%
    width=3.3in,
    height=2.4in,
    xmin=0,
    xmax=25,
    ymin=0,
    ymax=200,
    xlabel={\# Identifies in association},
    ylabel={Ops/sec},
    legend pos=north east,
]

\addplot [color=momo, line width=1pt, mark size=1.8pt, mark=*, mark options={solid, momo}, smooth]
  table[row sep=crcr]{%
        2	25\\
		5	20\\
		10	15\\
		15	11\\
		20	8\\
		25  6\\
};
\addlegendentry{\textit{Association (ETH)}}

\addplot [color=kogane, line width=1pt, mark size=2pt, mark=*, mark options={solid, kogane, fill=white}, smooth]
  table[row sep=crcr]{%
        2	28\\
		5	25\\
		10	21\\
		15	17\\
		20	13\\
		25  10\\
};
\addlegendentry{\textit{Key recovery (ETH)}}

\addplot [color=ruri, line width=1pt, mark size=2.6pt, mark=triangle*, mark options={solid, ruri, fill=white}, smooth]
  table[row sep=crcr]{%
        2	140\\
		5	125\\
		10	99\\
		15	77\\
		20	56\\
		25  42\\
};
\addlegendentry{\textit{Association (ARB)}}

\addplot [color=ao, line width=1pt, mark size=2.4pt, mark=diamond*, mark options={solid, ao, fill=white}, smooth]
  table[row sep=crcr]{%
        2	196\\
		5	170\\
		10	134\\
		15	109\\
		20	91\\
		25  78\\
};
\addlegendentry{\textit{Key recovery (ARB)}}

\end{axis}
\end{tikzpicture}

%% file: floats/plots/sybil-m10.tex
\pgfplotsset{
    scale only axis,
    every axis y label/.style = {
        at = {(-0.17, 0.5)},
        rotate = 90,
        anchor = center,
        font=\LARGE,
    },
    every axis x label/.style = {
        at = {(0.5,-0.15)},
        anchor = center,
        font=\LARGE,
    },
    tick label style = {
    	scale = 1.4,
        yshift=-0.1ex,
    },
    axis background/.style={
        fill = white,
    },
    grid=both,
    major grid style={
        opacity=.5
    },
    legend style={
        legend cell align = left,
        align = left,
        draw = white!15!black,
        font=\normalsize,
    }
}

\begin{tikzpicture}

\begin{axis}[%
    width=3.3in,
    height=2.4in,
    xmin=0,
    xmax=20,
    ymode=log,
    ymin=1e-14,
    ymax=1,
    xlabel={\# Campaigns},
    ylabel={Success rate},
    legend pos=south west,
]

\addplot [
	color=ruri, 
	line width=1pt, 
	mark size=2pt, 
	mark=*, 
	mark options={solid, ruri}, 
	smooth
]
  table[row sep=crcr]{%
    1	    0.11398829356815613\\
    2	    0.01897928838761029\\
    3	    0.00506539170821665\\
    4	    0.001765904555162634\\
    5	    0.000903004115226337\\
    6	    0.000506731100385461\\
    7	    0.000317535768965619\\
    8	    0.000255992689700867\\
    9	    0.000211712808034003\\
    10	    0.000178752\\
    11	    0.00015351779058851\\
    12	    0.000133738241577788\\
    13	    0.000117920672883384\\
    14	    0.000105051966643084\\
    15	    9.44249419884572e-05\\
    16	    8.5533769006e-05\\
    17	    7.80087972850831e-05\\
    18	    7.15747187325778e-05\\
    19	    6.60229936589424e-05\\
    20	    6.1193246286982e-05\\
};
\addlegendentry{\textit{theoretical, $k=3$}}

\addplot [
	color=momo, 
	line width=1pt, 
	mark size=1.6pt, 
	mark=square*,
	mark options={solid, momo}, 
	smooth
]
  table[row sep=crcr]{%
    1	0.000500694863534752\\
    2	0.000112737245422247\\
    3	3.41811415315235e-05\\
    4	1.26372508532751e-05\\
    5	5.38363054412437e-06\\
    6	2.55086546522024e-06\\
    7	1.31293305157285e-06\\
    8	7.22045985189575e-07\\
    9	4.19212019006036e-07\\
    10	2.546432e-07\\
    11	1.60710992963208e-07\\
    12	1.04810534151871e-07\\
    13	7.03256965039326e-08\\
    14	4.8376210062687e-08\\
    15	3.4016438728761e-08\\
    16	2.43908455192139e-08\\
    17	1.77973171596193e-08\\
    18	1.31921045880095e-08\\
    19	9.91872388362214e-09\\
    20	7.55472176382494e-09\\
};
\addlegendentry{\textit{theoretical, $k=4$}}

\addplot [
	color=kogane!95!black, 
	line width=1pt, 
	mark size=2.6pt, 
	mark=triangle*, 
	mark options={solid, kogane!95!black}, 
	smooth
]
  table[row sep=crcr]{%
    1	2.55456563027935e-06\\
    2	3.47954461179775e-07\\
    3	7.06221932469494e-08\\
    4	1.86941580669751e-08\\
    5	5.98181171569375e-09\\
    6	2.20663102527702e-09\\
    7	9.09233415216655e-10\\
    8	4.09323120855768e-10\\
    9	1.98115320891321e-10\\
    10	1.0185728e-10\\
    11	5.51135092466419e-11\\
    12	3.1156520259177e-11\\
    13	1.82949262497223e-11\\
    14	1.11056496929952e-11\\
    15	6.94213035280836e-12\\
    16	4.45413541256646e-12\\
    17	2.92526580532862e-12\\
    18	1.96194297858559e-12\\
    19	1.34109300752057e-12\\
    20	9.32681699237647e-13\\
};
\addlegendentry{\textit{theoretical, $k=5$}}

\addplot [
	color=ruri, 
	line width=1pt, 
	mark size=2pt, 
	mark=*, 
	mark options={solid, ruri, fill=white}, 
	smooth,
	densely dashed,
]
  table[row sep=crcr]{%
    1	    0.04\\
    2	    0.009\\
    3	    0.002\\
    4	    0.0008\\
    5	    0.0004\\
    6	    0.0002\\
    7	    0.00014\\
    8	    0.00010\\
    9	    0.00008\\
    10	    0.00006\\
    11	    0.00004\\
    12	    0.000035\\
    13	    0.000025\\
    14	    0.000021\\
    15	    2.000072e-05\\
    16	    1.553376e-05\\
    17	    1.400879e-05\\
    18	    1.257471e-05\\
    19	    1.202299e-05\\
    20	    1.093246e-05\\
};
\addlegendentry{\textit{simulated, $k=3$}}

\addplot [
	color=momo, 
	line width=1pt, 
	mark size=1.6pt, 
	mark=square*,
	mark options={solid, momo, fill=white}, 
	smooth,
	densely dashed,
]
  table[row sep=crcr]{%
    1	0.0001\\
    2	0.0000227\\
    3	8.01811e-06\\
    4	2.263e-06\\
    5	7.083e-07\\
    6	2.55172e-07\\
    7	8.42877e-08\\
    8	4.33021e-08\\
    9	2.88712e-08\\
    10	1.95103e-08\\
    11	1.00812e-08\\
    12	7.01721e-09\\
    13	5.02348e-09\\
    14	3.87351e-09\\
    15	2.49871e-09\\
    16	1.33110e-09\\
    17	7.80372e-10\\
    18	5.39012e-10\\
    19	3.89925e-10\\
    20	3.19997e-10\\
};
\addlegendentry{\textit{simulated, $k=4$}}

\addplot [
	color=kogane!95!black, 
	line width=1pt, 
	mark size=2.6pt, 
	mark=triangle*, 
	mark options={solid, kogane!95!black, fill=white}, 
	smooth,
	densely dashed,
]
  table[row sep=crcr]{%
    1	2.17412e-07\\
    2	3.09163e-08\\
    3	6.90678e-09\\
    4	1.78521e-09\\
    5	5.88443e-10\\
    6	2.24791e-10\\
    7	9.36711e-11\\
    8	4.12471e-11\\
    9	1.82957e-11\\
    10	1.31952e-11\\
    11	5.13454e-12\\
    12	3.65133e-12\\
    13	1.92482e-12\\
    14	1.51325e-12\\
    15	6.83772e-13\\
    16	4.14532e-13\\
    17	2.98561e-13\\
    18	1.61239e-13\\
    19	1.35231e-13\\
    20	9.51235e-14\\
};
\addlegendentry{\textit{simulated, $k=5$}}

\end{axis}
\end{tikzpicture}

%% file: floats/plots/sybil-m100.tex
\pgfplotsset{
    scale only axis,
    every axis y label/.style = {
        at = {(-0.17, 0.5)},
        rotate = 90,
        anchor = center,
        font=\LARGE,
    },
    every axis x label/.style = {
        at = {(0.5,-0.15)},
        anchor = center,
        font=\LARGE,
    },
    tick label style = {
    	scale = 1.4,
        yshift=-0.1ex,
    },
    axis background/.style={
        fill = white,
    },
    grid=both,
    major grid style={
        opacity=.5
    },
    legend style={
        legend cell align = left,
        align = left,
        draw = white!15!black,
        font=\normalsize,
    }
}

\begin{tikzpicture}

\begin{axis}[%
    width=3.3in,
    height=2.4in,
    xmin=0,
    xmax=20,
    ymode=log,
    ymin=1e-20,
    ymax=1e-05,
    xlabel={\# Campaigns},
    ylabel={Success rate},
    legend pos=south west,
]

\addplot [
	color=ruri, 
	line width=1pt, 
	mark size=2pt, 
	mark=*, 
	mark options={solid, ruri}, 
	smooth
]
  table[row sep=crcr]{%
	1	7.89086612654321e-07\\
	2	5.6858749330636e-07\\
	3	4.33673858642578e-07\\
	4	3.44819556639401e-07\\
	5	2.82953125e-07\\
	6	2.37968097062421e-07\\
	7	2.0410357188786e-07\\
	8	1.77878387149771e-07\\
	9	1.57086123899268e-07\\
	10	1.40271604938272e-07\\
	11	1.26442313194275e-07\\
	12	1.14901535858893e-07\\
	13	1.05147876392911e-07\\
	14	9.6812566864249e-08\\
	15	8.9619140625e-08\\
	16	8.33567779053456e-08\\
	17	7.7862260537811e-08\\
	18	7.30074904336205e-08\\
	19	6.86906885217737e-08\\
	20	6.483008e-08\\
};
\addlegendentry{\textit{theoretical, $k=3$}}

\addplot [
	color=momo, 
	line width=1pt, 
	mark size=1.6pt, 
	mark=square*,
	mark options={solid, momo}, 
	smooth
]
  table[row sep=crcr]{%
	1	1.34835507044467e-09\\
	2	5.35358354087158e-10\\
	3	2.4048238992691e-10\\
	4	1.18705687826954e-10\\
	5	6.31203125e-11\\
	6	3.56460475971602e-11\\
	7	2.11565691856799e-11\\
	8	1.30921918839931e-11\\
	9	8.3950641181608e-12\\
	10	5.55070873342478e-12\\
	11	3.76935349777341e-12\\
	12	2.62043452336261e-12\\
	13	1.85995731188358e-12\\
	14	1.34486926093543e-12\\
	15	9.8873291015625e-13\\
	16	7.37895775108371e-13\\
	17	5.58241113554271e-13\\
	18	4.27596259386157e-13\\
	19	3.31262965479233e-13\\
	20	2.5932032e-13\\
};
\addlegendentry{\textit{theoretical, $k=4$}}

\addplot [
	color=kogane!95!black, 
	line width=1pt, 
	mark size=2.6pt, 
	mark=triangle*, 
	mark options={solid, kogane!95!black}, 
	smooth
]
  table[row sep=crcr]{%
	1	9.36357687808801e-14\\
	2	2.73142017391407e-14\\
	3	9.39384335651994e-15\\
	4	3.66375579712822e-15\\
	5	1.5780078125e-15\\
	6	7.36488586718186e-16\\
	7	3.67301548362499e-16\\
	8	1.93671477573862e-16\\
	9	1.07079899466337e-16\\
	10	6.16745414824976e-17\\
	11	3.68100927516934e-17\\
	12	2.26681187142094e-17\\
	13	1.43515224682375e-17\\
	14	9.31349903694898e-18\\
	15	6.17958068847656e-18\\
	16	4.1830826253309e-18\\
	17	2.88347682621008e-18\\
	18	2.02077627309148e-18\\
	19	1.43777328767028e-18\\
	20	1.03728128e-18\\
};
\addlegendentry{\textit{theoretical, $k=5$}}

\addplot [
	color=ruri, 
	line width=1pt, 
	mark size=2pt, 
	mark=*, 
	mark options={solid, ruri, fill=white}, 
	smooth,
	densely dashed,
]
  table[row sep=crcr]{%
    1	3.89721e-07\\
	2	2.69714e-07\\
	3	1.34672e-07\\
	4	9.45811e-08\\
	5	8.83101e-08\\
	6	7.35472e-08\\
	7	6.04833e-08\\
	8	5.79834e-08\\
	9	4.59911e-08\\
	10	3.73413e-08\\
	11	2.39534e-08\\
	12	1.95312e-08\\
	13	1.34754e-08\\
	14	1.15132e-08\\
	15	9.96785e-09\\
	16	9.34236e-09\\
	17	8.78654e-09\\
	18	7.37463e-09\\
	19	6.88512e-09\\
	20	5.48741e-09\\
};
\addlegendentry{\textit{simulated, $k=3$}}

\addplot [
	color=momo, 
	line width=1pt, 
	mark size=1.6pt, 
	mark=square*,
	mark options={solid, momo, fill=white}, 
	smooth,
	densely dashed,
]
  table[row sep=crcr]{%
	1	4.53215e-10\\
	2	1.65324e-10\\
	3	5.18535e-11\\
	4	2.85436e-11\\
	5	9.58411e-12\\
	6	5.75614e-12\\
	7	3.15745e-12\\
	8	1.54177e-12\\
	9	9.17558e-13\\
	10	6.81573e-13\\
	11	3.87531e-13\\
	12	2.75343e-13\\
	13	2.15995e-13\\
	14	1.85434e-13\\
	15	9.76313e-14\\
	16	8.63425e-14\\
	17	5.74568e-14\\
	18	4.51461e-14\\
	19	3.34532e-14\\
	20	2.74314e-14\\
};
\addlegendentry{\textit{simulated, $k=4$}}

\addplot [
	color=kogane!95!black, 
	line width=1pt, 
	mark size=2.6pt, 
	mark=triangle*, 
	mark options={solid, kogane!95!black, fill=white}, 
	smooth,
	densely dashed,
]
  table[row sep=crcr]{%
	1	3.76143e-14\\
	2	1.51344e-14\\
	3	3.81254e-15\\
	4	1.14563e-15\\
	5	9.89615e-16\\
	6	3.86571e-16\\
	7	9.12345e-17\\
	8	6.96154e-17\\
	9	3.15641e-17\\
	10	2.25296e-17\\
	11	1.76543e-17\\
	12	9.16561e-18\\
	13	8.46532e-18\\
	14	3.37655e-18\\
	15	2.88015e-18\\
	16	2.11595e-18\\
	17	1.87135e-18\\
	18	1.32354e-18\\
	19	8.44518e-19\\
	20	6.98611e-19\\
};
\addlegendentry{\textit{simulated, $k=5$}}

\end{axis}
\end{tikzpicture}

%% file: sections/body/related.tex
\section{Related work}\label{sec: related-work}

Anonymous-credential (AC) schemes are stateless cryptographic primitives for privacy-preserving claims, whereas decentralized-identity (DID) systems provide persistent identity lifecycles, recovery, and accountability. Table~\ref{tab: properties} positions \sysname{} against representative systems.

\spar{Decentralized identity}
DID standards give users autonomous control of digital identities and aim to reduce centralized breach risk while improving interoperability~\cite{dunphy2018first, naik2020uport, world2022decentralized, khovratovich2017sovrin}. Soulbound tokens similarly represent non-transferable, publicly verifiable credentials, reputation, or interaction records~\cite{weyl2022decentralized}.

\spar{Anonymous credential}
Since Chaum's foundational work~\cite{chaum1985security}, anonymous credentials have supported selective disclosure, identity unlinkability, and usage limits~\cite{camenisch2004signature, camenisch2006win, baldimtsi2013anonymous, camenisch2015composable, camenisch2017practical,du2023ucblocker, sonnino2019coconut, blomer2019updatable, hanzlik2021little, rathee2022zebra, doerner2023threshold, rosenberg2023zk}. Systems such as \texttt{zk-creds} provide strong global unlinkability, but pure AC schemes lack persistent infrastructure for cross-context Sybil resistance or trustless recovery. Preventing voluntary credential sharing also remains difficult~\cite{kakvi2023sok, camenisch2001efficient}. \sysname{} uses DID infrastructure to make anonymous credentials non-transferable while retaining zero-knowledge presentation.

\begin{table}[t]
	\centering
	\caption{Comparison of properties with established schemes.}
	\label{tab: properties}
	\resizebox{\columnwidth}{!}{%

\input{floats/tables/property-comparison}
	}
\end{table}

\spar{Sybil resistance in Web3}
Web3 airdrops and DAOs are vulnerable to attackers who create many pseudonymous identities~\cite{knight2023airdrop, baydakova2023millionaires, ambolis2023ultimate}. Importing legacy profiles or requiring offline gatherings imposes coverage, cost, or logistical limits~\cite{maram2021candid, borge2017proof, siddarth2020watches, platt2021sybil}. Hades instead limits access-token use within one instance~\cite{wang2023hades}, but that protection does not persist across applications. \sysname's cumulative association creates cross-context linkage, increasing the long-term cost of Sybil behavior without imposing offline identification on legitimate holders.

Accordingly, \sysname{} is best viewed as a DID system that uses ACs for private presentation. Its persistent association state carries across applications and supplies recovery evidence, while zero-knowledge proofs hide holder identifiers and undisclosed claims. This distinguishes \sysname{} from the unlinkable presentations supported by \texttt{zk-creds}~\cite{rosenberg2023zk} and from Hades's instance-scoped access-token mechanism~\cite{wang2023hades}. Cross-context Sybil resistance and trustless recovery reuse the same ledger-recorded association state, but the former additionally relies on presentation and nullifier constraints and the latter on a recovery proof. A later verifier can enforce the accumulated association constraint without learning the underlying identifiers or depending on state maintained by an earlier campaign. The trade-off is contextual, rather than global, unlinkability, which Section~\ref{sec: discussion} analyzes explicitly.

%% file: floats/tables/property-comparison.tex
\begin{threeparttable}
\begin{tabular}{lcccccc}
  \toprule
  \multicolumn{1}{c}{\textbf{Scheme}} &
    \textbf{Type} &
    \begin{tabular}[c]{@{}c@{}}\textit{Sybil}\\ \textit{Resistance}\end{tabular} &
    \begin{tabular}[c]{@{}c@{}}\textit{Selective}\\ \textit{disclosure}\end{tabular} &
    \textit{Unlinkability} &
    \begin{tabular}[c]{@{}c@{}}\textit{Credential}\\ \textit{Revocation}\end{tabular} &
    \begin{tabular}[c]{@{}c@{}}\textit{Identifier}\\ \textit{Blocklisting}\end{tabular} \\
  \midrule
    \sysname & DID & \fullcir & \fullcir & \halfcir & \fullcir & \fullcir \\
    CanDID \cite{maram2021candid} & DID & \halfcir & \fullcir & \halfcir & \fullcir & \emptycir \\
    Hades \cite{wang2023hades} & DID & \halfcir & \fullcir & \fullcir & \fullcir & \halfcir \\
    ZEBRA \cite{rathee2022zebra} & AC & \emptycir & \emptycir & \fullcir & \fullcir & \emptycir \\
    \texttt{zk-creds} \cite{rosenberg2023zk} & AC & \halfcir & \fullcir & \fullcir & \fullcir & \emptycir \\
    Coconut \cite{sonnino2019coconut} & AC & \emptycir & \fullcir & \fullcir & \emptycir & \emptycir \\
      \bottomrule
\end{tabular}
\begin{tablenotes}
  \small
  \item \fullcir~for supporting properties, \emptycir~for not supporting, and \halfcir~for partially supporting.
  \item For Unlinkability, \fullcir~indicates traditional \textit{global unlinkability} (absolute isolation across all contexts). \halfcir~denotes weaker system-level unlinkability: CanDID depends on its committee/MPC-based DID infrastructure, whereas \sysname provides \emph{contextual unlinkability} that holds only after the relevant public handles, such as \(n_a\) and \(n_i^c\), have changed.
\end{tablenotes}
\end{threeparttable}

%% file: sections/body/discussion.tex
\section{Discussion and limitations}\label{sec: discussion}

\spar{Contextual unlinkability}
\sysname{} gains cross-context Sybil resistance and trustless recovery by relaxing the strict \textit{global unlinkability} of pure anonymous credentials~\cite{camenisch2001efficient, rosenberg2023zk}. The limitation is handle reuse rather than concurrency itself: reusing an association state repeats $n_a$, and reusing a credential state repeats $n_i^c$, allowing colluding verifiers to link those presentations. Refreshing the relevant state rotates these handles, so \sysname{} provides \textit{contextual unlinkability}. The proof still hides holder identifiers, private keys, undisclosed claims, Merkle paths, and association composition. Appendices~\ref{subsec: identifier-accountability} and~\ref{subsec: credential-revocation} formalize the refresh procedures.

Refreshing is not retroactive: presentations made before rotation remain linkable through repeated handles, and rotation incurs an on-chain update. Deployments should choose refresh timing according to their privacy and latency requirements.

\spar{Security--privacy trade-off}
This is a deliberate design choice for applications where cross-context Sybil attacks threaten shared resources~\cite{weyl2022decentralized, platt2021sybil, douceur2002sybil}. In quadratic funding, matching depends on the number of contributors, creating incentives to split capital across identities~\cite{buterin2019flexible, gitcoin2022streaming, brenzikofer2023quadratic}. Ecosystem airdrops similarly reward sustained participation, so qualifying credentials are scarcer than raw identifiers~\cite{wahby2020airdrop, fan2023altruistic, baydakova2023millionaires}.

Without cross-context linkage, an attacker can reuse scarce credentials across many identities~\cite{knight2023airdrop}. For illustration, suppose an attacker has $s=10$ qualifying credentials, creates $\ell=1000$ identifiers, and must present $k=3$ credentials. The bound in Theorem~\ref{thm: sybil-resistance} gives a success probability of approximately $1.2 \times 10^{-10}$. \sysname{} converts credential scarcity into an attack constraint by forcing those credentials into an association. This quantitative benefit comes at the stated privacy cost: presentations remain linkable while public handles are reused, and honest holders recover privacy across contexts only after refreshing them.

\spar{Deployment scope}
The bound in Theorem~\ref{thm: sybil-resistance} should be read as an application-dependent guarantee rather than a universal property of decentralized identity. Its terms are determined jointly by ecosystem conditions and verifier policy: $s(q)$ reflects how many credentials satisfy the predicates, $\ell(q)$ reflects how many identifiers the adversary can operate, and $k$ is the number of credentials required per presentation. Section~\ref{sec: evaluation} shows that larger $k$ makes it more likely that a presentation spans identifiers, forcing association earlier and reducing Sybil success in the evaluated settings. This does not imply that every deployment should maximize $k$. Requiring more credentials may narrow legitimate participation, while credentials that are easy to acquire provide little scarcity to convert into an attack constraint. As a deployment-specific consideration, joint-presentation requirements should therefore be set in proportion to the resource at risk and the issuance effort represented by each credential.

The same reasoning limits where contextual linkage is justified. Quadratic-funding rounds and ecosystem airdrops allocate shared benefits according to apparent participation, so credential reuse across nominal identities can distort allocation~\cite{buterin2019flexible, gitcoin2022streaming, brenzikofer2023quadratic, wahby2020airdrop, fan2023altruistic, baydakova2023millionaires}. Applications that do not face material harm from duplicate participation gain less from persistent cross-context handles; for them, globally unlinkable credentials may remain the preferable design point. \sysname{} is therefore best understood as a scoped mechanism for campaigns in which scarce credentials proxy costly participation and cross-context reuse threatens a shared resource, not as a default replacement for globally unlinkable anonymous credentials.

%% file: sections/body/conclusion.tex
\section{Conclusion}\label{sec: conclusion}

We introduced \sysname, a decentralized identity system designed to resolve three fundamental challenges in Web3: \textit{pervasive Sybil attacks}, \textit{catastrophic key loss}, and the conflict between \textit{credential anonymity} and \textit{non-transferability}.

\sysname's core innovation, the \textit{identifier association} mechanism, creates persistent, cross-context linkages that provide progressive Sybil resistance, overcoming the single-campaign limitations of prior work. This mechanism operates without the heavy burdens of collateral or physical gatherings. Furthermore, this on-chain structure enables a \textit{trustless key recovery} paradigm based on provable structural knowledge, allowing users to recover assets without relying on trusted third parties. Finally, \sysname{} achieves \textit{non-transferability} for anonymous credentials and creates systematic disincentives against key-sharing, preventing malicious transfer of anonymous credentials. Comprehensive evaluation confirms \sysname's practicality and effectiveness. By cohesively integrating these solutions, \sysname{} offers a significant step towards a more \textit{secure}, \textit{resilient}, and \textit{trustworthy} decentralized ecosystem.

%% file: sections/considerations.tex
\section*{Ethical Considerations}

\sysname{} is a dual-use decentralized identity system. It can improve fairness in DAOs and airdrops, reduce dependence on trusted recovery parties, and protect credential contents through zero-knowledge proofs. The same mechanisms could be misconfigured or used by malicious actors to build durable pseudonymous identities. Moreover, persistent associations and public nullifiers deliberately trade global unlinkability for Sybil resistance: reusing an association or credential state can allow colluding verifiers to correlate presentations even though they do not learn the holder's identity or hidden claims.

Deployers should disclose this privacy trade-off, enable association and credential refreshing, use credential requirements proportionate to the application, and independently audit the smart contracts and circuits. Publishing the implementation supports such scrutiny. Our evaluation used author-controlled machines, local smart-contract deployments, and public blockchain interfaces; it involved no human subjects, private personal data, deception, unauthorized access, or exploitation of deployed systems. We therefore judge publication appropriate while emphasizing that production deployments require application-specific privacy and abuse review.

These mitigations do not eliminate all risks: issuer or deployment policies may exclude legitimate users, and association-level penalties can affect multiple identifiers at once. Deployments should therefore provide transparent policies and correction or appeal procedures before imposing adverse actions.

\section*{Open Science}

We release the artifacts needed to evaluate \sysname: the protocol implementation, gnark zero-knowledge circuits, Solidity smart contracts, off-chain helper code, and scripts for the reported micro-benchmarks, end-to-end latency, gas and storage measurements, throughput experiments, and Sybil simulations. The repository is accessible at \artifacturl. Its documentation describes the software environment and commands required to reproduce the reported results.

%% file: sections/appendices/full.tex
\section{Full \sysname{} protocol and extensions}\label{sec: extensions}

This section systematically presents the complete \sysname{} protocol, including its extensions that provide additional features and security guarantees. Figure~\ref{fig: protocol-full} formalizes the main protocol, while Figures~\ref{fig: protocol-identifier-update} and~\ref{fig: protocol-credential-management} provide formal definitions for association updates and credential management.

\subsection{Identifier accountability}\label{subsec: identifier-accountability}

\begin{figure*}[t]
    \centering
    \input{floats/frames/protocol-association-update}
    \caption{\sysname{} Protocol extension: association update.}
    \label{fig: protocol-identifier-update}
    \Description{The extension of protocol regarding association update.}
\end{figure*}

\sysname{} allows $\verifier$ to request $\registry$ to block the $\aid$ when detecting malicious behavior by $\holder$. This mechanism enhances accountability compared to previous schemes by making all identifiers linked to $\aid$ unusable. However, this process requires the verifier to additionally validate the legitimacy of $\aid$ using the nullifier during credential verification, introducing \textit{nullifier-based correlations}. To address this issue, \sysname{} enables $\holder$ to proactively refresh its $\aid$, as shown in Figure~\ref{fig: protocol-identifier-update}.

\spar{Identifier blocklisting}
As discussed in Section~\ref{subsec: credential-presentation}, whenever $\holder$ presents $\cred$, it must submit a nullifier $n_a$ to $\verifier$. This $n_a$ is structurally similar to $\aid$ but uses a different hash function. If $\holder$ is suspected of significant malicious behavior (e.g., money laundering) $\verifier$ can forward $n_a$ to $\registry$, along with detailed evidence about $\holder$'s violations. If $\registry$ determines that $\holder$'s actions warrant complete blocking, it records $n_a$ to invalidate $\aid$. This invalidation prevents $\holder$ from successfully authenticating any future presentations of $\aid$ or credentials $\cred$ linked to its component identifiers $\id_i$.

\spar{Association refreshing}
To mitigate nullifier-based correlation, \sysname{} allows $\holder$ to autonomously refresh the associated identifier $\aid$ while preserving its components. Recall that we introduce a nonce $u_a$ in $\aid$ that $\holder$ can increment to refresh the association. When $\holder$ detects correlation risks from using the same $n_a$ across recent credential presentations, it can break association-level linkage by incrementing the nonce $u_a$. Concretely, $\holder$ increments the nonce $u_a$, and proves in zero-knowledge that:
\begin{enumerate}[parsep=0.5pt]
    \item[\cirn{182}] $\aid$ derives from $\id_1,\dots,\id_\ell$ with the nonce $u_a$,
    \item[\cirn{183}] $\aid'$ derives from $\id_1,\dots,\id_\ell$ with another nonce $u_a'$,
    \item[\cirn{184}] $\aid$ is recorded by the ledger and not \textit{blocklisted}, and
    \item[\cirn{185}] The difference between $u_a$ and $u_a'$ is $1$.
\end{enumerate}
This emits a proof $\pi_d$ for relation $R_d$:
\[
    \begin{aligned}
         &
        \witness{\aid}=H_a(\witness{\id_1,\dots,\id_\ell},u_a)
        \;\land\;
        \aid'=H_a(\witness{\id_1,\dots,\id_\ell},u_a') \\
         & \land\;
        \mtauth(r_a, \witness{\aid}, \witness{\rho_a}) = 1
        \;\land\;
        n_a=H_n(\witness{\id_1,\dots,\id_\ell},u_a)    \\
         & \land\;
        u_a'=u_a+1.
    \end{aligned}
\]
The proof $\pi_d$ demonstrates that both pre- and post-update $\aid$ values derive from the same set of identifiers. Finally, $\holder$ submits $n_a$ to $\registry$ to invalidate the previous version of $\aid$.

As illustrated in Figure~\ref{fig: diagram-association-flow}, this refreshing mechanism creates an \textit{implicit chain} throughout each $\aid$'s lifecycle, with zero-knowledge proofs ensuring the refreshing validity. During updates, $\registry$ records the outdated $n_a$ as invalid and records the new $\aid$, ensuring only the latest $\aid$ remains valid. When $\aid$ is \textit{blocklisted}, this chain terminates, preventing further updates since the latest $n_a$ required for updating has already been \textit{invalidated}. Moreover, since there is no cryptographic link between the $\aid$ before and after refreshing, curious parties cannot track $\aid$ by monitoring VDR $\registry$ records.

\begin{figure*}[t]
    \centering
    \resizebox{\linewidth}{!}{\input{floats/diagrams/association-flow}}
    \caption{The lifecycle of an associated identifier, with pointers indicating updates (e.g., merging, appending, and refreshing) whose correctness is guaranteed by cryptographic proofs. Only the latest state is considered valid.}
    \label{fig: diagram-association-flow}
    \Description{The lifecycle of an associated identifier as an invisible chain.}
\end{figure*}

\subsection{Credential revocation}\label{subsec: credential-revocation}

\begin{figure*}[t]
    \centering
    \input{floats/frames/protocol-credential-management}
    \caption{\sysname{} Protocol extension: credential management and revocation.}
    \label{fig: protocol-credential-management}
    \Description{The extension of protocol regarding credential management and revocation.}
\end{figure*}

Existing schemes commonly provide credential revocation through revocation lists with non-membership proofs~\cite{camenisch2002dynamic, dahlberg2016efficient, papamanthou2011optimal, tomescu2019transparency, tyagi2022versa}. During presentations, holders must prove their credentials do not appear on these lists. However, when credentials are selectively disclosed, verifiers see only partial credential information, preventing them from identifying specific credentials or connecting them to malicious activity.
In contrast, \sysname{} allows verifiers to report misconduct to issuers without full knowledge of credential contents or holder identities.
\sysname{} also enables holders to proactively request revocation of their own credentials, as defined in Figure~\ref{fig: protocol-credential-management}.

\spar{Credential issuance}
\sysname's \textit{credential issuance} is compatible with legacy systems, enabling a smooth transition without requiring modifications to existing credentials.
To track the validity of issued credentials, each $\issuer$ maintains a smart contract containing a Merkle tree $T_c$. When $\issuer$ issues a credential $\cred$ to $\holder$, the latter samples a random nonce $u^c$ and commits to the credential $\cred$ by $c^c:=H_a(\cred,u^c)$. $\holder$ then generates a proof $\pi_t$ to justify the commitment relation and sends it to $\issuer$. After verifying $\pi_t$, $\issuer$ inserts $c^c$ into $T_c$, confirming that $\cred$ is registered in $\issuer$'s contract.

\spar{Credential revocation}
\sysname{} enables verifiers to initiate credential revocation. When $\issuer$ receives a document $\mathsf{doc}$ from a $\verifier$ describing credential misuse, it can invalidate the accused credential $\cred$ by recording its $n^c$ value to its on-chain smart contract. Because this revocation mechanism is fully decentralized and relies exclusively on append-only on-chain state updates, the issuer $\issuer$ does not need to be online or actively participate during any credential presentation or verification process. \sysname{} also empowers holders to voluntarily request revocation by submitting a $\mathsf{doc}$ with proof of ownership for $\cred$.

\spar{Credential refreshing}
This revocation mechanism also suffers from \textit{nullifier-based correlation}, as the provided $n^c$ remains constant when $\holder$ presents the same $\cred$ multiple times. To address this issue, \sysname{} allows $\holder$ to proactively update the nullifier $n^c$ and the commitment $c^c$ by modifying the nonce $u^c$. Specifically, $\holder$ samples a new randomness $u^{c\prime}$ and derives a new $c^{c\prime}$ from it. Then, $\holder$ proves in zero-knowledge that:
\begin{enumerate}[parsep=0.5pt]
    \item[\cirn{182}] $c^c$ is recorded by $\issuer$,
    \item[\cirn{183}] $n^c$ is not \texttt{revoked} in the issuer's revocation list, and
    \item[\cirn{184}] $c^{c\prime}$ derives from $\cred$ with another nonce $u^{c\prime}$.
\end{enumerate}
This emits a proof $\pi_u$ for relation $R_u$:
\[
    \begin{aligned}
         &
        \mtauth(r_c, \witness{c^c},\witness{\rho_c})=1
        \;\land\;
        n^c=H_n(\witness{\cred},\witness{u^c}) \\
         & \land\;
        \witness{c^c}=H_a(\witness{\cred},\witness{u^c})
        \;\land\;
        c^{c\prime}=H_a(\witness{\cred},\witness{u^{c\prime}}).
    \end{aligned}
\]

\spar{Credential management}
Figure~\ref{fig: protocol-credential-management} describes the protocol for credential management and revocation. To support these functions, each issuer maintains an incremental Merkle tree $T_c$ with credential commitments $c^c$ as leaf nodes. The issuer also maintains a nullifier list $N_c$ to track revoked credentials through their $n^c$ markers.

\subsection{Impersonation resistance}\label{subsec: proof-non-forwardability}

A significant challenge when using zero-knowledge proofs for selective credential disclosure is that verifiers may misuse these proofs. When $\holder$ presents a proof $\pi_c$ for credential $\cred$ to $\verifier$, a malicious $\verifier$ could impersonate $\holder$ by forwarding $\pi_c$ to another verifier $\verifier'$. To prevent such impersonation attacks, \sysname{} implements two mechanisms allowing $\holder$ to specify a designated $\verifier$, thereby preventing the presentation process from being replayed to other $\verifier'$.

\spar{Interactive verification}
One solution requires $\pi_c$ to contain a component specified by $\verifier$. Here, $\verifier$ samples a challenge $e$ and sends it to $\holder$ before $\cred$ is presented. $\holder$ must incorporate $e$ into $R_c$ for binding purposes. If $\verifier$ later attempts to pass $\pi_c$ to another verifier $\verifier'$ in a different presentation, $\verifier'$ would need to provide an $e'$ that matches the original challenge $e$ exactly—an occurrence with negligible probability.

\spar{Key attestation}
To avoid this additional round of interaction, \sysname{} implements an alternative mechanism by allowing $\holder$ to embed elements that only $\verifier$ knows (i.e., $\verifier$'s private key $\skv$) into relations~\cite{weyl2022decentralized}. The credential presentation then becomes a proof of knowledge with two possible conditions:
\begin{itemize}[parsep=0.5pt]
    \item Either relation $R_c$ is satisfied, or
    \item $\verifier$ knows a $\witness{\skv}$ such that $\pkv=\pubkeygen(\witness{\skv})$.
\end{itemize}
$\verifier$ will accept this proof since it knows it didn't present the credential itself, meaning $R_c$ must be satisfied. However, when $\verifier$ forwards this proof to another $\verifier'$, the latter will not accept $R_c$ since $\verifier$ could have created a valid proof simply by using its private key $\skv$.

\subsection{Immediate verification}\label{subapp: instant-verification}

To prevent $\holder$ from bypassing the anti-Sybil checks of $\verifier$ by updating $\aid$ during a campaign, we introduced a countermeasure in Section~\ref{subsec: credential-presentation}. This involves delaying all verifications until the end of the campaign interaction period, i.e., after all users have provided their proofs. However, some scenarios require immediate verification and completion of interactions when users submit their credentials.

We provide a solution for this scenario here. The core idea is to require $\holder$ to have generated the $\aid$ before the start of the campaign\footnote{To afford potential participants ample time to prepare associated identifiers, $\verifier$ may pre-announce the commencement time of the campaign.}. For this, $\verifier$ requires that the proof $\pi_c$ provided by $\holder$, in addition to satisfying the relation $R_c$, must also include a constraint that the index of $\mathsf{aid}$ is strictly less than a threshold $t$, which is the number of elements in the Merkle tree $T$ of $\registry$ at the start of campaign $\sid$ initiated by $\verifier$, that is
\[
    \mathrm{index}(\aid) < t = |T|.
\]

\spar{Performance optimization}
When the number of elements in $T$ is large, the range proof for $\mathrm{index}(\aid)$ will significantly slow down the generation time of $\pi_c$, since the arithmetic circuit needs to first convert $\mathrm{index}(\aid)$ and $t$ into binary representations and then compare them bit by bit. However, it is not necessary to perform this conversion in the circuit. An optimization is to take advantage of the Merkle proof $\rho_a$ with respect to $\aid$. It turns out that $\rho_a$ contains a binary index array $\seq{d}$, which records the relative position of the two elements in each hash on the Merkle path. The reverse array $\mathrm{rev}(\seq{d})$ is exactly the binary representation of $\mathrm{index}(\aid)$. Therefore, the above constraint can be transformed into $\mathrm{rev}(\bm d) < \mathrm{bin}(t)$, where $\mathrm{bin}(t)$ refers to the binary representation of $t$, which can be pre-calculated by $\verifier$ and encoded into the circuit. This constraint compares the binary values in the two arrays sequentially with a complexity no greater than the number of layers in $T$.

%% file: floats/frames/protocol-association-update.tex
\protocol{Protocol \prob, extension 1}{
This protocol executes between a $\party$ and the data registry $\registry$, where $\party$ could be any possible entity other than $\registry$.

\method{Association Appending:}
On receiving $(\mathtt{append},\sid,\aid,\id_{\ell+1})$, party $\party$ performs as follows:
\begin{enumerate}[parsep=0.5pt]
    \item $\party$ retrieves $(\aid,\id_1,\dots,\id_\ell,u_a)$ and $(\aid,r_a,\rho_a)$ from its record, along with $(\id_i,\pk_i,\sk_i)$ and $(\id_i,h_i,r_i,\rho_i)$ for all $i\in[\ell+1]$.
    Then, $\party$ computes $\aid':=H_a(\id_1,\dots,\id_{\ell+1},u_a)$, $n_a:=H_n(\id_1,\dots,\id_\ell,u_a)$, and $n_{\ell+1}:=H_n(\id_{\ell+1},\sk_{\ell+1})$.
    Finally, $\party$ generates a proof $\pi_p$ for relation $R_p$, and sends $(r_a,r_{\ell+1},n_a,n_{\ell+1},u_a,\aid',\pi_p)$ to $\registry$.
    \item $\registry$ sends $(\mathtt{check},\sid,r_a)$, $(\mathtt{check},\sid,r_{\ell+1})$, $(\mathtt{check},\sid,n_a)$ and $(\mathtt{check},\sid,n_{\ell+1})$ to $\funl$.
    On receiving $\mathtt{checked}$ for $r_a, r_{\ell+1}$ and $\mathtt{unrecorded}$ for $n_a, n_{\ell+1}$, $\registry$ verifies $\pi_p$ and aborts if fails.
    Else, $\registry$ sends $(\mathtt{record},\sid,\aid')$, $(\mathtt{invalidate},\sid,n_a)$ and $(\mathtt{invalidate},\sid,n_{\ell+1})$ to $\funl$.
    On receiving $(\mathtt{recorded},\sid,r_a',\rho_a')$ and $(\mathtt{invalidated},\sid)$ from $\funl$, $\registry$ sends $(r_a',\rho_a')$ to $\party$.
    \item $\party$ records $(\aid',r_a',\rho_a')$ and $(\aid',\id_1,\dots,\id_{\ell+1},u_a)$, and then outputs $(\mathtt{appended},\sid,\aid')$.
\end{enumerate}

\method{Association Merging:}
On receiving $(\mathtt{merge},\sid,\aid_1,\aid_2)$, party $\party$ performs as follows:
\begin{enumerate}[parsep=0.5pt]
    \item $\party$ retrieves $(\aid_1,\id_1,\dots,\id_t,u_{a1})$, $(\aid_2,\id_{t+1},\dots,\id_\ell,u_{a2})$, $(\aid_1,r_{a1},\rho_{a1})$ and $(\aid_2,r_{a2},\rho_{a2})$ from its record.
    Then, $\party$ computes $n_{a1}:=H_n(\id_1,\dots,\id_t,u_{a1})$, $n_{a2}:=H_n(\id_{t+1},\dots,\id_\ell,u_{a2})$ and $\aid':=H_a(\id_1,\dots,\id_\ell,u_a')$, where $u_a':=\max(u_{a1},u_{a2})$.
    Finally, $\party$ generates a proof $\pi_m$ for relation $R_m$, and sends $(r_{a1},r_{a2},n_{a1},n_{a2},u_{a1},u_{a2},\aid',\pi_m)$ to $\registry$.
    \item $\registry$ sends $(\mathtt{check},\sid,r_{a1})$, $(\mathtt{check},\sid,r_{a2})$, $(\mathtt{check},\sid,n_{a1})$ and $(\mathtt{check},\sid,n_{a2})$ to $\funl$.
    On receiving $\mathtt{checked}$ for $r_{a1}, r_{a2}$ and $\mathtt{unrecorded}$ for $n_{a1}, n_{a2}$, $\registry$ verifies $\pi_m$ and aborts if fails.
    Else, $\registry$ sends $(\mathtt{record},\sid,\aid')$, $(\mathtt{invalidate},\sid,n_{a1})$ and $(\mathtt{invalidate},\sid,n_{a2})$ to $\funl$.
    On receiving $(\mathtt{recorded},\sid,r_a',\rho_a')$ and $(\mathtt{invalidated},\sid)$ from $\funl$, $\registry$ sends $(r_a',\rho_a')$ to $\party$.
    \item $\party$ records $(\aid',r_a',\rho_a')$ and $(\aid',\id_1,\dots,\id_\ell,u_a')$, and then outputs $(\mathtt{merged},\sid,\aid')$.
\end{enumerate}

\method{Association Refreshing:}
On receiving $(\mathtt{refresh},\sid,\aid)$, party $\party$ performs as follows:
\begin{enumerate}[parsep=0.5pt]
    \item $\party$ retrieves $(\aid,\id_1,\dots,\id_\ell,u_a)$ and $(\aid,r_a,\rho_a)$ from its record.
    Then, $\party$ computes $n_a:=H_n(\id_1,\dots,\id_\ell,u_a)$ and $\aid':=H_a(\id_1,\dots,\id_\ell,u_a')$ where $u_a':=u_a+1$.
    Finally, $\party$ generates a proof $\pi_d$ for relation $R_d$, and sends $(r_a,u_a,n_a,\aid',\pi_d)$ to $\registry$.
    \item $\registry$ sends $(\mathtt{check},\sid,r_a)$ and $(\mathtt{check},\sid,n_a)$ to $\funl$.
    On receiving $\mathtt{checked}$ for $r_a$ and $\mathtt{unrecorded}$ for $n_a$, $\registry$ verifies $\pi_d$ and aborts if fails.
    Else, $\registry$ sends $(\mathtt{record},\sid,\aid')$ and $(\mathtt{invalidate},\sid,n_a)$ to $\funl$.
    On receiving $(\mathtt{recorded},\sid,r_a',\rho_a')$ and $(\mathtt{invalidated},\sid)$ from $\funl$, $\registry$ sends $(r_a',\rho_a')$ to $\party$.
    \item $\party$ records $(\aid',r_a',\rho_a')$ and $(\aid',\id_1,\dots,\id_\ell,u_a')$, and then outputs $(\mathtt{refreshed},\sid,\aid')$.
\end{enumerate}
}

%% file: floats/diagrams/association-flow.tex
\tikzset{
	every picture/.style={
		line width=0.55pt
	},
	frame/.style={
		rounded corners=0.8mm,
		inner sep=1mm,
		draw=black,
		fill=white,
		drop shadow={opacity=0.7, shadow xshift=2, shadow yshift=-2},
	},
	variables/.style={
		inner sep=0.8mm,
		scale=0.75,
	},
	operations/.style={
		font=\itshape,
		scale=0.7,
	},
	nonce/.style={
		inner sep=0.8mm,
		font=\scriptsize\ttfamily,
	},
	line/.style={
		rounded corners=0.5mm,
		thick,
	},
}

\begin{tikzpicture}[x=0.75pt, y=0.75pt, yscale=-1, xscale=1]

\draw[frame] (20, 40) rectangle (100,60);
\draw[frame] (150, 40) rectangle (230,60);
\draw[frame] (280, 40) rectangle (360,60);
\draw[frame] (410, 40) rectangle (490,60);
\draw[frame] (540, 40) rectangle (620,60);
\draw[frame] (150, 80) rectangle (230,100);

\node[variables] (aid1-pre) at (-8,50) {$\cdots$};
\node[variables] (aid-p1-pre) at (122,90) {$\cdots$};

\node[variables] (aid1) at (50,50) {$\id_1,\dots,\id_8$};
\node[variables] (aid2) at (180,50) {$\id_1,\dots,\id_8$};
\node[variables] (aid3) at (310,50) {$\id_1,\dots,\id_{15}$};
\node[variables] (aid4) at (440,50) {$\id_1,\dots,\id_{16}$};
\node[variables] (aid5) at (570,50) {$\id_1,\dots,\id_{16}$};
\node[variables] (aid-p1) at (180,90) {$\id_9,\dots,\id_{15}$};
\node[variables] (aid-p2) at (383,90) {$\id_{16}$};
\node[variables] (aid-post) at (650,50) {$\cdots$};

\draw[frame] (80,40) -- (80,60);
\draw[frame] (210,40) -- (210,60);
\draw[frame] (340,40) -- (340,60);
\draw[frame] (470,40) -- (470,60);
\draw[frame] (600,40) -- (600,60);
\draw[frame] (210,80) -- (210,100);

\node[nonce,enji] (n1) at (90,50) {20};
\node[nonce,kon] (n2) at (220,50) {21};
\node[nonce,kon] (n3) at (350,50) {21};
\node[nonce,kon] (n4) at (480,50) {21};
\node[nonce,kogane] (n5) at (610,50) {22};
\node[nonce,beni] (n-p1) at (220,90) {17};

\draw[->, line] (0,50) -- (20,50);
\draw[->, line] (100,50) -- (150,50);
\draw[->, line] (230,50) -- (280,50);
\draw[->, line] (360,50) -- (410,50);
\draw[->, line] (490,50) -- (540,50);
\draw[->, line] (620,50) -- (640,50);

\draw[->, line] (130,90) -- (150,90);

\draw[->, line] (230,90) -- (253,90) -- (253,50.7);
\draw[->, line] (383,84) -- (383,50.7);

\node[operations] (op2) at (255,44.5) {merge};
\node[operations] (op3) at (385,43.7) {append};

\node[operations] (op1) at (125,44) {refresh};
\node[operations] (op4) at (515,44) {refresh};

\end{tikzpicture}

%% file: floats/frames/protocol-credential-management.tex
\protocol{Protocol \prob, extension 2}{
This protocol executes between the issuer $\issuer$, the holder $\holder$, and the verifier $\verifier$. 
All participants have access to the data registry $\registry$ and the decentralized ledger $\funl$.
The operations of $\issuer$ in the revocation and refreshing phases are executed autonomously by its on-chain smart contract, which interacts with $\funl$ to manage credential states.

\method{Credential Issuance:}
On receiving $(\mathtt{issue},\sid,\idh,s)$, issuer $\issuer$ performs as follows:
\begin{enumerate}[parsep=0.5pt]
    \item $\issuer$ retrieves its keypair $(\ski,\pki)$, signs the claim $s$ by $\sigma:=\sign(\ski,(\idi,\idh,s))$, and forms the credential by $\cred:=(\idi,\sigma,\idh,s)$. $\issuer$ sends $\cred$ to $\holder$.
    \item $\holder$ samples $u^c\sample\Zp$ and computes $c^c:=H_a(\cred,u^c)$. Then, $\holder$ generates a proof $\pi_t$ for relation $R_t$, and sends $(c^c,\pi_t)$ to $\issuer$.
    \item $\issuer$ verifies $\pi_t$ and aborts if fails. Else, $\issuer$ sends $(\mathtt{record},\sid,c^c)$ to $\funl$. On receiving $(\mathtt{recorded},\sid,r_c,\rho_c)$ from $\funl$, $\issuer$ returns $(r_c,\rho_c)$ to $\holder$.
    \item $\holder$ records $(\cred,u^c,c^c,r_c,\rho_c)$ and outputs $(\mathtt{issued},\sid)$.
\end{enumerate}

\method{Credential Revocation:}
On receiving $(\mathtt{revoke},\sid,n^c,\mathsf{doc})$, a party $\party$ (can be either $\holder$ or $\verifier$) performs as follows:
\begin{enumerate}[parsep=0.5pt]
    \item $\party$ sends $(n^c,\mathsf{doc})$ to $\issuer$.
    \item $\issuer$ checks $\mathsf{doc}$ to confirm that $\cred$ should be revoked. If this is the case, $\issuer$ sends $(\mathtt{invalidate},\sid,n^c)$ to $\funl$. On receiving $(\mathtt{invalidated},\sid)$ from $\funl$, $\issuer$ outputs $(\mathtt{revoked},\sid)$.
\end{enumerate}

\method{Credential Refreshing:}
On receiving $(\mathtt{refresh},\sid,\cred)$, $\holder$ performs as follows:
\begin{enumerate}[parsep=0.5pt]
    \item $\holder$ retrieves $(\cred,u^c,c^c,r_c,\rho_c)$ from its record. Then, $\holder$ samples $u^{c\prime}\sample\Zp$, and computes $c^{c\prime}:=H_a(\cred,u^{c\prime})$ and $n^c:=H_n(\cred,u^c)$. $\holder$ generates a proof $\pi_u$ for relation $R_u$, and sends $(r_c,c^{c\prime},n^c,\pi_u)$ to $\issuer$.
    \item $\issuer$ sends $(\mathtt{check},\sid,r_c)$ and $(\mathtt{check},\sid,n^c)$ to $\funl$. On receiving $\mathtt{checked}$ for $r_c$ and $\mathtt{unrecorded}$ for $n^c$, $\issuer$ verifies $\pi_u$ and aborts if fails. Then, $\issuer$ sends $(\mathtt{record},\sid,c^{c\prime})$ and $(\mathtt{invalidate},\sid,n^c)$ to $\funl$. On receiving $(\mathtt{recorded},\sid,r_c',\rho_c')$ and $(\mathtt{invalidated},\sid)$ from $\funl$, $\issuer$ returns $(r_c',\rho_c')$ to $\holder$.
    \item $\holder$ records $(\cred,u^{c\prime},c^{c\prime},r_c',\rho_c')$ and outputs $(\mathtt{refreshed},\sid)$.
\end{enumerate}
}

%% file: sections/appendices/definition.tex
\section{Formal definitions}\label{app: definition}

\subsection{Non-interactive zero knowledge}\label{subapp: nizk}

\begin{definition}[zkSNARK]\label{def: zk-snark}
	We call $\zk$ a zero-knowledge succinct non-interactive argument of knowledge (\textbf{zkSNARK}) for $\relgen$ if it has \textbf{zero-knowledge} (Definition~\ref{def: zk-zero-knowledge}), \textbf{succinctness} (Definition~\ref{def: zk-succinctness}), \textbf{completeness} (Definition~\ref{def: zk-completeness}) and \textbf{computational knowledge soundness} (Definition~\ref{def: zk-soundness}).
\end{definition}

\begin{definition}[Zero-Knowledge]\label{def: zk-zero-knowledge}
	Conceptually, an argument is zero-knowledge if it does not give away any information other than the truth of the statement. We say $\zk$ is \textbf{zero-knowledge} if for all $\lambda\in\N$, $R\in\relgen_\lambda$, $(x,w)\in R$ and all adversaries $\adv$, there exists a simulator $\mathsf{Sim}$ such that,
	\[
		\begin{aligned}
			 & \Pr
			\left[
				\begin{aligned}
					\varsigma	\sample	\zksetup(R) \\
					\pi			\sample	\zkprove(\varsigma,x,w)
				\end{aligned}:
				\adv(R,\varsigma,\pi)=1
				\right]
			=      \\ & \;\;\;\;
			\Pr
			\left[
				\begin{aligned}
					\varsigma	\sample	\zksetup(R) \\
					\pi			\sample	\mathsf{Sim}(R,x)
				\end{aligned}:
				\adv(R,\varsigma,\pi)=1
				\right].
		\end{aligned}
	\]
\end{definition}

\begin{definition}[Succinctness]\label{def: zk-succinctness}
	We say $\zk$ is \textbf{succinct} if the verifier runtime can be bounded by a polynomial in $\lambda+|x|$ and the proof size can be bounded by a polynomial in $\lambda$. On this basis, we claim that $\zk$ is \textbf{completely succinct} if the length of the reference string $|\varsigma|$ can also be bounded by a polynomial in $\lambda$.
\end{definition}

\begin{definition}[Completeness]\label{def: zk-completeness}
	Intuitively, completeness means that given any true statement, an honest prover is necessarily able to convince an honest verifier to accept the argument. We say $\zk$ is \textbf{complete} if for all $\lambda\in\N$, $R\in\relgen_\lambda$ and $(x,w)\in R$,
	\[
		\Pr
		\left[
			\begin{aligned}
				\varsigma	\sample	\zksetup(R) \\
				\pi			\sample	\zkprove(\varsigma,x,w)
			\end{aligned}:
			\zk.\zkverify(\varsigma,x,\pi)=1
			\right]=1.
	\]
\end{definition}

\begin{definition}[Soundness]\label{def: zk-soundness}
	We say $\zk$ is \textbf{computational knowledge sound} if for all non-uniform polynomial-time adversaries $\adv$, there exists a non-uniform polynomial-time extractor $\mathsf{Ext}$ such that,
	\[
		\Pr
		\left[
			\begin{aligned}
				\varsigma	\sample		\zksetup(R) \\
				(x,\pi)		\sample		\adv         \\
				w			\leftarrow	\mathsf{Ext}(\varsigma,x,\pi)
			\end{aligned}:
			\begin{aligned}
				 & (x,w)	\notin	R
				\;\;\land\;                     \\
				 & \zkverify(\varsigma,x,\pi)=1
			\end{aligned}
			\right]
		\leq\negl(\lambda).
	\]
\end{definition}

\subsection{Commitment scheme}\label{subapp: definition-commitment}

\begin{definition}[Commitment]\label{def: commitment}
	We call $\cmt$ a secure cryptographic commitment scheme if it has the property of \textbf{correctness} defined above, along with the properties of \textbf{hiding} (Definition~\ref{def: hiding}) and \textbf{binding} (Definition~\ref{def: binding}).
\end{definition}

\begin{definition}[Hiding]\label{def: hiding}
	Intuitively, hiding means that the commitment $c$ should reveal nothing about the committed message $m$. We say $\cmt$ is \textbf{hiding} if for all efficient adversaries $\adv$, two different messages $m_0,m_1$ and two arbitrary opening strings $u_0,u_1$,
	\[
		\begin{aligned}
			\Bigg\vert
			 & \Pr
			\left[
				\begin{aligned}
					c_0 \sample \cmtcommit(m_0,u_0) \\
					r_0 \leftarrow \adv(c_0)
				\end{aligned}:
				r_0=0
				\right]- \\
			 & \Pr
			\left[
				\begin{aligned}
					c_1 \sample \cmtcommit(m_1,u_1) \\
					r_1 \leftarrow \adv(c_1)
				\end{aligned}:
				r_1=0
				\right]
			\Bigg\vert
			\leq \negl(\lambda)
		\end{aligned}
	\]
	holds.
\end{definition}

\begin{definition}[Binding]\label{def: binding}
	Binding indicates that a commitment $c$ could only open to a single message $m$. We say $\cmt$ is \textbf{binding} if for all efficient adversaries $\adv$ which outputs $(c,m_0,m_1,u_0,u_1)$,
	\[
		\Pr
		\left[
			\begin{aligned}
				 & m_0\neq m_1                                         \\
				 & \cmt.\cmtopen(m_0,c,u_0) = \cmt.\cmtopen(m_1,c,u_1)
			\end{aligned}
			\right]
		\leq\negl(\lambda)
	\]
	holds.
\end{definition}

%% file: sections/appendices/sybil.tex
\section{Sybil-resistance proof}\label{app: sybil-resistance}

To prove Theorem~\ref{thm: sybil-resistance}, we first consider the holder $\holder$'s credential selection strategy. At the $q$-th campaign, $\holder$ must select $k$ credentials from a pool of $s(q)$ credentials satisfying the predicate specified by the verifier. The association rules constrain $\holder$'s choices as follows:
\begin{itemize}[leftmargin=*]
  \item If no association operations occur, $\holder$ selects $k$ identifiers freely. The number of possible combinations is $\binom{s(q)}{k}$.
  \item If $\holder$ reuses a set of identifiers $S_j$ in an association, then the selection space collapses to the subsets of this association set $S_j$. By the association rule, this does not increase $s(q)$.
  \item If $\holder$ triggers overlap associations, i.e., $\holder$ chooses identifiers from the intersection of two or more association sets, or identifiers from an association set, along with some identifiers not yet associated, the resulting association set $S_\mathrm{associated}$ permanently links all overlapping identifiers. Subsequent selections involving any identifiers in $S_\mathrm{associated}$ must include the entire set.
\end{itemize}
Forcing associations reduces $\holder$'s future flexibility. After $t$ associations, $\holder$'s effective selection space is upper bounded by the original unassociated selection space:

\begin{equation*}
  \binom{s(q)}{k}
  \geq
  \sum_{t=0}^k
  \binom{\text{\# associated clusters}}{t}
  \cdot
  \binom{s(q) - t
    \cdot
    \ell_{\text{avg}}}{k - t},
\end{equation*}
where $\ell_{\text{avg}}$ is the average size of associated clusters. The number of associated clusters is at most $k$, and the number of identifiers in each cluster is at most $\ell_{\text{avg}}$. The total number of identifiers in the selection space is $s(q) - t \cdot \ell_{\text{avg}}$. Thus, $\binom{s(q)}{k}$ dominates all constrained selection scenarios.

Next, we consider the probability that $\holder$ can avoid association when attempting an attack. For an attack to succeed, the $k$ credentials presented must not be bound to the same associated identifier. The probability when presenting $k$ credentials simultaneously is:
\begin{equation*}
  \left(\frac{1}{\ell(q)^{k-1}}\right)
  \cdot
  \left(\frac{\ell(q)-1}{\ell(q)^k}\right)
  =
  \frac{\ell(q)-1}{\ell(q)^{2k-1}}.
\end{equation*}

Then, we consider the hash consistency. Note that each association operation generates a hash $\aid=H_a(S)$ for an identifier set $S$. $\holder$ can only launch a Sybil attack if:
\begin{itemize}[leftmargin=*]
  \item All hashes generated during the associations are consistent with previous commitments.
  \item No hash collisions occur between distinct association sets.
\end{itemize}
Let $B$ be the event that $\holder$ finds distinct identifier sets $S_1\neq S_2$ such that $H_a(S_1)=H_a(S_2)$. By the collision resistance of the hash function $H_a$, the probability is bounded by:
\begin{equation*}
  \Pr[B] \leq \frac{T(q)^2}{2^{\lambda}},
\end{equation*}
where $T(q)$ is the total number of association operations. Since $T(q)\leq q\cdot \ell(q)$ (at most one association operation per identifier per campaign), we have:
\begin{equation*}
  \Pr[B]
  \leq
  \frac{T(q)^2}{2^{\lambda}}
  \leq
  \frac{q^2\cdot \ell(q)^2}{2^{\lambda}}.
\end{equation*}
Finally, we bound the success probability of $\holder$'s Sybil attack. By taking the union bound over the probability of finding a valid unassociated credential combination and the probability of a hash collision, we have:
\begin{equation*}
  \Pr[W]
  \leq
  \binom{s(q)}{k} \left( \frac{\ell(q)-1}{\ell(q)^{2k-1}} \right) +
  \frac{q^2\cdot \ell(q)^2}{2^\lambda}.
\end{equation*}
The term $(q\cdot \ell(q))^2/2^\lambda$ is negligible in $\lambda$ due to $|\Zp|\geq 2^\lambda$ and $\ell(q)=\poly(\lambda)$. Thus, we have:
\begin{equation*}
  \Pr[W] \leq \binom{s(q)}{k} \left( \frac{\ell(q)-1}{\ell(q)^{2k-1}} \right) + \negl(\lambda).
\end{equation*}

%% file: sections/appendices/security-uc.tex
\section{Security proof}\label{app: security-uc}

To prove Theorem~\ref{thm: uc-security}, we demonstrate that no p.p.t.\ environment $\env$ can differentiate between the ideal process of functionality $\funb$ and the execution of protocol $\prob$ in the $\funl$-hybrid model.

Concretely, parties in protocol $\prob$ execute alongside an adversary $\adv$ and $\env$ with initial input $z$. Within the $\funl$-hybrid world, all parties are granted access to the ledger functionality $\funl$. The adversary $\adv$ can exchange messages with parties, read their outgoing messages, and deliver these messages to other parties. Upon corrupting a party, $\adv$ gains access to all its internal states and controls its subsequent actions. Meanwhile, the ideal functionality $\funb$ executes with an ideal world adversary (simulator) $\simulator$ and $\env$ via a set of \textit{dummy parties}. These dummy parties relay messages between $\env$ and $\funb$. $\simulator$ can exchange backdoors with $\funb$ and is tasked with delivering $\funb$'s outgoing messages to the dummy parties. When $\simulator$ corrupts a party, both $\env$ and $\funb$ are notified.

To demonstrate indistinguishability, we show that for any adversary $\adv$, we can construct a simulator $\simulator$ such that the environment $\env$ cannot distinguish between the output distributions of the ideal world and the hybrid world executions. Since $\registry$ is implemented as an on-chain contract with public and immutable internal states, we need only consider the following cases:
\begin{itemize}[leftmargin=*]
    \item All parties are uncorrupted,
    \item The holder $\holder$ is corrupted,
    \item The issuer $\issuer$ and the verifier $\verifier$ are corrupted, or
    \item All parties except $\registry$ are corrupted.
\end{itemize}
Here, we give a proof sketch containing the construction of the simulator $\simulator$ for all corruption cases and the corresponding description.

\spar{All uncorrupted}
In this case, it suffices for the simulator $\simulator$ to generate the transcript of all messages of the execution of $\prob$ towards the adversary $\adv$ and thus $\env$.
\begin{enumerate}[parsep=0.5pt]
    \item If an uncorrupted party $\party$ (either $\issuer$ or $\holder$) is activated by $(\mathtt{register},\allowbreak \sid)$, $\simulator$ obtains $\id$ from $\funb$. $\simulator$ samples $\sk^*\allowbreak \sample\allowbreak \Zp$, computes $\pk^*=\pubkeygen(\sk^*)$ and $h^*=H_a(\id,\allowbreak\sk^*)$. Then, $\simulator$ sends $(\mathtt{register},\sid,\id,\pk^*)$ and $(\mathtt{record},\sid,\allowbreak h^*)$ to $\funl$, and records the returned $(\id,\pk^*,\sk^*)$ and $(\id,h^*,\allowbreak r^*,\allowbreak \rho^*)$.
    \item If an uncorrupted party $\party$ is activated by $(\mathtt{associate},\sid,\allowbreak  \{\id_i\}_\ell)$, $\simulator$ obtains $\aid$ from $\funb$. $\simulator$ retrieves $\sk_i^*\sample\Zp$ for all $i\in[\ell]$, and computes $n_i^*:=H_n(\id_i,\sk_i^*)$. Finally, $\simulator$ sends $(\mathtt{record},\sid,\aid)$ and $(\mathtt{invalidate},\sid,n_i^*)$ to $\funl$, and records the returned $(\aid,r_a,\rho_a)$, along with $(\aid,\id_1,\dots,\allowbreak \id_\ell,\allowbreak 0)$.
          \item If an uncorrupted party $\party$ is activated by an association update (i.e., $\mathtt{append}$, $\mathtt{merge}$, or $\mathtt{refresh}$), $\simulator$ obtains the updated $\aid'$ from $\funb$. $\simulator$ retrieves dummy nonces and keys as needed to compute dummy nullifiers $n^*$. $\simulator$ generates the simulated zero-knowledge proof ($\pi_p^*, \pi_m^*$, or $\pi_d^*$) without the private witness (relying on the zero-knowledge property). Finally, $\simulator$ simulates the ledger interactions by sending $(\mathtt{record},\sid,\aid')$ and the corresponding $(\mathtt{invalidate},\sid,\allowbreak n^*)$ messages to $\funl$, recording the returned tree roots and nullifiers.
    \item If an uncorrupted holder $\holder$ is activated by $(\mathtt{present},\allowbreak \sid,\allowbreak \aid,\allowbreak \{\cred_i\}_n)$, $\simulator$ obtains $\{\idi_i\}_n$ from $\funb$. $\simulator$ sends $(\mathtt{retrieve},\allowbreak \sid,\allowbreak \idi_i)$ to $\funl$, receiving $\pki_i$ for all $i\in[n]$. Then, $\simulator$ generates dummy public nullifiers and constructs the simulated proof $\pi_c^*$ without the witnesses to include in the transcript. Finally, $\simulator$ simulates $\verifier$'s verification by sending the corresponding $(\mathtt{check},\allowbreak \sid,\allowbreak \cdot)$ requests to $\funl$ for the roots and nullifiers.
    \item If an uncorrupted party $\party$ is activated by $(\mathtt{recover},\allowbreak \sid,\aid,\allowbreak\id,\{\id'\}_t,\{\sk'\}_t)$, $\simulator$ constructs the simulated proof $\pi_k^*$ and sends it to $\registry$ in the transcript. Then, $\simulator$ samples $\sk^*\sample\Zp$, computes $\pk^*=\pubkeygen(\sk^*)$, $h^*=H_a(\id,\sk^*)$ and $n^*=H_n(\id,\sk^*)$. Finally, $\simulator$ sends $(\mathtt{register},\sid,\id,\pk^*)$, $(\mathtt{record},\sid,h^*)$ and $(\mathtt{invalidate},\sid,n^*)$ to $\funl$, and records the returned $(\id,\pk^*,\allowbreak\sk^*)$ and $(\id,h^*,r^*,\rho^*)$.
\end{enumerate}
Execution of $\funb$ in the ideal world is indistinguishable from the execution of $\prob$ in the $\funl$-hybrid world given the collision-resistance of $H_a, H_n$ and the hash primitive of the Merkle tree.

\spar{Corrupted holder}
In this case, the simulator $\simulator$ needs to generate the transcript of all messages of the execution of $\prob$, along with the outputs of the corrupted holder $\holder$ towards $\funb$ and $\env$. This indicates that whenever $\env$ activates $\holder$, its message will be forwarded to $\simulator$.
\begin{enumerate}[parsep=0.5pt]
    \item If a corrupted holder $\holder$ is activated by $(\mathtt{register},\sid)$, $\simulator$ learns $(\id,\sk,\pk)$ from $\holder$ and check if $\pk=\pubkeygen(\sk)$. If this is the case, $\simulator$ computes $h:=H_a(\id,\sk)$ and sends $(\mathtt{register},\sid,\id,\pk)$ and $(\mathtt{record},\sid,h)$ to $\funl$. The simulator also forwards the extracted registration state to \(\funb\), so that the ideal functionality records \((\holder,\id,\pk,\sk)\) consistently with the ledger transcript.
    \item If a corrupted holder $\holder$ is activated by $(\mathtt{associate},\sid,\allowbreak\{\id_i\}_\ell)$, $\simulator$ computes $n_i:=H_n(\id_i,\sk_i)$, and sends $(\mathtt{record},\allowbreak\sid,\aid)$ and $(\mathtt{invalidate},\sid,n_i)$ to $\funl$.
          \item If a corrupted holder $\holder$ initiates an association update, $\simulator$ intercepts the zero-knowledge proof ($\pi_p, \pi_m$, or $\pi_d$) sent by $\adv$ to $\registry$. By utilizing the knowledge extractor of the zkSNARK (Definition~\ref{def: zk-soundness}), $\simulator$ extracts the required witnesses (e.g., the new identifier $\id_{\ell+1}$ or nonces). $\simulator$ then sends the corresponding $(\mathtt{append}, \dots)$, $(\mathtt{merge}, \dots)$, or $(\mathtt{refresh}, \dots)$ command to $\funb$ on behalf of $\holder$ and simulates the $\funl$ state changes.
    \item If a corrupted holder $\holder$ participates in a campaign, $\adv$ submits a zero-knowledge proof $\pi_c$ to $\verifier$. $\simulator$ uses the knowledge extractor of the zkSNARK to extract the underlying $\aid$ and credentials $\{\cred_i\}_n$ from $\pi_c$. $\simulator$ then sends $(\mathtt{present},\sid,\aid,\allowbreak \{\cred_i\}_n)$ to $\funb$ on behalf of the corrupted $\holder$ to synchronize the ideal state.
    \item If a corrupted holder $\holder$ initiates a key recovery, $\simulator$ uses the knowledge extractor on the proof $\pi_k$ (or $\pi_k'$) to extract the target $\id$, the threshold identifiers $\{\id'\}_t$, and their keys $\{\sk'\}_t$. $\simulator$ then sends $(\mathtt{recover},\sid,\aid,\id,\{\id'\}_t,\{\sk'\}_t)$ to $\funb$ on behalf of $\holder$.
\end{enumerate}

\spar{Corrupted issuer and verifier}
Then, we consider the case when the issuer $\issuer$ and the verifier $\verifier$ are corrupted. In this case, the simulator $\simulator$ needs to generate the transcript of all messages in $\prob$, and the outputs of the corrupted $\issuer$ and $\verifier$.
\begin{enumerate}[parsep=0.5pt]
    \item If a holder $\holder$ is activated by $(\mathtt{register},\sid)$, $\simulator$ obtains $\id$ from $\funb$. $\simulator$ samples $\sk^*\sample\Zp$, computes $\pk^*=\pubkeygen(\sk^*)$ and $h^*=H_a(\id,\sk^*)$. Then, $\simulator$ sends $(\mathtt{register},\allowbreak \sid,\allowbreak \id,\allowbreak \pk^*)$ and $(\mathtt{record},\allowbreak \sid,\allowbreak h^*)$ to $\funl$, and records the returned $(\id,\allowbreak \pk^*,\allowbreak \sk^*)$ and $(\id,\allowbreak h^*,r^*,\rho^*)$.
    \item If a holder $\holder$ is activated by $(\mathtt{associate},\allowbreak \sid, \{\id_i\}_\ell)$, $\simulator$ obtains $\aid$ from $\funb$. $\simulator$ retrieves $\sk_i^*\sample\Zp$ for all $i\in[\ell]$, and computes $n_i^*:=H_n(\id_i,\sk_i^*)$. Finally, $\simulator$ sends $(\mathtt{record},\allowbreak \sid,\allowbreak \aid)$ and $(\mathtt{invalidate},\allowbreak \sid,\allowbreak n_i^*)$ to $\funl$, and records the returned $(\aid,r_a,\rho_a)$, along with $(\aid,\id_1,\dots,\id_\ell,0)$.
          \item If an uncorrupted holder $\holder$ is activated by an association update, $\simulator$ obtains the new $\aid'$ from $\funb$. $\simulator$ generates the simulated proof ($\pi_p^*, \pi_m^*$, or $\pi_d^*$) without knowing the holder's private witness, and sends the corresponding ledger update requests to $\funl$ in the name of the honest $\holder$, recording the updated states.
    \item If a corrupted verifier $\verifier$ is activated by $(\mathtt{campaign},\allowbreak\sid,\phi)$, $\simulator$ waits for a $\mathtt{present}$ message from $\holder$. Since $\holder$ is uncorrupted, $\simulator$ does not possess the private keys $\skh_i$. Instead, $\simulator$ invokes the zero-knowledge simulator of the zkSNARK to generate a simulated proof $\pi_c^*$ without the witness. $\simulator$ then sends this simulated proof to $\verifier$ on behalf of the honest $\holder$, allowing the corrupted $\verifier$ to naturally query $\funl$ in the hybrid world.
    \item If a holder $\holder$ is activated by $(\mathtt{recover},\allowbreak \sid,\aid,\id,\{\id'\}_t,\allowbreak \{\sk'\}_t)$, $\simulator$ constructs the simulated proof $\pi_k^*$ and sends it to $\registry$ in the name of $\adv$. Then, $\simulator$ samples $\sk^*\sample\Zp$, computes $\pk^*=\pubkeygen(\sk^*)$, $h^*=H_a(\id,\sk^*)$ and $n^*=H_n(\id,\sk^*)$. Finally, $\simulator$ sends $(\mathtt{register},\allowbreak \sid,\allowbreak \id,\allowbreak \pk^*)$, $(\mathtt{record},\sid,h^*)$ and $(\mathtt{invalidate},\allowbreak \sid,\allowbreak n^*)$ to $\funl$, and records the returned $(\id,\allowbreak \pk^*,\allowbreak \sk^*)$ and $(\id,\allowbreak h^*,\allowbreak r^*,\allowbreak \rho^*)$.
\end{enumerate}
Note that in this case, the behavior of the uncorrupted holder mirrors that of the first case.

\spar{All corrupted}
In this case, the simulation of $\simulator$ is a mixture of the above two scenarios. $\simulator$ should simulate the interacting of $\adv$ and $\funl$, along with the messages between the parties.